\documentclass[12pt,a4paper]{article}
\usepackage[pdfborder={0 0 0}]{hyperref}
\usepackage{multirow}
\usepackage{amsmath}
\usepackage{xspace}
\usepackage{graphicx}
\usepackage{feynmf}
\makeatletter
\newif\if@preliminary
\@preliminaryfalse
\def\preliminary{\@preliminarytrue}

\def\preprintno#1{\def\@preprintno{#1}}
\def\address#1{\def\@address{#1}}
\def\email#1#2{\thanks{\tt #1@{}#2}}
\def\abstract#1{\def\@abstract{#1}}
\renewcommand\abstractname{ABSTRACT}
\newlength\preprintnoskip
\newlength\abstractwidth
\@titlepagetrue
\renewcommand\maketitle{\begin{titlepage}%
  \let\footnotesize\small
  \hfill\parbox{\preprintnoskip}{%
  \begin{flushright}\@preprintno\end{flushright}}\hspace*{1cm}
  \vskip 60\p@
  \begin{center}%
    {\Large\bf\boldmath \@title \par}\vskip 1cm%
    {\sc\@author \par}\vskip 3mm%
    {\@address \par}%
    \if@preliminary
      \vskip 2cm {\large\sf PRELIMINARY DRAFT \par \@date}%
    \fi
  \end{center}\par
  \@thanks
  \vfill
  \begin{center}%
    \parbox{\abstractwidth}{\centerline{\abstractname}%
    \vskip 3mm%
    \@abstract}
  \end{center}
  \end{titlepage}%
  \setcounter{footnote}{0}%
  \let\thanks\relax\let\maketitle\relax
  \gdef\@thanks{}\gdef\@author{}\gdef\@address{}%
  \gdef\@title{}\gdef\@abstract{}\gdef\@preprintno{}
}%
\def\@citex[#1]#2{\if@filesw\immediate\write\@auxout{\string\citation{#2}}\fi
  \def\@citea{}\@cite{\@for\@citeb:=#2\do
    {\@citea\def\@citea{,\penalty\@m}\@ifundefined
       {b@\@citeb}{{\bf ?}\@warning
       {Citation `\@citeb' on page \thepage \space undefined}}%
\hbox{\csname b@\@citeb\endcsname}}}{#1}}
\def\citerange{\@ifnextchar [{\@tempswatrue\@citexr}{\@tempswafalse\@citexr[]}}
\def\@citexr[#1]#2{\if@filesw\immediate\write\@auxout{\string\citation{#2}}\fi
  \def\@citea{}\@cite{\@for\@citeb:=#2\do
    {\@citea\def\@citea{--\penalty\@m}\@ifundefined
       {b@\@citeb}{{\bf ?}\@warning
       {Citation `\@citeb' on page \thepage \space undefined}}%
\hbox{\csname b@\@citeb\endcsname}}}{#1}}
\long\def\@makecaption#1#2{%
  \vskip\abovecaptionskip
  \sbox\@tempboxa{#1: \emph{#2}}%
  \ifdim \wd\@tempboxa >\hsize
    #1: \emph{#2}\par
  \else
    \hbox to\hsize{\hfil\box\@tempboxa\hfil}%
  \fi
  \vskip\belowcaptionskip}
\def\fmslash{\@ifnextchar[{\fmsl@sh}{\fmsl@sh[0mu]}}
\def\fmsl@sh[#1]#2{%
  \mathchoice
    {\@fmsl@sh\displaystyle{#1}{#2}}%
    {\@fmsl@sh\textstyle{#1}{#2}}%
    {\@fmsl@sh\scriptstyle{#1}{#2}}%
    {\@fmsl@sh\scriptscriptstyle{#1}{#2}}}
\def\@fmsl@sh#1#2#3{\m@th\ooalign{$\hfil#1\mkern#2/\hfil$\crcr$#1#3$}}
\makeatother

\newcommand{\sherpa}{S\scalebox{0.8}{HERPA}\xspace}
\newcommand{\order}{\mathcal{O}}		
\newcommand{\dsl}{\!\!\not\!}			

\newcommand{\vp}{\vphantom{\frac{|}{|}}}
\newcommand{\vP}{\vphantom{\frac{|^|}{|^|}}}
\newcommand{\VP}{\vphantom{\int\limits_a^b\frac{|^|}{|^|}}}
\newcommand{\done}{{\rm d}}

\newcommand{\keV}{{\ensuremath\rm keV}}
\newcommand{\MeV}{{\ensuremath\rm MeV}}
\newcommand{\GeV}{{\ensuremath\rm GeV}}

\newcommand{\m}{{\ensuremath\rm m}}

\newcommand{\Sherpa}{S\scalebox{0.8}{HERPA}\xspace}

\newcommand{\PythiaEight}{P\scalebox{0.8}{YTHIA8}\xspace} 
 
\newcommand{\Herwigpp}{H\scalebox{0.8}{ERWIG++}\xspace} 
\newcommand{\Photos}{P\scalebox{0.8}{HOTOS}\xspace}  
\newcommand{\Photonspp}{P\scalebox{0.8}{HOTONS++}\xspace}  
\newcommand{\Amegicpp}{A\scalebox{0.8}{MEGIC++}\xspace}  
\newcommand{\Sophty}{S\scalebox{0.8}{OPHTY}\xspace}  
\newcommand{\Windec}{W\scalebox{0.8}{INDEC}\xspace}  
\newcommand{\Horace}{H\scalebox{0.8}{ORACE}\xspace}

\newcommand{\bt}{\begin{tabular}}
\newcommand{\et}{\end{tabular}}
\def\be{\begin{equation}}
\def\ee{\end{equation}}
\def\bc{\begin{center}}
\def\ec{\end{center}}
\newcommand{\nnb}{\nonumber}
\newcommand{\bea}{\begin{eqnarray}}
\newcommand{\eea}{\end{eqnarray}}
\newcommand{\bit}{\begin{itemize}}
\newcommand{\eit}{\end{itemize}}
\newcommand{\balclcl}{\begin{array}{lclcl}}
\newcommand{\ba}{\begin{array}}
\newcommand{\ea}{\end{array}}

\begin{document}

\title{Soft Photon Radiation in Particle Decays in \protect\Sherpa}
\preprintno{DCPT/07/96\\ IPPP/07/48\\ MCNET 08/15}

\author{%
 M.~Sch{\"o}nherr\email{marek.schoenherr}{tu-dresden.de}$^a$,
 F.~Krauss\email{frank.krauss}{durham.ac.uk}$^b$
}
\address{\it%
$^a$Institut f\"ur Kern- und Teilchenphysik, TU Dresden, D--01062 Dresden, Germany\\
$^b$Institute for Particle Physics Phenomenology, Durham University, Durham DH1 3LE, UK
\\[.5\baselineskip]
}
\abstract{
In this paper the Yennie--\-Frautschi--\-Suura approach is used to simulate real and 
virtual QED corrections in particle decays.  It makes use of the universal structure 
of soft photon corrections to resum the leading logarithmic QED corrections to all 
orders, and it allows a systematic correction of this approximate result to exact 
fixed order results from perturbation theory.  The approach has been implemented as a 
Monte Carlo algorithm, which a posteriori modifies decay matrix elements through the 
emission of varying numbers of photons.  The corresponding computer code is 
incorporated into the \Sherpa event generator framework.}
\maketitle

\tableofcontents
\newpage



\section{Introduction}

The next round of collider-based experiments in particle physics will 
start with the LHC finally becoming fully operational, and producing
particle collision data at unprecedented rate and energy.  In the
preparation for this huge enterprise, a new generation of Monte Carlo
simulation tools, like \PythiaEight \cite{Sjostrand:2007gs}, 
\Herwigpp \cite{Bahr:2008pv} and \Sherpa \cite{Gleisberg:2003xi} has 
been constructed, to meet the increasingly complicated experimental 
situation and the demand for an improved description of data at a higher 
level of accuracy.  In addition, it was anticipated that by moving to 
the new programming paradigm of object orientation and modularization 
the more mundane software management task of code validation and 
maintenance could be addressed in a more transparent and alleviated way.
This leads to the typical strategy of event generators, to dissect the 
simulation of full events into different physics aspects, being better 
reflected in the modular structure of the emerging new codes.

In this paper the construction of a new physics module for the \Sherpa 
framework is discussed, which deals with the simulation of QED radiation in 
particle decays.  Up to now, this has typically been left to the \Photos
\cite{Barberio:1990ms,Nanava:2006vv} programme.  However, there have been 
two reasons for replacing \Photos:  First of all, \Photos builds on a 
parton-shower like collinear approximation for the simulation of photon 
emissions, which intrinsically has some shortcomings when the mass of the 
decaying particle is not much larger than the masses of its decay products.  
This has already been noted in \cite{Hamilton:2006xz,Was:2008zu}
and triggered the development of the module \Sophty \cite{Hamilton:2006xz} 
in the framework of the \Herwigpp event generator.  It also has become a 
wide-spread belief among the authors of the main event generators that the 
maintenance of interface structures to additional codes such as \Photos, 
supplementing QED radiation to the simulation, overwhelms the burden of 
constructing and maintaining corresponding modules directly in the event 
generators.  

Similar to the case of \Sophty in \Herwigpp, the construction of the new 
module, \Photonspp, in \Sherpa bases on the approach of Yennie, Frautschi
and Suura (YFS) \cite{Yennie:1961ad} for the calculation of higher order 
QED corrections to arbitrary processes.  This approach resides on the idea 
of resumming the leading soft logarithms to all orders, rather than 
focusing on the leading collinear terms.  These soft logarithms are largely 
independent of the inner process characteristics and can be calculated 
from the external particles and their four-momenta only.  The big advantage 
of the YFS formalism is that in addition it allows for a systematic 
improvement of this eikonal approximation, order-by-order in the QED 
coupling constant.  This explains why a good fraction of the most precise 
tools for the simulation of QED radiation root in this algorithm 
\cite{Jadach:1999vf,Jadach:2000eu,Jadach:2000kw,Jadach:2001mp}.  

In contrast to the \Sophty implementation the aim of this implementation
is to address also decays with more than two final state particles.
This leads to different strategies of enforcing four-momentum conservation
after the soft photons are reconstructed.  In addition, some correction 
terms to restore precise results for the first order in the electromagnetic 
coupling constant are employed, improving the formal accuracy of the results 
of \Photonspp.  This also has not been included in \Sophty.  

The outline of this paper is as follows: After briefly reviewing the YFS formalism 
in Sec.~\ref{Section_YFS_Exponentiation} in the framework of particle decays, the 
Monte Carlo algorithm adopted here is detailed in Sec.~\ref{Sec:Algorithm}.  Then, 
some higher order corrections are discussed in Sec.~\ref{Sec:Higher_Order}.  
Finally, in Sec.~\ref{Sec:Results} the new code is validated through a detailed
comparison with \Horace and \Windec \cite{CarloniCalame:2004qw} for the 
case of leptonic $Z$ and $W$ decays before some results relating to other 
particle decays are presented.

\section{YFS-Exponentiation}\label{Section_YFS_Exponentiation}

In this section, the YFS approach \cite{Yennie:1961ad} for an approximative 
description of real and virtual QED corrections to arbitrary scattering or 
decay processes will shortly be reviewed, in the framework of particle decays.  
The virtue of this formalism is that it can systematically be improved, order 
by order in the electromagnetic coupling constant $\alpha$.  The YFS approach 
bases on the observation that the soft limits for matrix elements with real 
and/or virtual photons exhibit a universal behaviour, and on the fact that 
the corresponding soft divergences can be factorised into universal factors 
multiplying leading order matrix elements.  

When defining the final state as a configuration of primary decay products 
with momenta $p_f$ and any number of additional soft photons with momenta
$k$ the fully inclusive decay rate reads
\begin{eqnarray}
\Gamma = \frac{1}{2M}\!\sum_{n_R = 0}^\infty\frac{1}{n_R !}
\int\done\Phi_p\,\done\Phi_k\,
(2\pi)^4\delta\left(\sum p_i-\sum p_f-\sum k\right)
\left|\sum_{n_V=0}^\infty\mathcal{M}_{n_R}^{n_V+\frac{1}{2}n_R}\right|^2
\end{eqnarray}
where $p_i$ is the four-momentum of the decaying particle.  Here and in the 
following $n_V$ and $n_R$ are the numbers of additional virtual and real photons, 
respectively, that show up in the higher-order matrix element but not in the 
uncorrected zeroth order matrix element (thus labelled by $\mathcal{M}_0^0$).  
The starting point of the YFS algorithm is to approximate these dressed matrix 
elements through the zeroth order one times eikonal factors, which depend on 
the external particles only.  This effectively catches the leading logarithmic
QED corrections to the process.  The correct result can then be restored order by 
order in perturbation theory by supplementing the non-leading, process-dependent 
pieces.

In the case of one virtual photon this can be formalised as
\bea\label{Eq:virtual_factorisation}
 \mathcal{M}_0^1 = \alpha B \mathcal{M}_0^0 + M_0^1\,,
\eea
where $M_0^1$ is the infrared-subtracted matrix element including one virtual 
photon (with $M_0^1$ finite when $k\to 0$ due to the subtraction).  All soft 
divergences due to this virtual photon are contained in the process-independent, 
universal factor $B$, see Appendix \ref{Appendix_YFS} for a more thorough
discussion.  Here, and in the following, the sub- and superscripts denote
the number of real photons and the order of $\alpha$, respectively, both 
for the infrared-subtracted and for the original matrix elements.  

Similar to the one-photon case, YFS showed that the subsequent insertion 
of further virtual photons in all possible ways leads to
\bea
\mathcal{M}_0^0 & = & M_0^0 \nnb\\
\mathcal{M}_0^1 & = & \alpha B M_0^0 + M_0^1 \nnb\\
\mathcal{M}_0^2 & = & \frac{(\alpha B)^2}{2!}M_0^0+\alpha B M_0^1 + M_0^2
\eea
and so on.  Therefore, for a fixed order in $\alpha$,
\bea
\mathcal{M}_0^{n_V} = \sum_{r=0}^{n_V}M_0^{n_V-r}\frac{(\alpha B)^r}{r!}
\eea
and, summing over all numbers of virtual photons $n_V$, 
\bea
\sum_{n_V=0}^\infty\mathcal{M}_0^{n_V}  = 
\exp(\alpha B)\sum_{n_V=0}^\infty M_0^{n_V}\,.
\eea
Since photons do not carry any charge and because virtual photons inserted in 
closed charged loops do not produce any additional infrared singularity\footnote{
	A similar program cannot directly be translated to QCD, where the emitted 
	gluons act as parts of antennae emitting further gluons, thus modifying 
	the pattern of possible infrared poles and thus leading logarithms in each 
	emission. 
}, 
this can be generalised to any number of real photons, such that
\bea
\left|\sum_{n_V=0}^\infty\mathcal{M}_{n_R}^{n_V+\frac{1}{2}n_R}\right|^2 & = &
\exp(2\alpha B)\left|\sum_{n_V=0}^\infty 
	M_{n_R}^{n_V+\frac{1}{2}n_R}\right|^2\,.
\eea
Hence, $M_{n_R}^{n_V+\frac{1}{2}n_R}$ is free of soft singularities due to virtual 
photons but it still may contain those due to real photons.

YFS showed in \cite{Yennie:1961ad} that the factorisation for real photon emission 
proceeds on the level of the squared matrix elements rather than on the amplitude 
level.  For a single photon emission therefore this yields
\bea
\frac{1}{2(2\pi)^3}\left|\sum_{n_V=0}^\infty 
              M_1^{n_V+\frac{1}{2}}\right|^2 
& = & 
\tilde{S}(k)\left|\sum_{n_V=0}^\infty M_0^{n_V}\right|^2 + 
\sum_{n_V=0}^\infty\tilde{\beta}_{1}^{n_V+1}(k)\,. 
\eea
Here, $\tilde{S}(k)$ is an eikonal factor containing the soft divergence
related to the real photon emission, see Appendix \ref{Appendix_YFS}.  
Denoting with $\tilde{\beta}_{n_R}^{n_V+n_R}$ the complete IR-finite 
(subtracted) squared matrix element for the basic process plus the emission 
of $n_R$ photons including $n_V$ virtual photons and using the abbreviation 
\bea
\tilde{\beta}_{n_R} = \sum_{n_V=0}^\infty\tilde{\beta}_{n_R}^{n_V+n_R}\,,
\eea
the squared matrix element for $n_R$ real emissions, summed over all
possible virtual photon corrections, can be written as
\bea\label{n_R_real_emissions}
\lefteqn{\left(\frac{1}{2(2\pi)^3}\right)^{n_R}
\left|\sum_{n_V=0}^\infty M_{n_R}^{n_V+\frac{1}{2}n_R}\right|^2}\nnb\\
& = & \tilde{\beta}_0 \prod_{i=1}^{n_R}\left[\vp\tilde{S}(k_i)\right]
    +\sum_{i=1}^{n_R}\left[\frac{\tilde{\beta}_1(k_i)}{\tilde{S}(k_i)}\right] 
	\prod_{j=1}^{n_R}\left[\vp\tilde{S}(k_j)\right]
    +\sum_{\genfrac{}{}{0pt}{}{i,j=1}{i\neq j}}^{n_R}\left[
	\frac{\tilde{\beta}_2(k_i,k_j) }{\tilde{S}(k_i)\tilde{S}(k_j)}\right]
	\prod_{l=1}^{n_R}\left[\vp\tilde{S}(k_l)\right]
    +\dots \nnb\\
&& {}+\sum_{i=1}^{n_R}\left[\vp
	\tilde{\beta}_{n_R-1}(k_1,\dots,k_{i-1},k_{i+1},\dots,k_{n_R})\,
	\tilde{S}(k_i)\right]
   +\tilde{\beta}_{n_R}(k_1,\dots,k_{n_R})\,.
\eea
Demanding agreement with the exact result up to $\mathcal{O}(\alpha)$, this 
expression thus contains only terms with $\tilde{\beta}_0^0$, 
$\tilde{\beta}_0^1$ and $\tilde{\beta}_1^1$.  Then
\bea
\lefteqn{\left(\frac{1}{2(2\pi)^3}\right)^{n_R}
	\left|\sum_{n_V=0}^\infty M_{n_R}^{n_V+\frac{1}{2}n_R}\right|^2}\nnb\\ 
& = & 
\left[\vP\tilde{\beta}_0^0+\tilde{\beta}_0^1\right]
	\prod_{i=1}^{n_R}\left[\vP\tilde{S}(k_i)\right]
+ \sum_{i=1}^{n_R}\left[
	\frac{\tilde{\beta_1^1}(k_i)}{\tilde{S}(k_i)}\right]
	\prod_{j=1}^{n_R}\left[\vP\tilde{S}(k_j)\hspace{2mm}\right]
+ \mathcal{O}(\alpha^2)\,.
\eea
Inserting this into the expression for the decay rate and expressing the
$\delta$-functions ensuring four-momentum conservation as exponentials yields,
\bea
2M\cdot\Gamma 
& = & \hspace{5.7mm}
\sum_{n_R}\frac{1}{n_R!}\int\done\Phi_{p_f}\left\{\VP\hspace{1mm}
	\exp\left[2\alpha B\right]
\int\done y\hspace{1mm}\exp\left[iy\left(\sum p_i-\sum p_f\right)\right]
\right.\nnb\\
&&\hspace{36mm}\left.\VP\times
\left(\int\frac{\done^3k}{k}\tilde{S}(k)e^{-iyk}\right)^{n_R}
\left(\tilde{\beta}_0^0+\tilde{\beta}_0^1\right)\right\} \nnb\\
&&{} + \sum_{n_R-1}\frac{1}{(n_R-1)!}\int\done\Phi_{p_f}\left\{\VP\hspace{1mm}
	\exp\left[2\alpha B\right]
	\int\done y\hspace{1mm}\frac{\done^3K}{K}\hspace{1mm}
	\exp\left[iy\left(\sum p_i-\sum p_f-K\right)\right]\right.\nnb\\
&&\hspace{50mm}\left.\VP\times
\left(\int\frac{\done^3k}{k}\tilde{S}(k)e^{-iyk}\right)^{n_R-1}
	\tilde{\beta}_1^1(K)\right\}
\hspace{3mm}+\hspace{3mm} \mathcal{O}(\alpha^2)\nnb\\
& = & \int \done^4y\int \done\Phi_{p_f}\left\{\VP\hspace{1mm}
	\exp\left[2\alpha B\right]
	\exp\left[iy\left(\sum p_i-\sum p_f\right)+
		\int\frac{\done^3k}{k}\tilde{S}(k)e^{-iyk}\right]\right. \nnb\\
&&\hspace{33mm}\left.\VP\times
\left[\tilde{\beta}_0^0+\tilde{\beta}_0^1
      +\int\frac{\done^3K}{K}\hspace{1mm}e^{-iyK}
		\hspace{1mm}\tilde{\beta}_1^1(K)
\hspace{3mm}+\hspace{3mm} \mathcal{O}(\alpha^2)\hspace{3mm}\right]\right\}\,.
\eea
As before, all singularities due to virtual photons are contained in $B$, 
while all singularities due to real emissions are incorporated in the 
integral over $\tilde{S}(k)$.  To restore the momentum conserving 
$\delta$-function the divergences have to be split off this integral.  This 
can be done by simply subtracting the terms that are divergent for $k\to 0$. 
To this end, a small ``soft'' region $\Omega$ is defined together with an
infrared-safe function $D(\Omega)$\footnote{
	Obviously $\Theta(k,\Omega)$ divides the phase space into two 
	regions.  While $\Omega$ comprises the region containing the 
	infrared divergence, $(1-\Omega)$ is completely free of those 
	divergences.  Hence, $\Theta(k,\Omega)=1$ if $k \notin \Omega$ and 
	zero otherwise.  Thus, $D(\Omega)$ is IR save and $\tilde{B}(\Omega)$ 
	contains the divergence.  
}, such that
\bea
\lefteqn{\int\frac{\done^3k}{k}\hspace{1mm}\tilde{S}(k)e^{-iyk}}\nnb\\
&=&\int\frac{\done^3k}{k}\hspace{1mm}\left\{\tilde{S}(k)
   \left[\left(\vp 1-\Theta(k,\,\Omega)\right)+e^{-iyk}\Theta(k,\,\Omega)
        +\left(\vp e^{-iyk}-1\right)\left(\vp 1-\Theta(k,\,\Omega)\right)
   \right]\right\}
\nnb\\
&=&  
2\alpha\tilde{B}(\Omega)+D(\Omega)
\eea
where 
\bea
D(\Omega) &=&
\int\frac{\done^3k}{k}\hspace{1mm}\tilde{S}(k)
	\left[\left(\vp e^{-iyk}-1\right)\left(\vp 1-\Theta(k,\Omega)\right) 
              +e^{-iyk}\Theta(k,\Omega)\right]\nnb\\
&\stackrel{\Omega\to 0}{\longrightarrow}&
\int\frac{\done^3k}{k}\hspace{1mm}\tilde{S}(k)\,e^{-iyk}\,\Theta(k,\Omega)
\eea
and
\bea\label{Eq:def_B_tilde}
2\alpha\tilde{B}(\Omega) & = & 
\int\frac{\done^3k}{k}\hspace{1mm}\tilde{S}(k)\left(1-\Theta(k,\Omega)\right)
=
\int\limits_\Omega\frac{\done^3k}{k}\hspace{1mm}\tilde{S}(k)\,.
\eea
Reinserting this into the cross section, executing the $y$-integration and 
reexpanding the exponen\-tiated integral yields
\bea\label{Eq:exact_distribution}
2M\;\Gamma
& = & 
\sum_{n_R}\frac{1}{n_R!}\int \done\Phi_{p_f}\done\Phi_k^\prime(2\pi)^4
	\delta^4\left(\sum p_i-\sum p_f-\sum k\right)
	e^{2\alpha (B+\tilde{B}(\Omega))} \nnb\\
&& \quad\quad\quad\quad
\times\prod_{i=1}^{n_R}\tilde{S}(k_i)\Theta(k_i,\Omega)
\left(\tilde{\beta}_0^0+\tilde{\beta}_0^1+
	\sum_{i=1}^{n_R}\frac{\tilde{\beta}_1^1(k_i)}{\tilde{S}(k_i)}
	+\hspace{3mm} \mathcal{O}(\alpha^2)\hspace{3mm}\right)\,.
\eea
The whole factorisation is independent of possible spin correlations in the 
``undressed'' matrix element.  Thus, the same result is obtained if the spin-summed 
and averaged matrix element squared $\left|\mathcal{M}\right|^2$ is replaced by 
$\left.\rho_{\alpha\beta}\mathcal{M}^\alpha\mathcal{M}^\beta\right.^\ast$ 
where $\rho_{\alpha\beta}$ is a spin density matrix.

The infrared subtracted squared matrix elements read, up to $\order(\alpha)$,
\bea
\tilde{\beta}_{0}^{0} & = & \left.M_0^0M_0^0\right.^\ast \nnb\\
\tilde{\beta}_{0}^{1} & = & \left.M_0^0M_0^1\right.^\ast+
			    \left.M_0^1M_0^0\right.^\ast \nnb\\
\tilde{\beta}_{1}^{1} & = & \frac{1}{2(2\pi)^3}
			\left.M_1^\frac{1}{2}M_1^\frac{1}{2}\right.^\ast-
				\tilde{S}(k)\left.M_0^0M_0^0\right.^\ast
\eea
or
\bea
\tilde{\beta}_0^0 
& = & \rho_{\alpha\beta}\left.M_0^{0\alpha}M_0^{0\beta}\right.^\ast \nnb\\
\tilde{\beta}_0^1 
& = & \rho_{\alpha\beta}\left(\left.M_0^{0\alpha}M_0^{1\beta}\right.^\ast
	+\left.M_0^{1\alpha}M_0^{0\beta}\right.^\ast\right) \nnb\\
\tilde{\beta}_0^1 
& = & \rho_{\alpha\beta}\left(\left.\frac{1}{2(2\pi)^3}M_1^{\frac{1}{2}\alpha}
	M_1^{\frac{1}{2}\beta}\right.^\ast
	-\tilde{S}(k)\left.M_0^{0\alpha}M_0^{0\beta}\right.^\ast\right) \,.
\eea
\section{The Algorithm}\label{Sec:Algorithm}

\subsection{The master formula}\label{Sec:MasterFormula}

The basic, undressed matrix element (no additional photons) reads
\be
2M\cdot\Gamma_0 = 
\int\done\Phi_q\,(2\pi)^4\delta^4(p_C+p_N-Q_C-Q_N) \left|\mathcal{M}\right|^2
\ee
where the differential phase-space element for the outgoing momenta 
$q\in \{Q_C,\,Q_N\}$ is given by
\bea
\done\Phi_q = \prod_{i=1}^n \frac{\done^3q_i}{(2\pi^3)2q_i^0}\,.
\eea
Here, and in the following, the initial and final state momenta have been 
classified to whether the respective particles are charged or neutral: 
the sums of all initial state momenta are labelled by $p_{C,N}$ for
charged and neutral particles, respectively, while $Q_{C,N}$ denotes the sums 
of all charged or neutral final state momenta.  After QED corrections, the 
$Q_C$ and $Q_N$ will become $P_C$ and $P_N$, respectively.  $K$ is the sum 
of all additional real, resolved Bremsstrahlungs-photons generated in the 
process, whereas photons already present in the core process are included in 
$P_N$ and $Q_N$, respectively (an example for this seemingly unlikely case
would be the rare decay $B^+\to K^{*+}\gamma$).

In the previous section the factorisation of infrared divergent terms and
the construction of infrared-finite expressions for cross sections with
all possible numbers of resolved photons has been discussed.  In these 
expressions the universal, process-independent parts of the QED corrections 
have been separated and exponentiated, the residual process dependence and
the effect of particle spins etc.\ has been absorbed in infrared-finite,
subtracted terms $\tilde{\beta}$, cf.\ Eq.~(\ref{Eq:exact_distribution}).  
With small changes in the notation this form of the cross section thus reads
\bea\label{Eq:master_equation}
2M\cdot\Gamma & = & 
\sum_{n_\gamma}\frac{1}{n_\gamma!}\int\done\Phi\; e^{Y(\Omega)}
	\prod_{i=1}^{n_\gamma}\tilde{S}(k_i)\Theta(k_i,\Omega)\,
	\tilde{\beta}_0^0\;\mathcal{C}\,.
\eea
Here, the phase space has been separated into a phase space element for
the particles of the ``core'' process and one for the additional 
$n_\gamma$ resolved real photons,
\bea\label{Eq:BasicPSelement}
\done\Phi = 
\done\Phi_p\,\done\Phi_k\,(2\pi)^4\delta\left(p_C+p_N-P_C-P_N-K\right).
\eea
with
\bea\label{Eq:BasicPSelementsPandK}
\done\Phi_p & = & \prod_{i=1}^n \frac{\done^3p_i}{(2\pi)^32p_i^0} \\
\done\Phi_k & = & \prod_{i=1}^{n_\gamma}\frac{\done^3k}{k^0}\,.
\eea
Note that the factor $\frac{1}{2(2\pi)^3}$, missing in the photon phase 
space element, has already been incorporated in the eikonal factor 
$\tilde{S}(k)$, in accordance with the choice made in \cite{Yennie:1961ad}.
In the equation above, Eq.~(\ref{Eq:master_equation}), the undressed matrix 
element $\tilde{\beta}_0^0$ has been factored out and the remainder of the 
perturbative expansion in $\alpha$ has been combined in the factor 
$\mathcal{C}$,
\bea\label{def_higher_order}
\mathcal{C} = 
1+\frac{1}{\tilde{\beta}_0^0}\left(\tilde{\beta}_0^1
 +\sum_{i=1}^{n_\gamma}\frac{\tilde{\beta}_1^1(k_i)}{\tilde{S}(k_i)}
 +\order(\alpha^2)\right) \,.
\eea
Furthermore, the YFS-Form-Factor has been introduced
\bea\label{YFS_Form_Factor}
Y(\Omega) = \sum_{i<j} Y_{ij}(\Omega) = 
\sum_{i<j} 2\alpha\left(B_{ij}+\tilde{B}_{ij}(\Omega)\right)
\eea
where the sum $i<j$ runs over all pairs of charged particles, taking into account
each pair only once.  The infrared factors $B_{ij}$ and $\tilde{B}_{ij}$ are 
defined as
\bea
B_{ij} 
&=& 
-\frac{i}{8\pi^3}Z_iZ_j\theta_i\theta_j\int\done^4k\frac{1}{k^2}
	\left(\frac{2p_i\theta_i-k}{k^2-2(k\cdot p_i)\theta_i}+
	      \frac{2p_j\theta_j+k}{k^2+2(k\cdot p_j)\theta_j}\right)^2 \\
\tilde{B}_{ij}(\Omega) 
&=& 
\frac{1}{4\pi^2}Z_iZ_j\theta_i\theta_j \int d^4k\delta(k^2)
	\left(1-\Theta(k,\Omega)\right)\left(\frac{p_i}{p_i\cdot k}-
		                             \frac{p_j}{p_j\cdot k}\right)^2\,.
\eea
They are the generalisation of the quantities defined in the last section, 
cf.\ Eqs.~(\ref{Eq:virtual_factorisation}) and (\ref{Eq:def_B_tilde}).  
Both contain the virtual and real infrared divergences, respectively.  
These divergences cancel according to the Kinoshita-Lee-Nauenberg theorem
\cite{Kinoshita:1962ur,Lee:1964is}.  Thus, each $Y_{ij}(\Omega)$ is guaranteed
to be finite, which is explicitely shown in Appendix \ref{Appendix_YFS}.  In
the terms above, $Z_i$ and $Z_j$ are the charges of the particles $i$ and $j$ 
in terms of the positron charge $e$, and the signature factors $\theta=\pm 1$ 
for particles in the final or initial state, respectively.  The symbol 
$\Theta$, already defined at the end of section \ref{Section_YFS_Exponentiation}, 
refers to a phase space constraint with $\Omega$ denoting the soft,
unresolvable region of photon radiation.  Hence, $\Theta(k,\Omega)=1$ if 
$k \notin \Omega$ and zero otherwise.  If this division is done by defining 
an energy cut-off, the definition of $\Omega$ is not Lorentz-invariant and 
the frame in which this cut-off forms a flat hypersurface also needs to be 
specified.  The advantage of splitting the photon phase space in that manner 
lies in the alleviation of integrating $\tilde{S}(k)$ over $k$.  If the 
cut-off is defined in the frame the photon generation and momentum 
reconstruction will be done in\footnote{
	In the algorithm presented here, this will be the rest frame of 
	the multipole, i.e.\ the combined rest frame of all charged particles 
	$p_C+P_C$.} 
then the integration over the photon energy separates from the angular 
integration (see Appendix \ref{Appendix_Generation}), leading to yet another
simplification of the calculation.

The eikonal factor $\tilde{S}(k)$ has already been introduced in the last 
section. It is defined as
\bea
\tilde{S}(k) = \sum_{i<j} \tilde{S}_{ij}(k) = 
\frac{\alpha}{4\pi^2}\sum_{i<j}Z_iZ_j\theta_i\theta_j
	\left(\frac{p_i}{p_i\cdot k}-\frac{p_j}{p_j\cdot k}\right)^2\,.
\eea

However, despite all terms being finite in Eq.~(\ref{Eq:master_equation}), 
it cannot be used straight away for Monte Carlo generation.  This is 
because it is written in terms of the already corrected final state momenta 
$p_i$ and not the original undressed momenta $q_i$.  The problem here is that
the undressed momenta are defined in an $n$-body phase space whereas the 
dressed momenta are part of an $(n+n_\gamma)$-body phase space.  This 
neccessitates a mapping procedure of the $n$-body onto the 
$(n+n_\gamma)$-body phase space.  In principle, details of this mapping 
procedure are irrelevant as long as it respects the soft limit of photon
radiation not altering the original kinematics, i.e.\ in this limit the 
momenta of the orinial particles in the $(n+n_\gamma)$-body phase space have 
to fall exactly onto those of the $n$-body phase space.

\subsection{Phase space transformation}\label{Sec:PSTrafo}

To solve this, consider the rest frame of all charged particles involved in 
the basic matrix element
\bea
P_M = p_C + P_C\,.
\eea
These particles form the multipole responsible for the Bremsstrahlung of the
additional photons.  In the rest frame of this multipole, a simple form of the 
mapping consists of a mere rescaling of the three-momenta of all final state 
particles by a common factor $u$ such that the additional photons are
accomodated.  Clearly, the initial state momenta cannot be altered, because 
they have already been fixed when the basic matrix element was calculated. 
So, the task is to rewrite Eq.~(\ref{Eq:BasicPSelement}), explicitely in 
the rest frame of the multipole in question.  The neccessary transformations 
are detailed in the appendix, cf.\ App.~\ref{Sec:RestframeTrafo}, here it
suffices to give the result.  It reads
\bea
\done\Phi 
&=& 
\done\Phi_p\,\done\Phi_k \,
(2\pi)^4\delta\left(p_C+p_N-P_C-P_N-K\right)\nnb\\
&=&
\prod_{i=1}^n\left[\frac{\done^3p_i}{(2\pi)^32p_i^0}\right]\,
\prod_{i=1}^{n_\gamma}\left[\frac{\done^3k}{k^0}\right]\,
(2\pi)^4\delta\left(p_C+p_N-P_C-P_N-K\right)\nnb\\
&=&
\done\Phi_p\,\done\Phi_k \,\frac{m_{M,p}^3}{M^2(P_C^0+P_N^0+K^0)}\,
(2\pi)^3\delta^3(\vec{P}_M)\,(2\pi)\delta\left(P_M^0 - P_C^0 - p_C^0\right)\,.
\eea
In a similar fashion, the phase space related to the zeroth order uncorrected 
cross section can be transformed to
\bea\label{Eq:PS_trafo_zero}
\done \Phi_0 & = & 
(2\pi)^4\done\Phi_q\,\delta^4\left(p_C+p_N-Q_C-Q_N\right) \nnb\\
& = & 
\frac{m_{M,q}^3}{M^2(Q_C^0+Q_N^0)}\,\done\Phi_q\, 
(2\pi)^3\delta^3(\vec{Q}_M)\,(2\pi)\delta\left(Q_M^0-Q_C^0-p_C^0\right)\,.
\eea
In both cases, $m_{M,p}$ ($m_{M,q}$) is the invariant mass of the corrected (uncorrected)
multipole and the vector components $P_C^0$ and $P_N^0$ ($Q_C^0$ and $Q_N^0$) 
are taken in the $P_M$ ($Q_M$) rest frame.  The Jacobian emerging in both cases 
will ultimately find its way into a correction weight in the Monte Carlo
realisation of the method.

\subsection{Mapping of momenta}\label{Sec:MomentaMapping}

As mentioned before, the mapping procedure still has to be defined in detail
to reconstruct the particles' momenta.  The basic ideas of the mapping
procedure suggested here are as follows: When representing all four-vectors 
in the rest frame of the multipole
\begin{itemize}
 \item treat all final state momenta equally
 \item scale their three-momenta by a common factor $u$
 \item distribute the photon momenta
 \item assign the energy-component of every vector such that momentum 
	conservation and all on-shell conditions are fullfilled
\end{itemize}
This will ultimately necessitate a change of the initial state momenta as 
well.  However, since they are already fixed for the calculation of the 
basic matrix element this change will reduce to employing another frame during 
the reconstruction procedure.

However, closer examination reveals that the mapping paradigm above in fact
enforces a different treatment for purely neutral and partially or fully 
charged initial state configurations.  The reason is that the momenta of the 
newly generated Bremsstrahlungs photons need to be balanced.  Furthermore, 
the phase space integral still has to be rewritten in terms of the undressed, 
original final state momenta defining the original matrix element and cross 
section without QED radiation.  This will be addressed in the next sections, 
Sec.~\ref{Sec:neut_init_mapping}-\ref{Sec:char_init_mapping}, where the case 
of decays, i.e. single initial state particles, either neutral or charged, 
will be discussed separately.  Formally, of course, both treatments will 
yield identical results, since only the soft limit of photon emission is 
defined from first principles and because both treatments respect this limit.  

\subsubsection{Neutral initial states: final state multipoles}
\label{Sec:neut_init_mapping}

The first case to be considered is the case of a neutral particle of mass $M$ 
decaying into a final state with charged particles.   The reconstruction 
paradigm above completely fixes the reconstruction procedure to a rescaling of 
all final state momenta, both charged and neutral, and balancing the summed 
photon momentum $K$ by moving the frame of the multipole and, hence, of the 
initial state\footnote{
	Note that it is not possible to distribute any fraction of the photon 
	momentum equally to all final states with the constraint that the 
	multipole remains in its rest frame.  It therefore is mandatory to 
	balance the photon momentum with the initial state.}.  
Denoting, again, with $q_i$ the undressed and with $p_i$ the dressed final 
state momenta, and denoting their respective sums by $Q_C$, $Q_N$, $P_C$ and 
$P_N$, as declared earlier, and using $K$ as the summed momentum of all 
Bremsstrahlungs photons, the reconstruction prescription reads as follows:
\begin{itemize}
\item The momenta of the $Q_M$ rest frame
	\bea
	p_N^\mu &=& \left(\sqrt{M^2+\vec{Q}_N^2},\vec{Q}_N\right) \nnb\\
	Q_C^\mu   &=& \left(Q_C^0,\vec{Q}_C=0\right) \nnb\\
	Q_N^\mu &=& \left(Q_N^0,\vec{Q}_N\right)
	\eea
	will be mapped onto
	\bea
 	p_N^\mu\longrightarrow {p_N^\prime}^\mu &=& 
		\left(\sqrt{M^2+(u\vec{Q}_N+\vec{K})^2},u\vec{Q}_N+\vec{K}=
			u\vec{p}_N+\vec{K}\right) \nnb\\
	P_C^\mu   &=& \left(P_C^0, u\vec{Q}_C=0\right) \\
	P_N^\mu &=& \left(P_N^0,u\vec{Q}_N\right) \\
	K^\mu   &=& \left(K^0,\vec{K}\right)
	\eea
	in the $P_M$ rest frame.

\item $p_N$ and $p_N^\prime$ are the same physical vector but in different 
	frames. The scaling parameter $u$ now is determined by momentum 
	conservation, i.e.\
	\bea\label{def_u_ff}
 	0 = \sqrt{M^2+\left(u\vec{Q}_N+\vec{K}\right)^2} - 
		\sum_{C}\sqrt{m_i^2+u^2\vec{q}_i^2} - 
		\sum_{N}\sqrt{m_i^2+u^2\vec{q}_i^2} - K^0\,,
	\eea
	where the subscripts $C$ and $N$ in the sums indicate a summation
	over charged and neutral particles, respectively.
\item The phase space element expressed in terms of the undressed final state 
	momenta then reads
	\bea
	\done\Phi &=& 
	(2\pi)^4\,\done\Phi_q\,\done\Phi_k\,
	\delta^3\left(\vec{Q}_M\right)\delta\left(Q_M^0-Q_C^0-p_C^0\right)
	\frac{m_{M,p}^3}{M^2\left(P_C^0+P_N^0+K^0\right)} \nnb\\ 
	&&\times\,u^{3n-4}\,
	\frac{\frac{\vec{p}_N^2}{p_N^0}-\sum_{C,N}\frac{\vec{q}_i^2}{q_i^0}}
	     {\frac{\vec{p}_N^\prime\vec{p}_N^{}}{p_N^{\prime 0}}-
				\sum_{C,N}\frac{\vec{p}_i\vec{q}_i}{p_i^0}}
	\hspace{2mm}\prod_{i=1}^n\left[\frac{q_i^0}{p_i^0}\right]\,.
	\eea
\end{itemize}

\subsubsection{Charged initial state: Mixed multipoles}
\label{Sec:char_init_mapping}

The other case of relevance in the framework of this publication is the decay 
of a charged particle of mass $M$, leading to multipoles containing both 
initial and final state particles emitting the photons.  Again the paradigm 
above completely fixes the reconstruction procedure.  Basically, the problem 
is to compensate the photon momentum after the final state momenta have been
rescaled.  This is achieved in the following way:
\begin{itemize}
\item The momenta of the $Q_M$ rest frame
	\bea
	p_C^\mu   &=& \left(\sqrt{M^2+\vec{Q}_C^2},-\vec{Q}_C\right) \nnb\\
	Q_C^\mu   &=& \left(Q_C^0,\vec{Q}_C\right) \nnb\\
	Q_N^\mu &=& \left(Q_N^0,\vec{Q}_N=-2\vec{Q}_C\right)
	\eea
	will be mapped onto
	\bea
	p_C^\mu\longrightarrow {p_c^\mu}^\prime 
	&=& \left(\sqrt{M^2+(u\vec{Q}_C-n_C\vec\kappa)^2},
		-u\vec{Q}_C+n_C\vec\kappa=
	   	u\vec{p}_C+n_C\vec\kappa\right) \nnb\\
	P_C^\mu &=& \left(P_C^0, u\vec{Q}_C-n_C\vec\kappa\right) \nnb\\
	P_N^\mu &=& \left(P_N^0,u\vec{Q}_N-n_N\vec\kappa = 
                    -2u\vec{Q}_C-n_N\vec\kappa\right) \nnb\\
	K^\mu   &=& \left(K^0,\vec{K}\right)
	\eea
	in the $P_M$ rest frame.  Here, $n_C$ and $n_N$ are the numbers of 
	charged and neutral final state particles, respectively, and the
	abbreviation
	\bea
	\vec\kappa = \frac{1}{2n_C+n_N}\vec{K}
	\eea
	has been introduced for a more compact notation.  Again, $p_C$ 
	and $p_C^\prime$ are the same physical vector represented in different 
	frames, thus specifying the relation between the $Q_M$ and the $P_M$ 
	rest frame.  In the soft limit, i.e.\ for $K\to 0$, the scaling 
	parameter $u\to 1$ and both vectors are identical, as required.  
\item In general, the scaling parameter is fixed through energy
	conservation as the solution of
	\bea\label{def_u_if}
	0 & = & \sqrt{M^2+\left(u\vec{Q}_C-n_C\vec\kappa\right)^2} -
		\sum_{C}\sqrt{m_i^2+\left(u\vec{q}_i-\vec\kappa\right)^2}
	-	\sum_{N}\sqrt{m_i^2+\left(u\vec{q}_i-\vec\kappa\right)^2} - 
		K^0\,.\nnb\\
	\eea
\item The phase space integral rewritten in terms of the $q_i$ reads  
	\bea
	\done\Phi 
	&=& 
	(2\pi)^4\,\done\Phi_q\,\done\Phi_k\,
	\delta^3\left(\vec{Q}_M\right)
	\delta\left(Q_M^0-Q_C^0-p_C^0\right)
	\frac{m_{M,p}^3}{M^2\left(P_C^0+P_N^0+K^0\right)} \nnb\\ 
	&&
	\times\,u^{3n-4}\,
	\frac{\frac{\vec{p}_C^2}{p_C^0}-\sum_{C,N}\frac{\vec{q}_i^2}{q_i^0}}
	{\frac{\vec{p}_C^\prime\vec{p}_C}{p_C^{\prime 0}}-\sum_{C,N}
	\frac{\vec{p}_i\vec{q}_i}{p_i^0}}
	\hspace{2mm}\prod_{i=1}^n \left[\frac{q_i^0}{p_i^0}\right]\,.
	\eea
	It is worth noting that this is identical to the case of a neutral
	particle in the initial state.   
\end{itemize}

\subsection{Event generation}
\label{Sec:EventGeneration}

Having transformed the phase space integrals allows to write the full decay 
rate including real and virtual QED radiation as
\bea
2M\cdot\Gamma &=& 
\sum_{n_\gamma}\frac{1}{n_\gamma!}
\int \done\Phi_q\,\done\Phi_k
(2\pi)^4\delta^3\left(\vec{Q}_M\right)\delta\left(Q_M^0-Q_C^0-p_C^0\right)\;
e^{Y(\Omega)}\;\tilde{\beta}_0^0\,\mathcal{C} \nnb\\
&&\times\,
\prod_{i=1}^{n_\gamma}\left[
\vphantom{\frac{\frac{\vec{p}^2}{p^0}}{\frac{\vec{p}^2}{p^0}}}
\tilde{S}(k_i)\Theta(k_i,\Omega)\right]\;
\frac{m_{M,p}^3\;u^{3n-4}}{M^2\left(P_C^0+P_N^0+K^0\right)}
\frac{\frac{\vec{p}^2}{p^0}-\sum_{C,N}\frac{\vec{q}_i^2}{q_i^0}}
           {\frac{\vec{p}^\prime\vec{p}}{p^{\prime 0}}-
			\sum_{C,N}\frac{\vec{p}_i\vec{q}_i}{p_i^0}}
\prod_{i=1}^n\left[\frac{q_i^0}{p_i^0}\right]\,,
\eea
where $p$ and $p^\prime$ now generally stand for the initial state particle.

The zeroth order differential decay rate $\done\Gamma_0$, which will be used 
by default in all decays in \Sherpa can easily be extracted and, employing 
Eq.\ (\ref{Eq:PS_trafo_zero}), reads
\bea\label{Eq:monte_carlo_equation}
\Gamma &=& 
\sum_{n_\gamma}\frac{1}{n_\gamma!}
\int\done\Gamma_0\,\done\Phi_k\, e^{Y(\Omega)}
\prod_{i=1}^{n_\gamma}\left[\tilde{S}(k_i)\Theta(k_i,\Omega)\right]\nnb\\
&&\times\,
\frac{m_{M,p}^3}{m_{M,q}^3}\frac{Q_C^0+Q_N^0}{P_C^0+P_N^0+K^0}\;u^{3n-4}
\frac{\frac{\vec{p}^2}{p^0}-
		\sum_{C,N}\frac{\vec{q}_i^2}{q_i^0}}
	     {\frac{\vec{p}^\prime\vec{p}}{p^{\prime 0}}-
		\sum_{C,N}\frac{\vec{p}_i\vec{q}_i}{p_i^0}}
\prod_{i=1}^n\left[\frac{q_i^0}{p_i^0}\right]
\mathcal{C}\,.
\eea
Up to here no approximations have been made at all. In order to generate the 
corresponding distribution with Monte Carlo techniques, however, this form
is not particularly useful.  To simplify Eq.~(\ref{Eq:monte_carlo_equation}) 
therefor, hit-or-miss and reweighting techniques are used, demanding upper 
bounds for the various pieces:
\begin{itemize}
\item All higher orders are neglected, thus setting $\mathcal{C}$ to one.
\item The maximum of all Jacobians is given for $K = 0$, coinciding with 
	the leading-order phase space. 
\item The dependences on the dressed momenta in the eikonal factors are 
	removed by approximating these factors by those depending on the 
	undressed variables from the generation.
\end{itemize} 
The resulting crude distribution reads
\bea
\Gamma_{\mbox{\scriptsize cr}} =
\sum_{n_\gamma=0}^\infty\frac{1}{n_\gamma!}
\int\done\Gamma_0\,\done\Phi_k\,e^{Y(\omega)}
\prod_{i=1}^{n_\gamma}\tilde{S}_q(k_i)\Theta(k_i,\Omega).
\eea
After executing all $k$-integrations giving
\bea
\int \prod_{i=1}^{n_\gamma}\frac{\done^3k_i}{k_i^0}
\tilde{S}_q(k_i)\Theta(k_i,\Omega) = \bar{n}^{n_\gamma}
\eea
the YFS-Form-Factor is estimated by
\bea
 Y(\Omega) \approx -\bar{n}
\eea
for suitable choices of $\Omega$ \footnote{
	In this publication (and in the code), this choice has been
	to limit the photon energies by setting an infrared energy 
	cut-off of $0.1 \GeV$, unless otherwise stated. 
}. 
Reinserting this into the crude estimate, the leading-order cross section 
can be seperated from the QED radiation, and
\bea
\Gamma_{\mbox{\scriptsize cr}} =
\Gamma_0\sum_{n_\gamma=0}^\infty\left[
\frac{1}{n_\gamma!}e^{-\bar{n}}\bar{n}^{n_\gamma}\right]\,.
\eea
The result is the undressed zeroth order cross section times a Poisson 
distribution with the avarage photon multiplicity $\bar{n}$.  In this 
factorised state the photon distribution can be separated from the 
generation of the basic matrix element.  Assuming the latter to be already 
generated it can {\it a posteriori} be corrected to the leading-\-logarithmic
all-\-order QED correction by generating the photon distribution as 
follows:
\begin{enumerate}
\item Generate the number of photons according to a Poissonian distribution 
	with mean $\bar{n}$.
\item Generate each photon's momentum according to $\tilde{S}_q(k)$.  This implies 
	\begin{itemize}
	\item that its energy $k^0$ is distributed according to 
		\bea
		\rho(k^0) \sim \frac{1}{k^0}
		\eea
	\item and that the azimutal and polar angles are distributed according 
		to
		\bea
		\rho(\theta,\,\phi) \sim 
		\sum\limits_{i<j}\left(\frac{q_i}{q_i\cdot e_k}-
					\frac{q_j}{q_j\cdot e_k}\right)^2\,, 
		\eea
		where $e_k$ is a null vector of unit length, 
		\bea\label{Eq:e_k}
		e_k^\mu = \frac{1}{k^0}k^\mu\;\;\;\mbox{\rm with}\;\;\;
		e_k^2 = 0\,.
		\eea  
	\end{itemize}
	It is possible that more than one hard photon is created such that the
	total energy of all photons exceeds the decaying system's energy.  
	Obviously, this has to be avoided to guarantee energy conservation.  
	A simple way of achieving this is a mere veto on such situations,
	accompanied with a repetition of photon generation, starting from
	step 1.
\item Reconstruct the momenta.
\item Calculate and apply all weights.  This yields a total weight, namely
	\bea
 	W = 
	W_{\mbox{\scriptsize dipole}}\times 
	W_{\mbox{\scriptsize YFS}}\times 
	W_{\mbox{\scriptsize J,L}}\times 
	W_{\mbox{\scriptsize J,M}}\times 
	W_{\mathcal{C}}\,,
	\eea
	where the individual weights are given by
	\bea
 	W_{\mbox{\scriptsize dipole}} 
	&=& 
	\prod_{i=1}^{n_\gamma}
	\frac{\tilde{S}(p_C,P_C,k_i)}{\tilde{S}(p_C,Q_C,k_i)} \\
	W_{\mbox{\scriptsize YFS}} 
	&=& 
	\exp{(Y(p_C,P_C,\Omega)+\bar{n})} \\
	W_{\mbox{\scriptsize J,L}} 
	&=& 
	\frac{m_{M,p}^3}{m_{M,q}^3}\frac{Q_C^0+Q_N^0}{P_C^0+P_N^0+K^0} \\
	W_{\mbox{\scriptsize J,M}} 
	&=& 
	u^{3n-4}
	\frac{\frac{\vec{p}^2}{p^0}-\sum_{C,N}
	\frac{\vec{q}_i^2}{q_i^0}}
	     {\frac{\vec{p}^\prime\vec{p}}{p^{\prime 0}}-
	\sum_{C,N}\frac{\vec{p}_i\vec{q}_i}{p_i^0}}
	\prod_{i=1}^n\left(\frac{q_i^0}{p_i^0}\right) \\
	W_{\mathcal{C}} &=& \mathcal{C} \,.
	\eea
	Here, $W_{\rm dipole}$ corrects the emitting dipoles from the
	unmapped to the mapped momenta, $W_{\rm YFS}$ accounts for the exact
	YFS form factor, $W_{\mbox{\scriptsize J,L}}$ essentially denotes 
	the Jacobian due to the Lorentz-transformation, 
	$W_{\mbox{\scriptsize J,M}}$ is the weight due the momenta-mapping,
	and $W_{\mathcal{C}}$ incorporates higher-order corrections,
	where available.

	The maximum of the combined weight indeed is smaller than the maximal
	weight employed for generating the distribution, $W < W(K=0)$.
	Hence application of the combined weight is just a realisation of a 
	hit-or-miss method.  The distribution obtained is now the exact 
	distribution of (\ref{Eq:master_equation}) or 
	(\ref{Eq:monte_carlo_equation}).
\end{enumerate}

\section{Higher Order Corrections}\label{Sec:Higher_Order}

In the last section, the procedure generating the QED corrections to cross
sections, following Eq.~(\ref{Eq:exact_distribution}), has been outlined.  
By construction, the algorithm yields exact all-orders results, if all matrix 
elements are known.  This, however, is never the case.  On the other hand, the
dominant universal soft photon contributions, real and virtual, are included 
to all orders in the YFS form factor, Eq.~(\ref{YFS_Form_Factor}).  Thus, if 
the zeroth order undressed matrix element only is known, i.e.\ if $\mathcal{C} = 1$, 
the photons will be solely generated according to a product of eikonal factors
$\tilde{S}(k_i)$.  Consequently, their distribution will be correct in the 
soft limit only.  Away from this limit, exact matrix elements at a given order
may be mandatory to yield satisfactory and sufficient accuracy.  For most 
applications on decay matrix elements - the topic of this publication - it 
will be sufficient to implement the matrix element correction to the first 
order in $\alpha$, i.e.\ for the emission of one additional real or virtual 
photon.  It should be noted here that hard photon emission predominantly 
occurs in situations where potential emitters are characterised by a large
energy-to-mass ratio and that in any case hard photon emissions tend to 
populate regions in phase space that are collinear w.r.t.\ the emitters.  
In contrast, large angle radiation has the tendency to be predominantly soft.

\subsection{Approximations for real emission matrix elements}
\label{Sec:ApproximateRealCorrections}

As already explained, the vast majority of hard photon emissions deserving
an improved description through corrections to the soft limit underlying
the YFS approach occurs in the collinear region of emission phase space.
In this region, the well-known collinear factorisation can be used to 
approximate $\tilde{\beta}_1^1$.  This amounts to an inclusion of the 
leading collinear logarithms arising in this limit, which are incorporated 
for instance in the Altarelli-Parisi evolution equation \cite{Altarelli:1977zs} 
and corresponding splitting kernels.  

Since masses are to be taken fully into account the quasi-collinear limit
defined in \cite{Dittmaier:1999mb,Catani:2000ef} replaces the more familiar
collinear one.  In this limit the matrix element factorises as
\bea\label{Eq:coll_approx_final}
\sum_{\lambda_\gamma}\left|\mathcal{M}_1^\frac{1}{2}(p_i,k)\right|^2 
\cong 
\left\{\begin{array}{ll}
e^2 Z_i^2 \mbox{\sl g}^{\mbox{\scriptsize (out)}}(p_i,k)
\left|\mathcal{M}_0^0(p_i+k)\right|^2 &
\mbox{\rm if $i\in$ F.S.}\\[2mm]
e^2 Z_i^2 \mbox{\sl g}^{\mbox{\scriptsize (in)}} (p_i,k)
\left|\mathcal{M}_0^0(x\cdot p_i)\right|^2&
\mbox{\rm if $i\in$ I.S..}\end{array}\right.
\eea
Here the $\mbox{\sl g}^{\mbox{\scriptsize (in,out)}}(p_i,k)$ denote massive 
splitting functions.  For instance, for the case of a fermion emitting a 
photon they are given by
\bea
 \mbox{\sl g}^{\mbox{\scriptsize (out)}}(p_i,k) 
&=& 
\frac{1}{(p_i\cdot k)}\left(P_{ff}(z)-\frac{m_i^2}{(p_i\cdot k)}\right) \\
\mbox{\sl g}^{\mbox{\scriptsize (in)}}(p_i,k) 
&=& 
\frac{1}{x(p_i\cdot k)}\left(P_{ff}(x)-\frac{x\,m_i^2}{(p_i\cdot k)}\right)\,,
\eea
where $x=\frac{p_i^0-k^0}{p_i^0}$ and $z=\frac{p_i^0}{p_i^0+k^0}$ are the 
fractions of the fermion energies kept after the emission of the photon,
and where $P_{ff}(y)$ is the well-known Altarelli-Parisi splitting function  
\bea
 P_{ff}(y) & = & \frac{1+y^2}{1-y}\,.
\eea

The dipole splitting functions of \cite{Dittmaier:1999mb} have been generalised
further in \cite{Catani:2002hc} to incorporate also polarisation.  Thus, in
principle they could directly be used in the framework of the YFS formulation 
replacing the original eikonal factors.  In the framework of this publication,
however, they are employed as universal correction factors, reweigthing 
explicit photon emission such that the correct collinear limit is recovered.
Since they interpolate smoothly between both limits they already include the
soft limit.  Therefore, in the correction weights, these soft terms have to 
be subtracted because they are already acounted for in the orginal YFS 
eikonals.  In addition, since the dipole splitting kernels refer to an
emitter and a spectator forming the dipole, for each dipole two such terms
have to be applied, such that the squared matrix element with the dipole
terms approximating the photon emission reads
\bea
 \left|\mathcal{M}_1^\frac{1}{2}\right|^2 
& \cong & 
-e^2\sum_{i\neq j}
\left[Z_iZ_j\theta_i\theta_j\mbox{\sl g}_{ij}(p_i,p_j,k)
	\left|\mathcal{M}_0^0\right|^2\right] \\
& \cong & 
-e^2\sum_{i<j}
\left[Z_iZ_j\theta_i\theta_j
	\left(\mbox{\sl g}_{ij}(p_i,p_j,k)+
		\mbox{\sl g}_{ji}(p_j,p_i,k)\right)
	\left|\mathcal{M}_0^0\right|^2\right]\,.
\eea
Here, charge conservation in the form $\sum Z_i\theta_i=0$ has been used.  
The second particle in each massive splitting function $\mbox{\sl g}_{ij}$ 
denotes the spectator of the emission process and accounts for the recoil,
thus ensuring four-momentum conservation.  It should also be noted that
the sum in the equations above runs over charged particles only.

In order to subtract the soft terms, it is useful to consider the soft
and quasi-collinear limits of the dipole splitting kernels 
$\mbox{\sl g}_{ij}(p_i,\,p_j,\,k)$:
\bea
\label{Eq:IR_limit}
\mbox{\sl g}_{ij}(p_i,p_j,k) 
& \stackrel{k\to 0}{\sim} & 
\frac{1}{(p_i\cdot k)}
	\left(\frac{2(p_i\cdot p_j)}{(p_i\cdot k)+(p_j\cdot k)}-
		\frac{m_i^2}{(p_i\cdot k)}\right) \\
\label{Eq:qc_limit}
\mbox{\sl g}_{ij}(p_i,p_j,k) 
& \stackrel{p\cdot k\to 0}{\sim} & 
\mbox{\sl g}^{\mbox{\scriptsize (out/in)}}\,.
\eea
Because the soft limit is universal and spin-independent, it is a 
straightforward exercise to define soft-subtracted dipole splitting kernels 
\bea\label{Eq:IR_sub_dsf}
\bar{\mbox{\sl g}}_{ij}(p_i,p_j,k) 
&=& 
\mbox{\sl g}_{ij}(p_i,p_j,k) - 
\mbox{\sl g}_{ij}^{\mbox{\scriptsize (soft)}}(p_i,p_j,k)\nnb\\
&=&
\mbox{\sl g}_{ij}(p_i,p_j,k) - 
\frac{1}{(p_i\cdot k)}
\left(\frac{2(p_i\cdot p_j)}{(p_i\cdot k)+(p_j\cdot k)}-
	\frac{m_i^2}{(p_i\cdot k)}\right)\,.
\eea
The soft-subtracted dipole splitting kernels $\bar{\mbox{\sl g}}_{ij}$ now 
have the correct (finite) soft limit while retaining the original 
quasi-collinear limit of $\mbox{\sl g}_{ij}$ (Eq.~(\ref{Eq:qc_limit})). 
Accordingly, the soft-subtracted matrix element can be approximated as
\bea
 \tilde{\beta}_1^1 & = & 
-\frac{\alpha}{4\pi^2}\sum_{i<j}Z_iZ_j\theta_i\theta_j
\left(\bar{\mbox{\sl g}}_{ij}(p_i,p_j,k)+
	\bar{\mbox{\sl g}}_{ji}(p_j,p_i,k)\right)\tilde{\beta}_0^0\,.
\eea
The exact form of the $\mbox{\sl g}_{ij}(p_i,p_j,k)$ for different 
emitter-spectator configurations will be given in Appendix \ref{Appendix_dsf}.

\subsection{Exact real emission matrix elements}
\label{Sec:RealCorrections}

In order to achieve an even higher precision, the implementation of exact
higher-order full matrix elements becomes mandatory.  It should be clear, however, 
that large differences with the approximated matrix elements above will occur only 
in non-singular regions of comparable hard, wide-angle emissions.  Since the module 
presented in this publication, \Photonspp, is embedded in the \sherpa framework it 
is easy to implement such infrared subtracted squared matrix elements, making use 
of tools and functions already provided within the framework.  In particular, some 
basic building blocks for the calculation of helicity amplitudes already used
in \cite{Krauss:2001iv,Gleisberg:2003ue} can be recycled to construct the neccessary, 
infrared-subtracted one-photon real emission matrix elements, which are then evaluated 
at momentum configurations generated by the algorithm of Section \ref{Sec:Algorithm}. 
These building blocks are listed in App.~\ref{App:XYZ-functions}.   Exact first-order
matrix elements have so been implemented for a number of relevant matrix elements, 
see below.  It is worthwile to stress that in principle also second-order precision
could be achieved, if neccessary.

In general, the infrared-subtracted squared matrix element can be written as
\bea
\tilde{\beta}_1^1 & = & 
\frac{1}{2(2\pi)^3}
\left.\mathcal{M}_1^\frac{1}{2}\mathcal{M}_1^\frac{1}{2}\right.^\ast - 
\tilde{S}(k)\left.\mathcal{M}_0^0\mathcal{M}_0^0\right.^\ast\,,
\eea
and it is only the amplitudes $\mathcal{M}$ that are process-specific and need to be 
listed for the different processes.  It should be noted that within the \sherpa framework
the real emission matrix elements are straightforward to implement, in contrast, the 
incorporation of loop matrix elements is somewhat more involved: in those cases the 
integral has to be calculated analytically and the divergences must be cancelled before 
implementation as a function of the outer momenta.

\subsubsection{Two-body decays of type $V\to FF$}

The matrix elements for two body decays where one neutral vector particle decays
into two charged fermions, $V\to FF$, read\footnote{
	All particles involved are considered to be point-like, 
	i.e.~their vertices do not contain form factors.
}
\bea
\mathcal{M}_0^0 
& = &
ie\;\varepsilon_\mu^V(p,\lambda)\;
\bar{u}(q_1,s_1)\gamma^\mu\left(c_L P_L + c_R P_R\right)v(q_2,s_2)\nnb\\
\mathcal{M}_1^\frac{1}{2} 
& = & 
ie^2\;\varepsilon_\mu^V(p,\lambda)\;
\bar{u}(p_1,s_1)\left[\gamma^\nu\frac{\dsl p_1+\dsl k+m}{(p_1+k)^2-m^2}\gamma^\mu
\left(c_L P_L + c_R P_R\right)\right. \nonumber\\
&&\hspace{20mm}\left.{}
-\gamma^\mu\left(c_L P_L + c_R P_R\right)
\frac{\dsl p_2+\dsl k-m}{(p_2+k)^2-m^2}\,\gamma^\nu\right]v(q_1,s_1)\;
\varepsilon_\nu^{\gamma^\ast}(k,\kappa) \,.
\eea
Of course, momentum conservation must hold, and therefore $p = q_1+q_2$ in 
the former and $p=p_1+p_2+k$ in the latter case. Hence, if $n_\gamma$ of the 
generated event exceeds the number of real photons in the respective infrared 
subtracted squared matrix element, a projection of the higher dimensional phase 
space onto the lower dimensional one has to be performed. 
In practise, this amounts to redoing the reconstruction procedure using only 
a subset of all photons generated in that run. Furthermore, 
\bea
c_L P_L + c_R P_R & = & c_L \frac{1-\gamma^5}{2} + c_R \frac{1+\gamma^5}{2}\, .
\eea
Thus, the generic matrix element is adjustable to various decays of neutral 
vector bosons. A few key examples of the couplings $c_L$ and $c_R$ to the left 
and right-handed fermionic currents are listed in Table \ref{Tab:c_L_c_R:V}.

The real-emission matrix elements depend on the polarisations, and they are expressed 
in terms of the $X$, $Y$ and $Z$ functions listed in Appendix \ref{App:XYZ-functions} 
as
\bea
\mathcal{M}_0^0  
&=& 
ie\;X\left(q_1,s_1;\varepsilon^V;q_2,\bar{s}_2;c_L,c_R\right)
\eea
and
\bea
\lefteqn{\mathcal{M}_1^\frac{1}{2} = } \nnb\\ &&ie^2\left[\frac{1}{2(p_a^2-m^2)}\left(1+\frac{m}{\sqrt{p_a^2}}\right)\sum_s X\left(p_1,s_1,\varepsilon^{\gamma\ast},p_a,s,1,1\right)X\left(p_a,s,\varepsilon^V,p_2,\bar{s}_2,c_L,c_R\right)\right. \nonumber\\
&&\hspace{5mm}\left.{}+\frac{1}{2(p_a^2-m^2)}\left(1-\frac{m}{\sqrt{p_a^2}}\right)\sum_s X\left(p_1,s_1,\varepsilon^{\gamma\ast},p_a,\bar{s},1,1\right)X\left(p_a,\bar{s},\varepsilon^V,p_2,\bar{s}_2,c_L,c_R\right)\right. \nonumber\\
&&\hspace{5mm}\left.{}-\frac{1}{2(p_b^2-m^2)}\left(1+\frac{m}{\sqrt{p_b^2}}\right)\sum_s X\left(p_1,s_1,\varepsilon^V,p_b,s,c_L,c_R\right)X\left(p_b,s,\varepsilon^{\gamma\ast},p_2,\bar{s}_2,1,1\right)\right. \nonumber\\
&&\hspace{5mm}\left.{}-\frac{1}{2(p_b^2-m^2)}\left(1-\frac{m}{\sqrt{p_b^2}}\right)\sum_s X\left(p_1,s_1,\varepsilon^V,p_b,\bar{s},c_L,c_R\right)X\left(p_b,\bar{s},\varepsilon^{\gamma\ast},p_2,\bar{s}_2,1,1\right)\right]\hspace{10mm}
\eea
where
\bea
p_a & = & p_1+k \nnb\\
p_b & = & p_2+k,
\eea
and where $p_1$ and $p_2$ are the momenta of the final state leptons. The 
bar over the fermion spin label $s_i$ signifies an anti-particle.

\begin{table}
 \bc
  \begin{tabular} {|c|c|c|}
   \hline
   Process & $c_L$ & $c_R$ \\
   \hline
   $Z\to\ell\bar\ell$ & $\frac{ie}{2s_Wc_W}2s_W^2$ & $\frac{ie}{2s_Wc_W}(2s_W^2-1)$ \\
   \hline
   $J/\psi\to\ell\bar\ell$ & $-ie$ & $-ie$ \\
   \hline
  \end{tabular}
  \caption{\label{Tab:c_L_c_R:V}Values of the coupling constants of different 
	    vector particles to the left- and right-handed leptonic currents.}
 \ec
\end{table}

\subsection{Virtual emission correction $\tilde{\beta}_0^1$}
\label{Sec:VirtualCorrections}

The only virtual corrections occuring to level $\order(\alpha)$ are
\bea
\tilde\beta_0^1
& = & \left.M_0^1M_0^0\right.^\ast+\left.M_0^0M_0^1\right.^\ast \nnb\\
& = & \left.\mathcal{M}_0^1\mathcal{M}_0^0\right.^\ast
	+\left.\mathcal{M}_0^0\mathcal{M}_0^1\right.^\ast
	-2\alpha B \tilde\beta_0^0\,.
\eea
For the above case of decays of the type $V\to FF$ they read 
\bea
 \tilde\beta_0^1
& = & \tfrac{\alpha}{\pi}
	\left[\ln\frac{m_V^2}{m_F^2}-A\right] 
	\tilde\beta_0^0
	\qquad\qquad
	m_V^2\gg\m_F^2 \,,
\eea
with
\bea
 A 
& = & \left\{\hspace{2mm}\ba{l}      1      \mbox{ in on-shell scheme}\\ 
                               \tfrac{7}{4} \mbox{ in $\overline{MS}$ scheme}\ea \right.
\eea
which agrees with \cite{Jadach:1988gb,Berends:1987ab}.
Effects of potentially different left- and right-handed couplings $c_L$ 
and $c_R$, cf.~Tab.\ref{Tab:c_L_c_R:V}, only enter in terms suppressed by 
$\frac{m_F^2}{m_V^2}$ and are curently neglected in \Photonspp.

For the process $W\to\ell\nu$, cf.\ \cite{Placzek:2003zg}, the first order 
virtual correction reads
\bea
 \tilde\beta_0^1
& = & \tfrac{\alpha}{\pi}
	\left[\ln\frac{m_W}{m_\ell}+\tfrac{1}{2}\right]
	\tilde\beta_0^0
	\qquad\qquad
	m_W^2\gg\m_\ell^2 \,.
\eea

 \section{Results}\label{Sec:Results}

In this section some of the results of the \Photonspp module, as it is 
implemented within the \Sherpa framework, are presented.  The focus lies 
on the central distribution produced by the preceding calculations, the 
total energy of {\bf all} photons radiated per event in the rest frame of the 
decaying particle.  In addition, angular distributions for dipole and 
multipole configurations will be shown.

\subsection{Validation: leptonic heavy gauge boson decays}

The leptonic decays of $W$ and $Z$ bosons, $W\to l \nu_l$ and $Z\to l\bar{l}$,
will play the central role in validating the accuracy of the \Photonspp
implementation of the YFS approach.  Before studying in more detail these 
processes and comparing the results obtained with \Photonspp with those from 
other codes, namely \Amegicpp \cite{Krauss:2001iv} and \Windec 
\cite{Placzek:2003zg}, it is worthwhile to discuss one of the key distributions, 
namely the total energy radiated off the decay.  

\subsubsection{Radiated photon energy}

\begin{figure}
 \bc
  \includegraphics[width = 230pt]{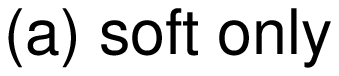}
  \includegraphics[width = 230pt]{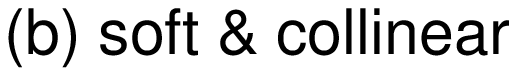}
  \includegraphics[width = 230pt]{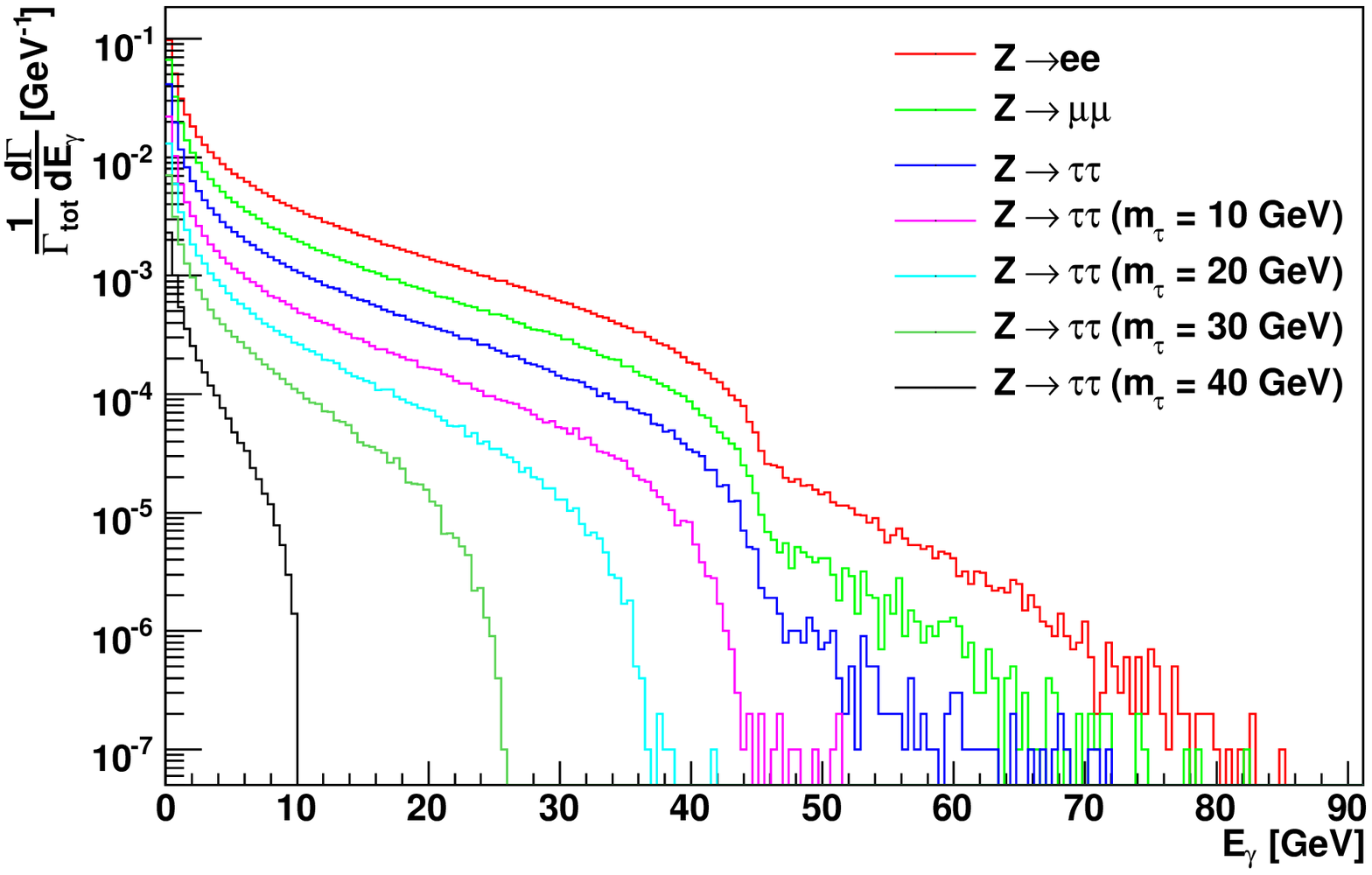}
  \includegraphics[width = 230pt]{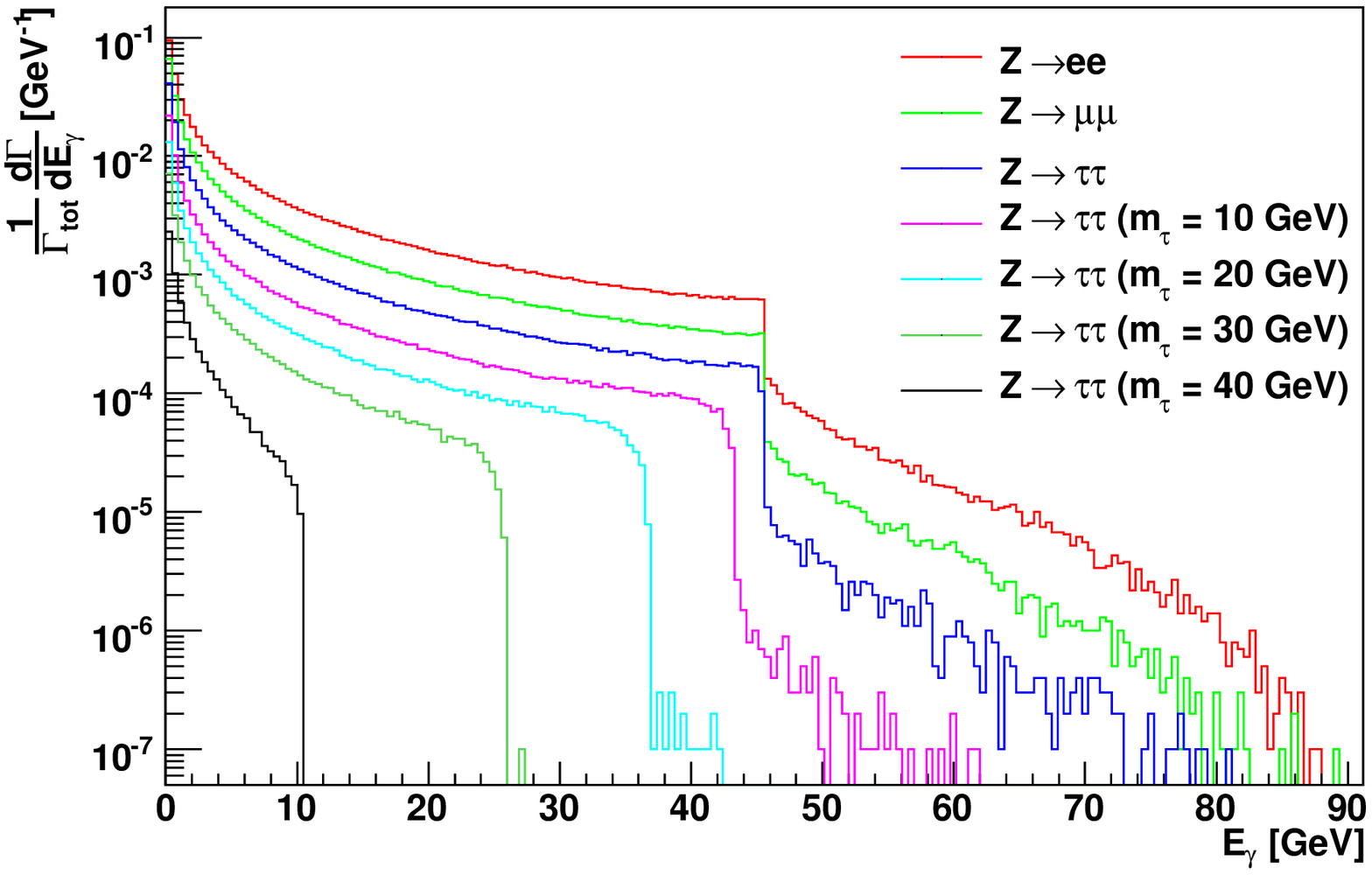}
  \includegraphics[width = 230pt]{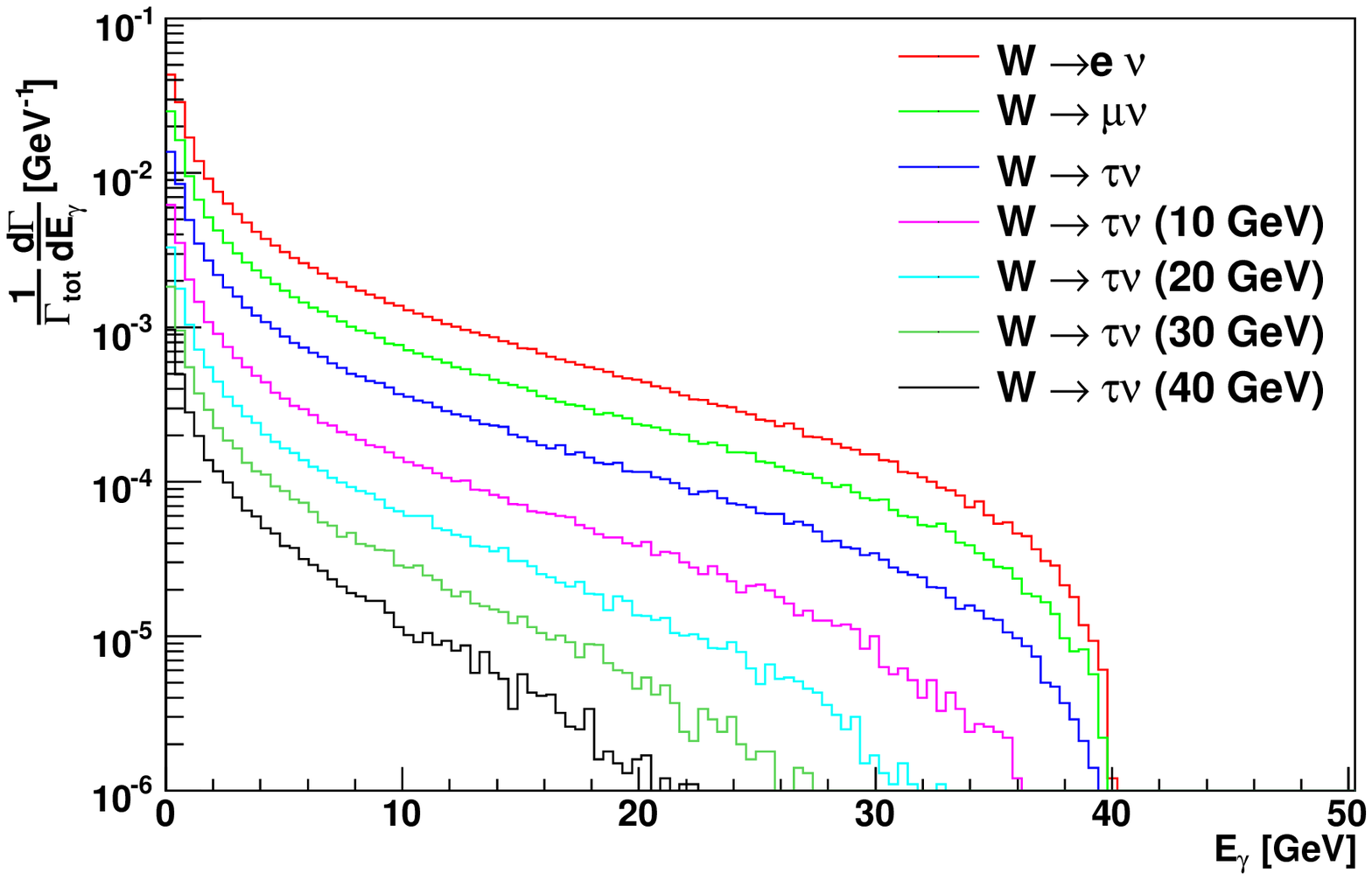}
  \includegraphics[width = 230pt]{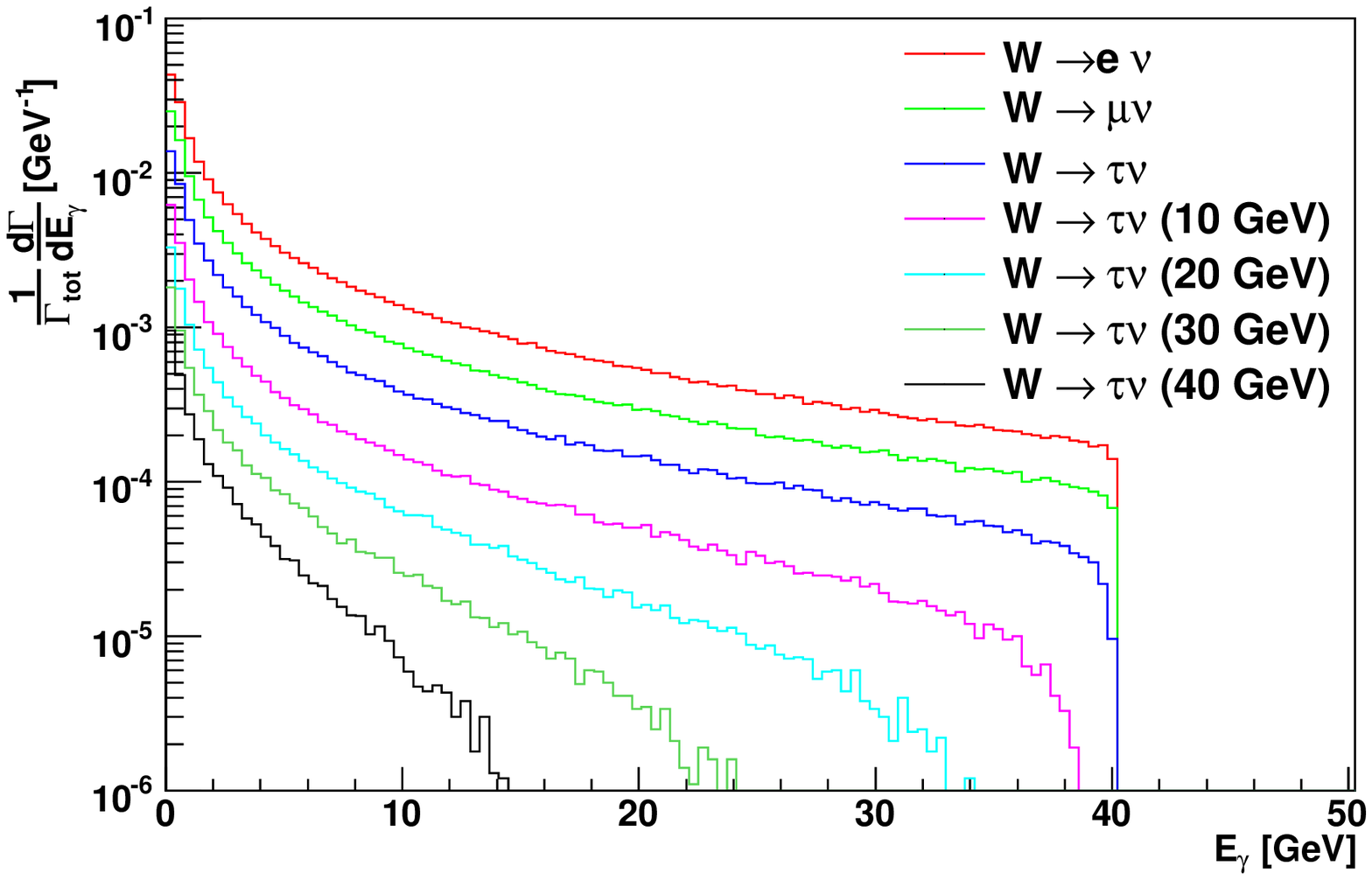}
  \caption{\label{Z_W_dist}Photon radiation in leptonic decays of
	$Z$ (upper panel) and $W$ bosons (lower panel) for different
	leptons, including fictional heavy $\tau$'s in a range of masses. 
	In the left panel, (a), $\mathcal{C} = 1$, i.e.\ photon generation
	according to the YFS form factor only is depicted, whereas in the
	right panel, (b), corrections up to $\order(\alpha)$ are included
	using the dipole splitting functions and the virtual corrections,
	cf.\ Sec.~\ref{Sec:VirtualCorrections} and 
	\ref{Sec:ApproximateRealCorrections}.  All distributions are 
	normalised on the total decay width of the decay into the respective 
	lepton and lepton-neutrino pair.  The infrared cut-off in all cases
	is set to $0.1 \GeV$.}
 \ec
\end{figure}

The result for this distribution, namely the total energy radiated off the
decay in form of photons, is presented in Figure \ref{Z_W_dist}.  For both 
processes, i.e.\ for both leptonic $Z$ and $W$ decays, different leptons
with different masses have been considered.  Clearly, radiation is most
important in final states involving electrons, being the lightest fermions 
taken into account, while radiation off heavier fermions is increasingly
suppressed.  One of the most prominent features of every radiated energy 
spectrum is the kink at around half the boson mass, which is due to kinematics.
It limits the energy involved in single photon emission off the final state 
fermion to its maximal energy, roughly half the boson mass.  This kink gets 
washed out and moves to the left with increasing fermion mass.  Events with
total radiated energy surpassing this limit must involve at least two sufficiently
hard photons, arranged such that they recoil, at least partially, against each 
other.  Naively, in the classical limit, such configurations are dominated by
photon emission off both fermions.  This motivates why radiation beyond the kink
is absent in the $W$-decay spectra.  Along the same lines of reasoning, such 
double hard photon emissions are decreasingly probable with rising lepton masses. 
However, since in the present state only approximated matrix elements up to 
$\order(\alpha)$ are included in the program these double emissions are not 
described correctly yet.

In Fig.~\ref{Z_W_dist} also different treatments of higher order matrix elements 
are exhibited: photons emitted solely according to the purely soft YFS eikonals
(left panel) are contrasted with corrections due to the approximated matrix 
elements presented in Sec.~\ref{Sec:ApproximateRealCorrections}.  The former 
distribution, labelled with \lq\lq soft\rq\rq, thus is correct in the soft limit 
but it is inadequate for the description of hard, collinear photon radiation.  
This, including virtual corrections of $\order(\alpha)$, is displayed in the 
panel labelled with \lq\lq soft \& collinear\rq\rq. 

The inclusion of these corrections gives reasonably good results as long as 
most photons are soft or if complicated correlations of hard photons are not
important.  

\subsubsection{Comparison with other codes}

After checking the physical sanity of the implementation in principle,
results obtained with \Photonspp are now to be compared to those from
other, established and dedicated Monte Carlo event generators capable of 
describing QED effects in the decays of $W$ and $Z$ bosons, in particular
with the \Windec package \cite{Placzek:2003zg}.  This program aims at the
description of the production and decay of $W$-bosons in hadronic collisions.
\Windec performs the decay of the $W$-boson into lepton-neutrino pairs 
including QED corrections summed in the YFS-approach and corrected by exact 
$\order(\alpha)$ real emission and virtual correction matrix elements.  
They are obtained for the decay only in the narrow width approximation,
i.e.\ only the $W\to\ell\nu$ decay is taken into account.  Furthermore, in 
this section, the results of \Photonspp are compared with the exact, fixed 
order, one-photon emission results of the \sherpa-inherent matrix element 
generator \Amegicpp.  However, comparisons with \Amegicpp are only sensible 
when the average photon number of the process under consideration is low, 
i.e.\ when multiphoton emission gives a negligible contribution to the 
differential cross section.  Additionally, it should be stressed that 
\Amegicpp lacks virtual corrections and therefore comparisons are sensible 
for normalised distributions only.  

The channel best suited for comparing all three generators is $W\to\tau\nu_\tau$. 
Besides the low avarage photon multiplicity (with an infrared cut-off of 0.1 
GeV multiphoton events make up for less than 3\% of all radiative events) 
virtual corrections merely amount to a 1\% correction of the zeroth order cross 
section.  Furthermore, as discussed earlier, the majority of multiphoton 
events will consist of at most one single hard photon and additional soft ones.  
Therefore, these events will be aproximately described by the hard emission only.

It should be stressed at this point, however, that there is one fundamental 
difference in the comparison of the various results, related to the way the
infrared cut-off is implemented: While in \Windec the energy cut-off is applied 
in the rest frame of the decaying $W$, it is applied in the rest-frame of the
decaying dipole in both \Photonspp and \Amegicpp.

\begin{figure}
 \bc
  \includegraphics[width = 230pt]{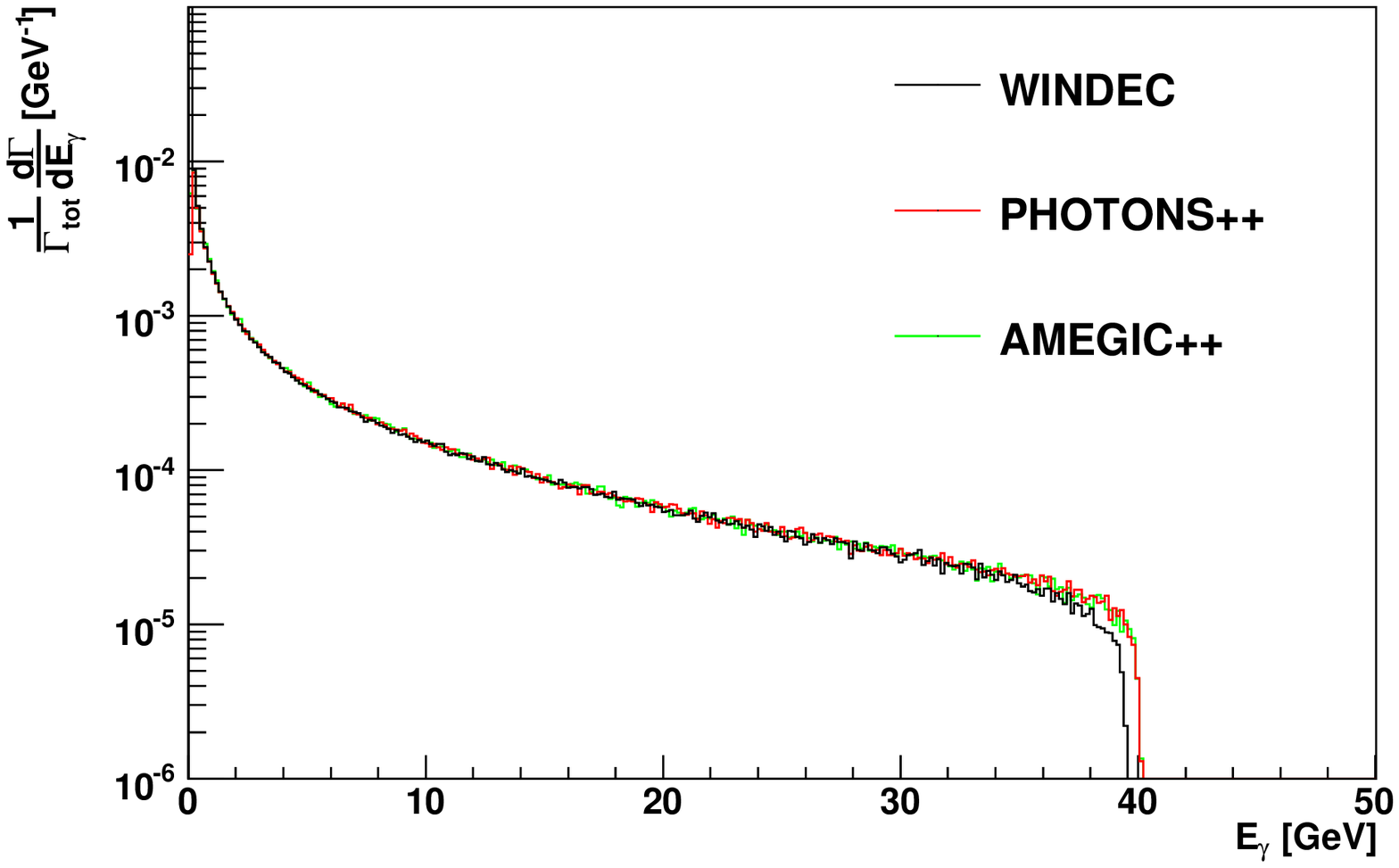}
  \includegraphics[width = 230pt]{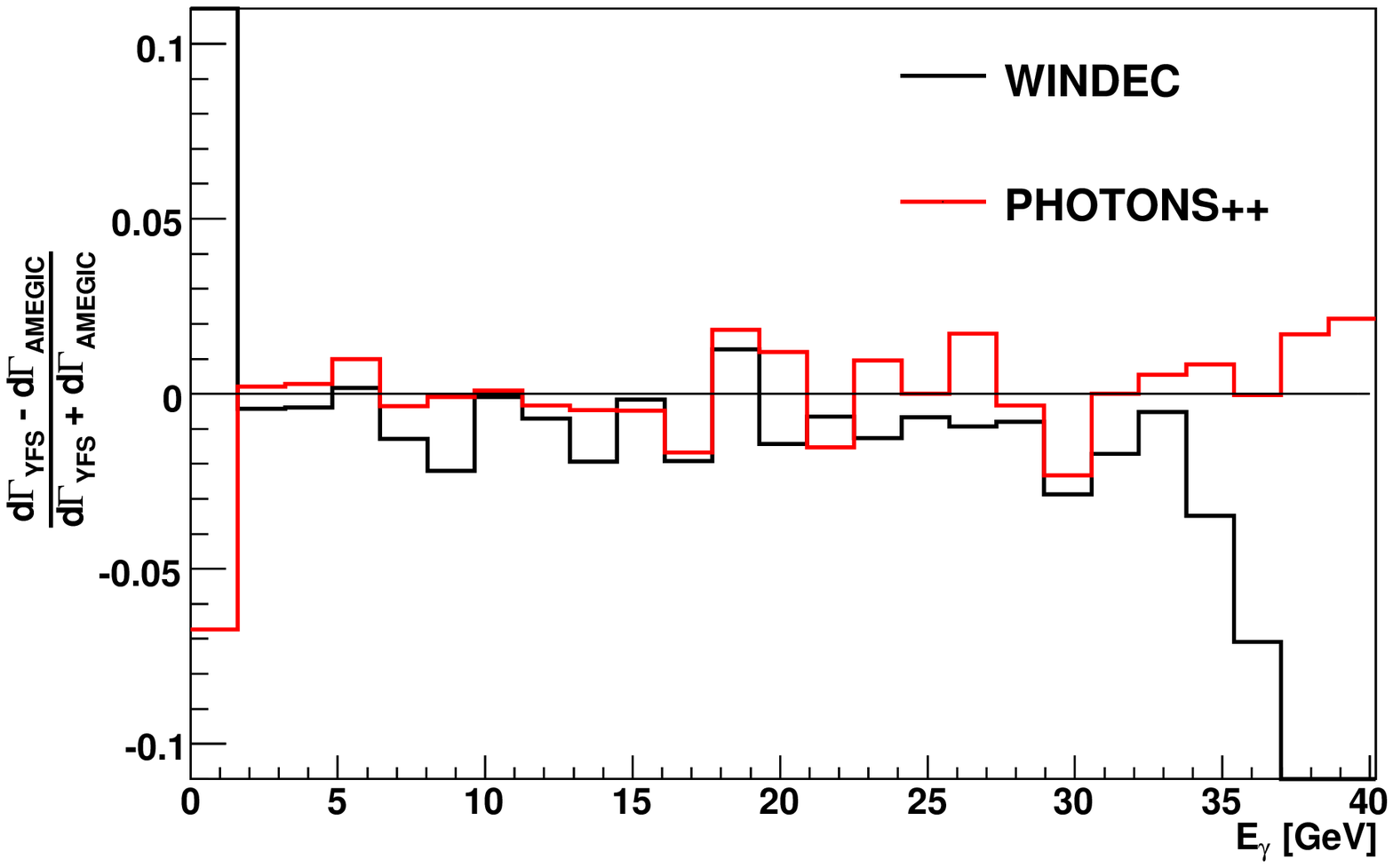}
  \caption{The total photon energy in the decay frame in $W\to\tau\nu_\tau$.  
	   In the left panel, (a), the distributions generated by \protect\Windec 
	   (black), \protect\Photonspp (red) and \protect\Amegicpp (green) are 
	   depicted, where the latter has been rescaled with the true average 
	   photon multiplicity.  In the right panel, the relative deviations of 
	   \protect\Windec (black) and \protect\Photonspp (red) with respect to 
	   the rescaled matrix element result of \protect\Amegicpp are displayed.}
\label{W_tau_comp}
 \ec
\end{figure}

\begin{figure}
 \bc
  \includegraphics[width = 230pt]{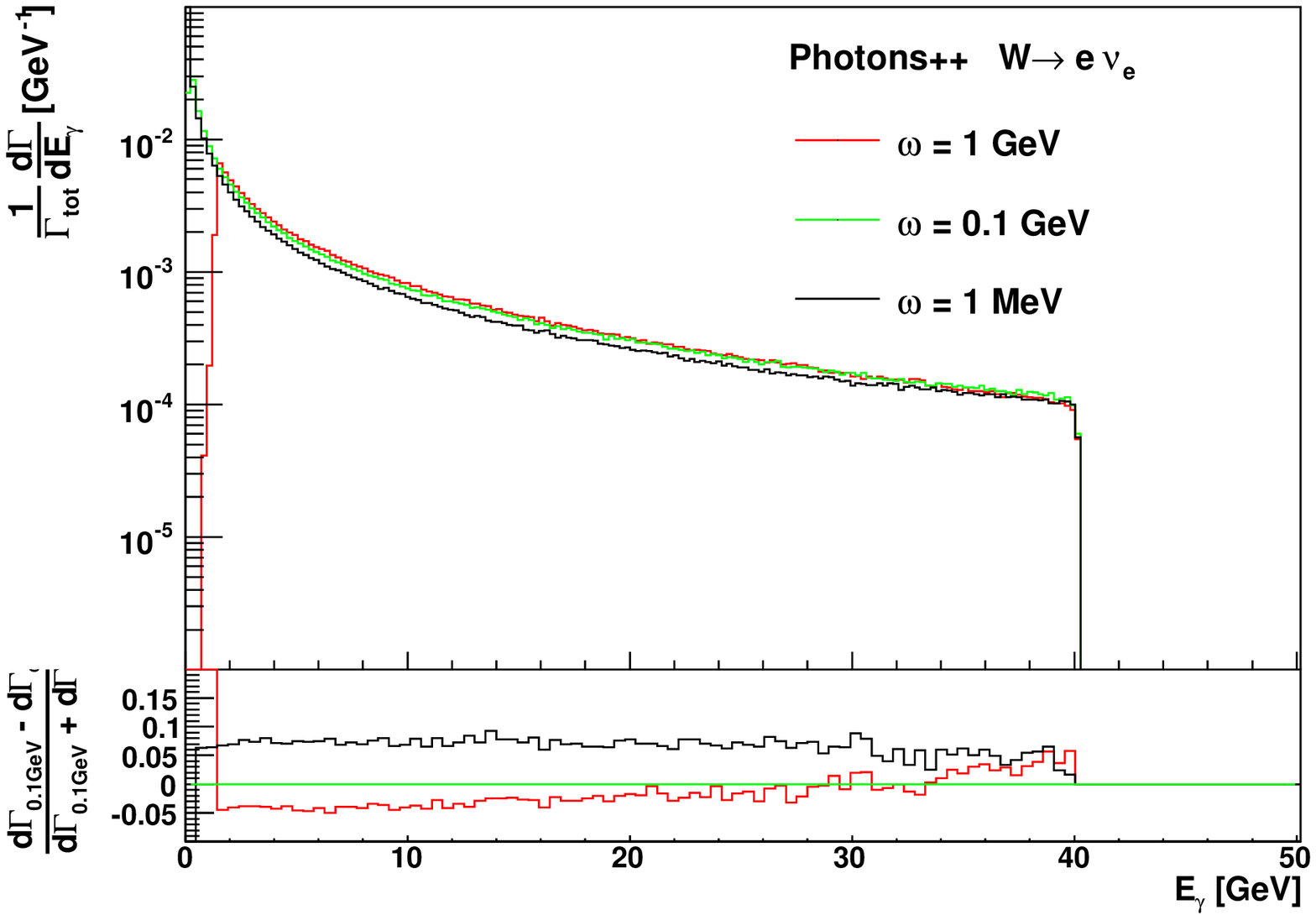}
  \includegraphics[width = 230pt]{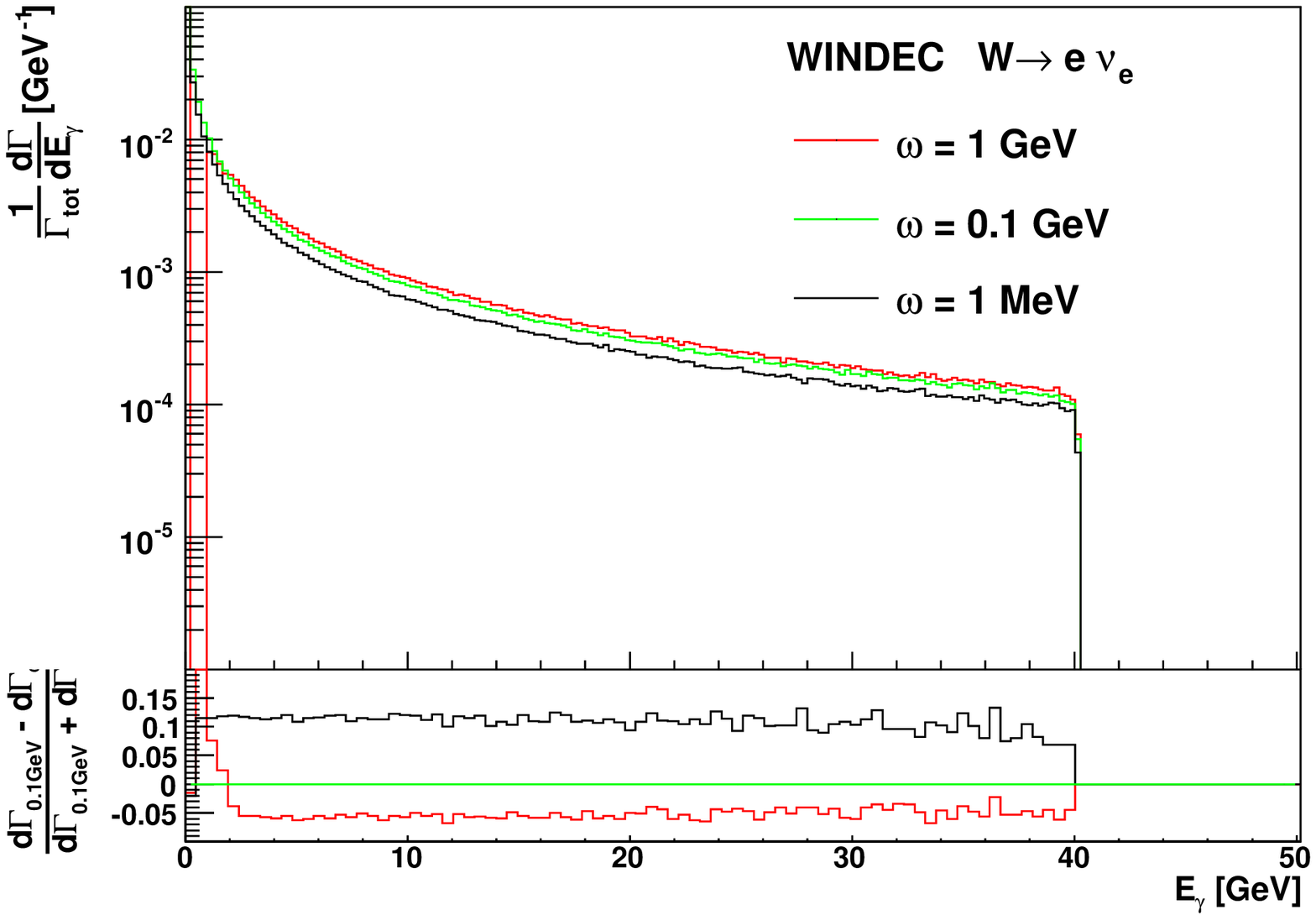}
  \includegraphics[width = 230pt]{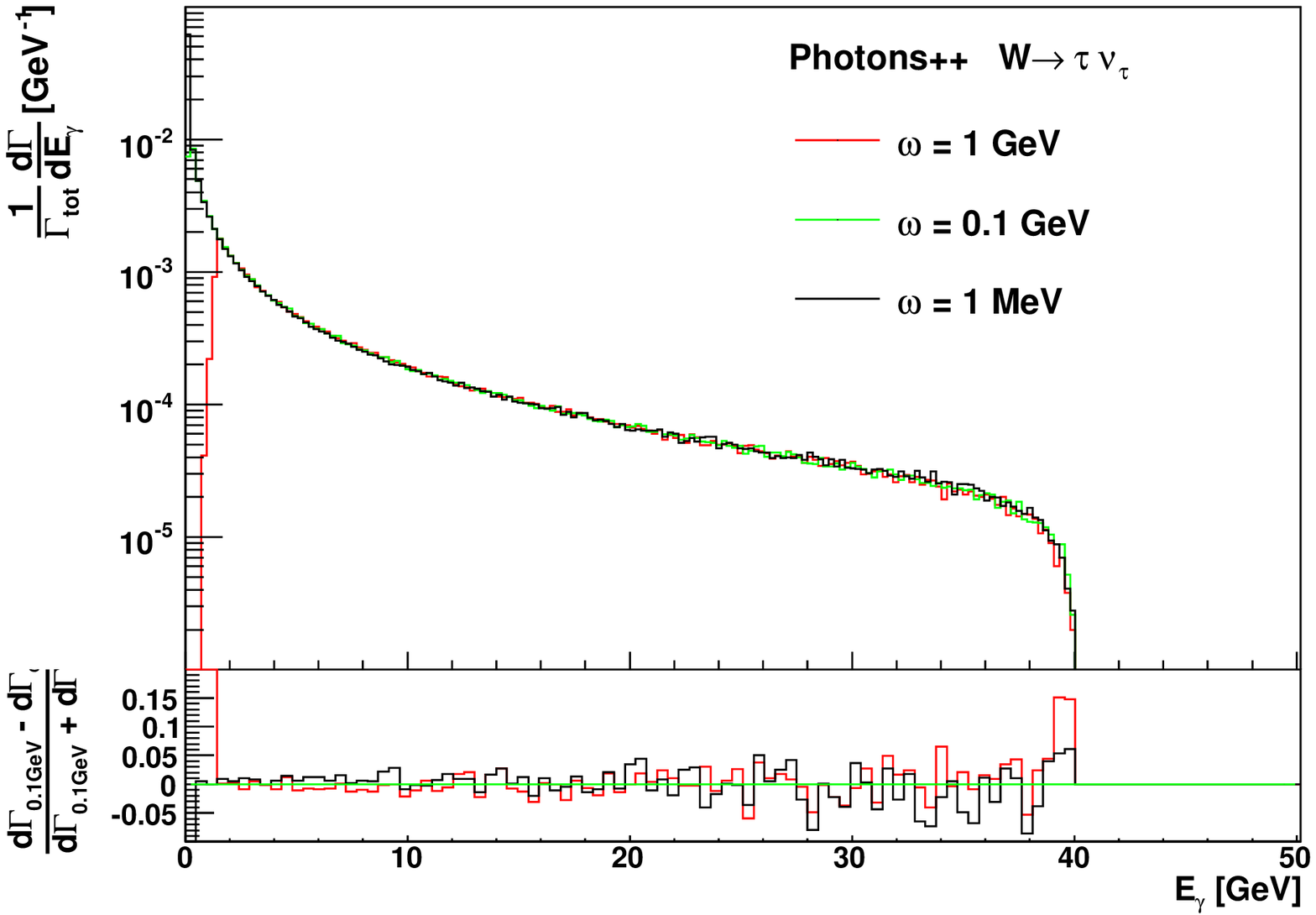}
  \includegraphics[width = 230pt]{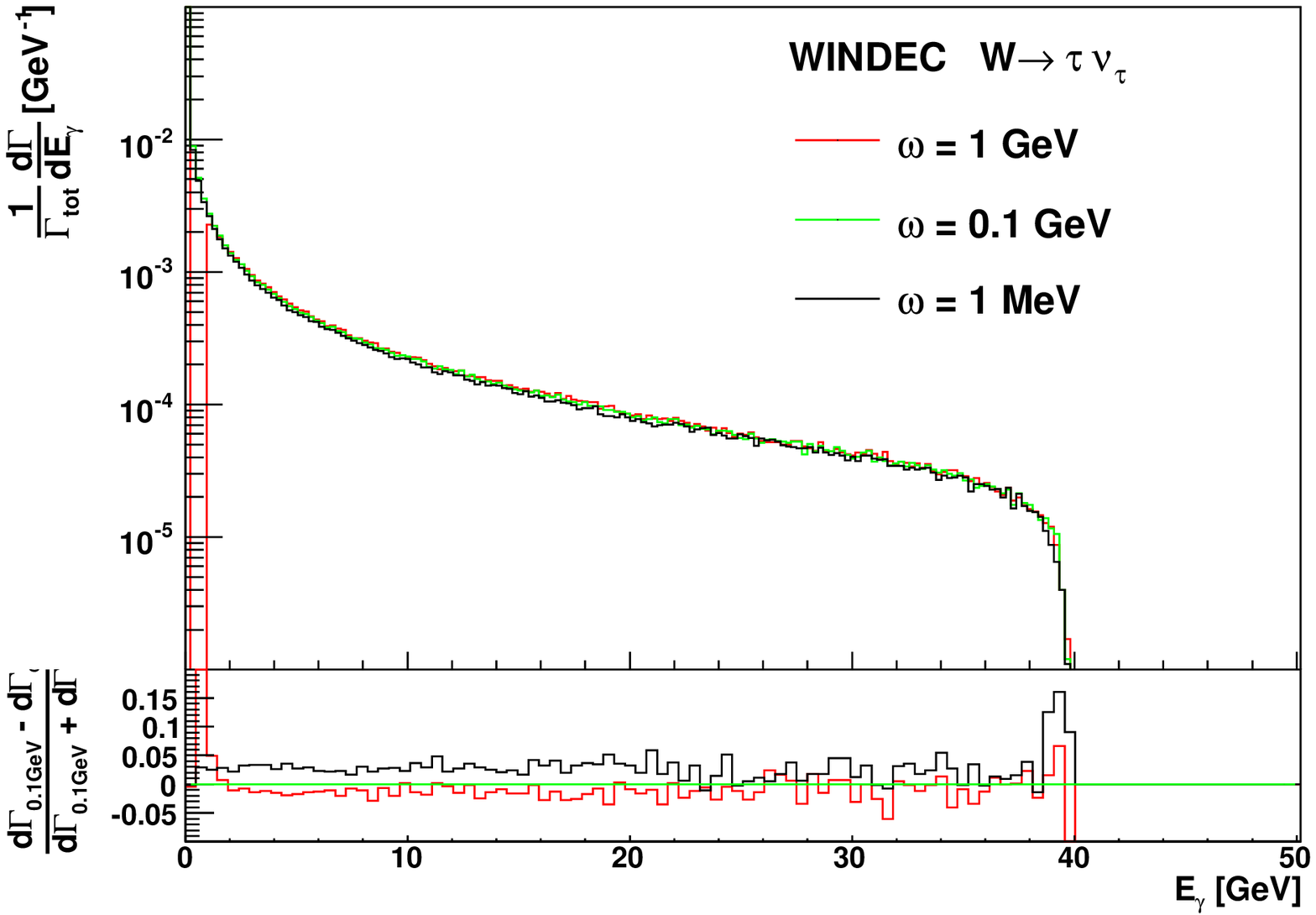}
  \caption{Dependence of the total energy of the radiated photons' on the 
	   infrared cut-off in \protect\Photonspp (left panel) and 
	   \protect\Windec (right panel) for $W\to e\nu_e$ (upper panel) 
	   and $W\to\tau\nu_\tau$ decays (lower panel). The relative 
	   difference to $\omega=0.1$GeV is shown.
	   }
\label{cut_off_dep}
 \ec
\end{figure}

The distributions generated by all three programs are shown in Fig.~\ref{W_tau_comp}. 
In general terms, the distributions agree reasonably well with each other.  There
is, however, a slight deviation in the region of large radiated energies, where
\Windec undershoots the results of the two other codes on the level of up to
10\%.  On the other hand, \Windec exhibits an overshoot in the very low energy 
bins, for radiated energies around or smaller 5 GeV, which is due to the different 
frames in which the infrared cut-offs are applied.  As already mentioned, in 
\Windec this is defined in the $W$ rest frame, hence resulting in a flat 
hypersurface in this frame.  In contrast, in \Photonspp it is applied in the rest 
frame of the $W$-$l$-dipole. Subsequently the surface of the region cut off by 
this definition forms a directionally dependent hypersurface in the rest frame of 
the $W$ (observable in Fig.~\ref{cut_off_dep} where the cut-off is set to $1\GeV$). 
The net result is that some photons having more than $0.1\GeV$ in this frame had 
less than $0.1GeV$ in the rest frame of the dipole, and vice versa.   Ultimately, 
different defintions of the infrared cut-off result in different behaviour 
in the vicinity of this cut-off in nearby frames. The differences are the larger 
the further both frames are apart.  On the other hand, the differences at high 
photon energies most likely stem from different mapping procedures in both codes. 
The mapping procedure in \Photonspp, cf.\ Sec.~\ref{Sec:char_init_mapping}, does
not involve neutral particles, in this case the neutrino.  It therefore ensures 
that the full phase space possible for the radiative decay can be mapped onto 
the leading order one.  

Another feature to study is the dependence of the distributions on the choice 
of the infrared cut-off $\omega$.  It is employed to separate the divergent region 
of real soft photon emission, which is exponentiated together with the virtual 
contributions, from the region of the phase space where resolvable photons will 
be generated.  Accordingly, this cut-off sets a limit on the number of photons to 
be generated.  In Figure \ref{cut_off_dep} the results of this variation on the 
spectrum of the radiated photons' total energy are exhibited for two different 
final states, electrons (upper panel) and $\tau$'s (lower panel).  In the case 
of the decays $W\to\tau\nu$ the two codes, \Photonspp (left panel) and \Windec 
(right panel) show a similar behaviour: Varying the cutoffs between 1 MeV and 
1 GeV yields stable results in large regions and especially also in the high-energy 
tail of the distribution, whereas differences appear only in the region of small 
energies, around 1-2 GeV.  However, in the case of the decays $W\to e\nu$ the
differences between the two codes are more pronounced.  Varying the cutoff there 
yields still comparably stable results for \Photonspp, but the results of \Windec 
show a significant dependence on the cut-off of the order of around 10\%.  This is 
due to the fact that with decreasing fermion mass the effect of the infrared 
cutoff on the avarage photon number increases\footnote{
	In fact, this feature was one of the reasons for preferring 
	the $\tau$ decay channel over the electron channel.}.

In order to choose an optimal value of the infrared cut-off $\omega$ there are
different considerations to be taken into account: On the one hand an efficient 
generation is desirable, pushing $\omega$ as high as physically sensible, e.g.\ 
the detector level energy resolution on soft photons or decay products.  Along 
the same lines it should be noted that all photons in the soft (unresolved) 
region will be assumed to yield a negligible combined momentum.  Therefore,
choosing a comparably large infrared cut-off will not have any effect on 
distributions involving the resolved Bremsstrahlung photons, but it will reduce 
the accuracy of results obtained for e.g.\ invariant masses of the primary 
decay products.  This consideration clearly demands a smaller cut-off.  On 
the other hand, when exponentiating the real soft photon emission a factor 
$\int\frac{d^3k}{k^0}\tilde{S}(k)(e^{-iyk}-1)\Theta(\omega-k^0)$ has been 
neglected, which is strictly true only for $\omega\to 0$.  Thus, some residual 
dependence is to be expected, even if infrared subtracted matrix element 
corrections were included to all orders.  This dependence is of course minimised
with small cut-offs.  

\subsubsection{Effects of inclusion of exact matrix elements}

\begin{figure}
\bc
\includegraphics[width = 230pt]{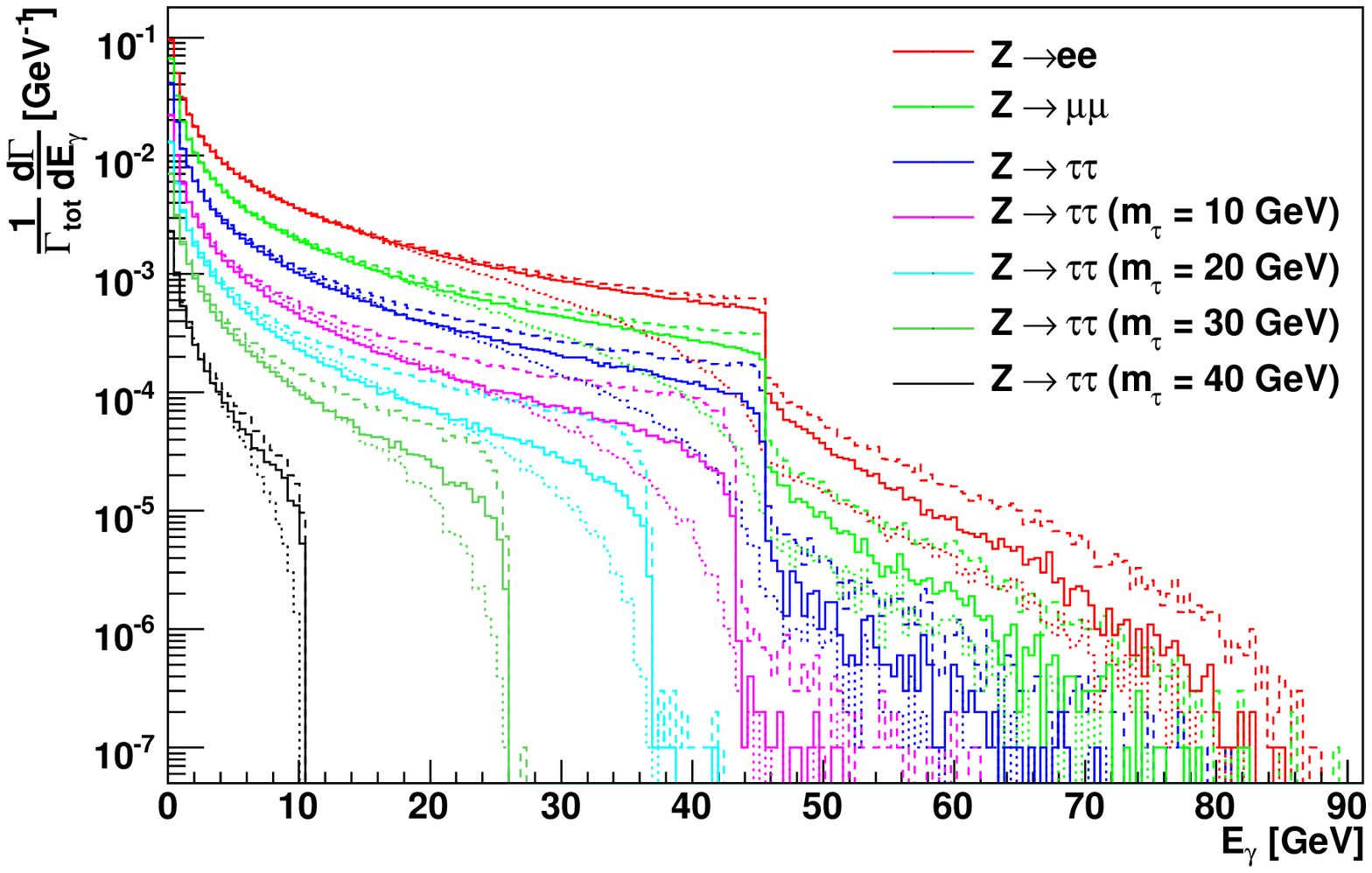}
\includegraphics[width = 230pt]{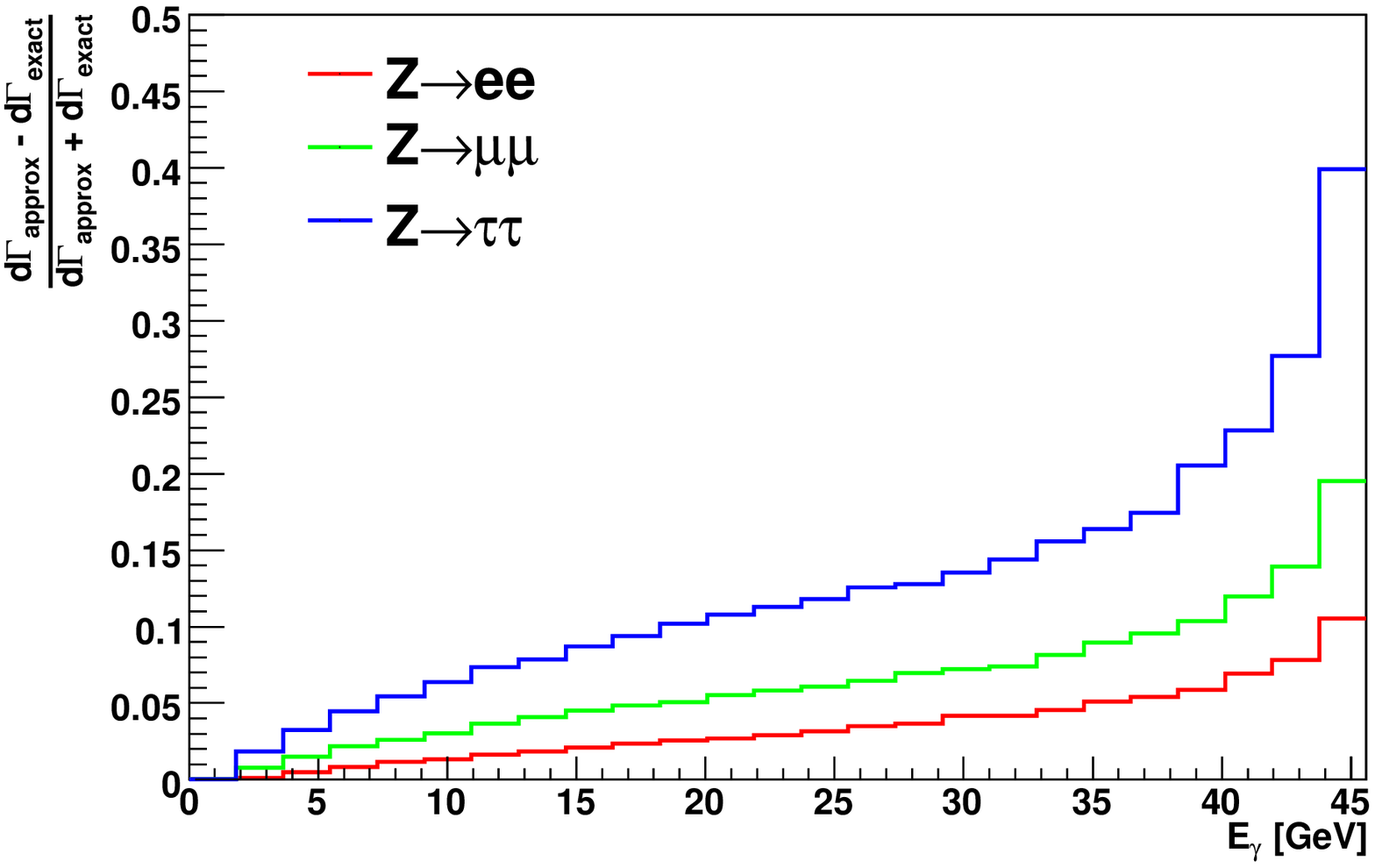}
\caption{The total energy of all photons radiated in $Z\to\ell\bar\ell$.  
	Left panel (a): The same plot as  in Fig.~\ref{Z_W_dist}(b), but this 
	time the correction is done by using the {\bf exact} matrix element 
	(solid)	instead of the {\bf approximated} one (dashed). The 
	distribution generated using the eikonals only is shown as a dotted 
	line. Right panel (b): The relative difference of the distributions 
	obtained using the exact and the approximated matrix elements. In both 
	cases again different fermion masses have been used.
	}
\label{Fig:Zll_ME}
 \ec
\end{figure}

\begin{figure}
 \bc
 \includegraphics[width = 230pt]{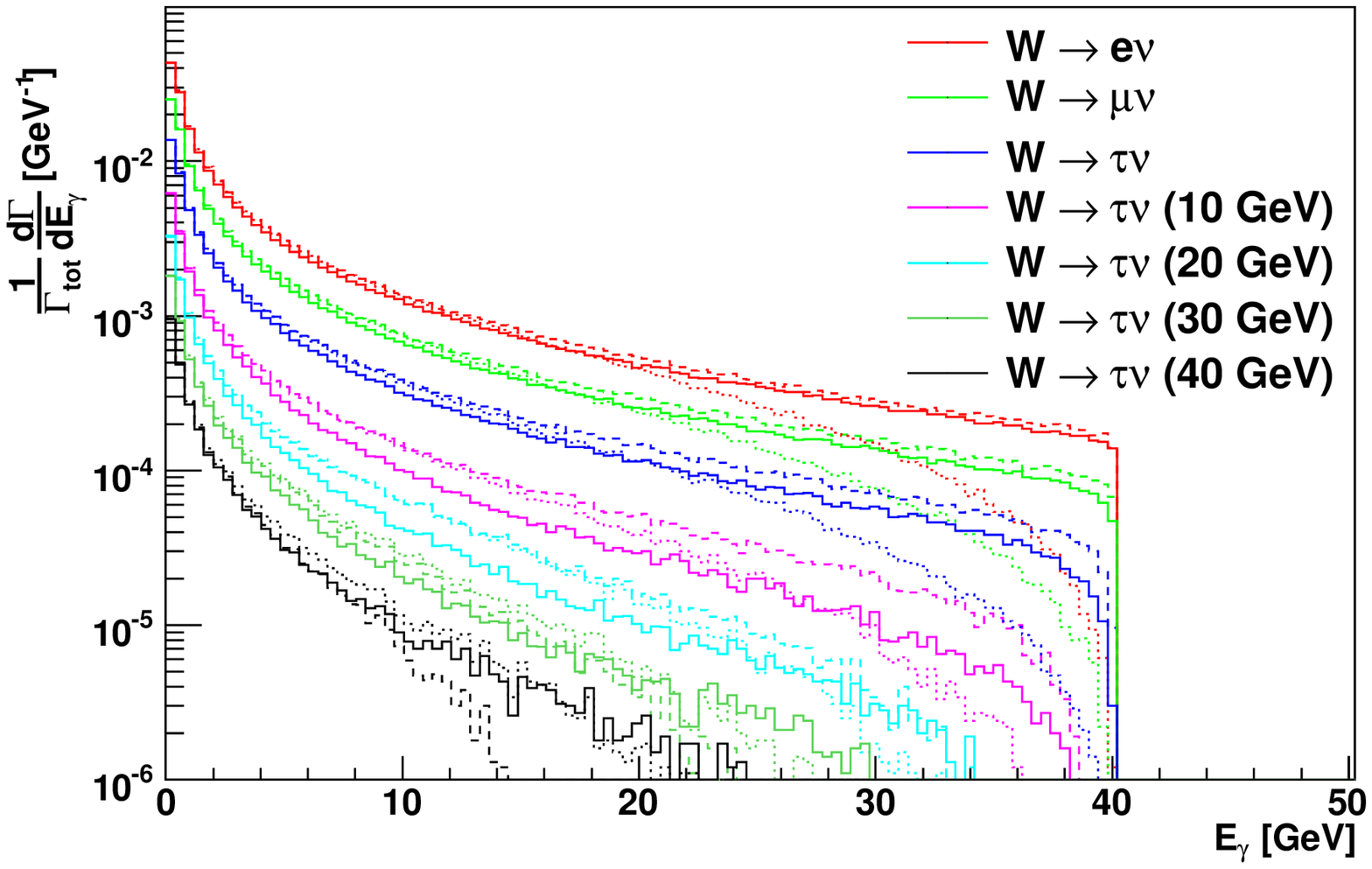}
 \includegraphics[width = 230pt]{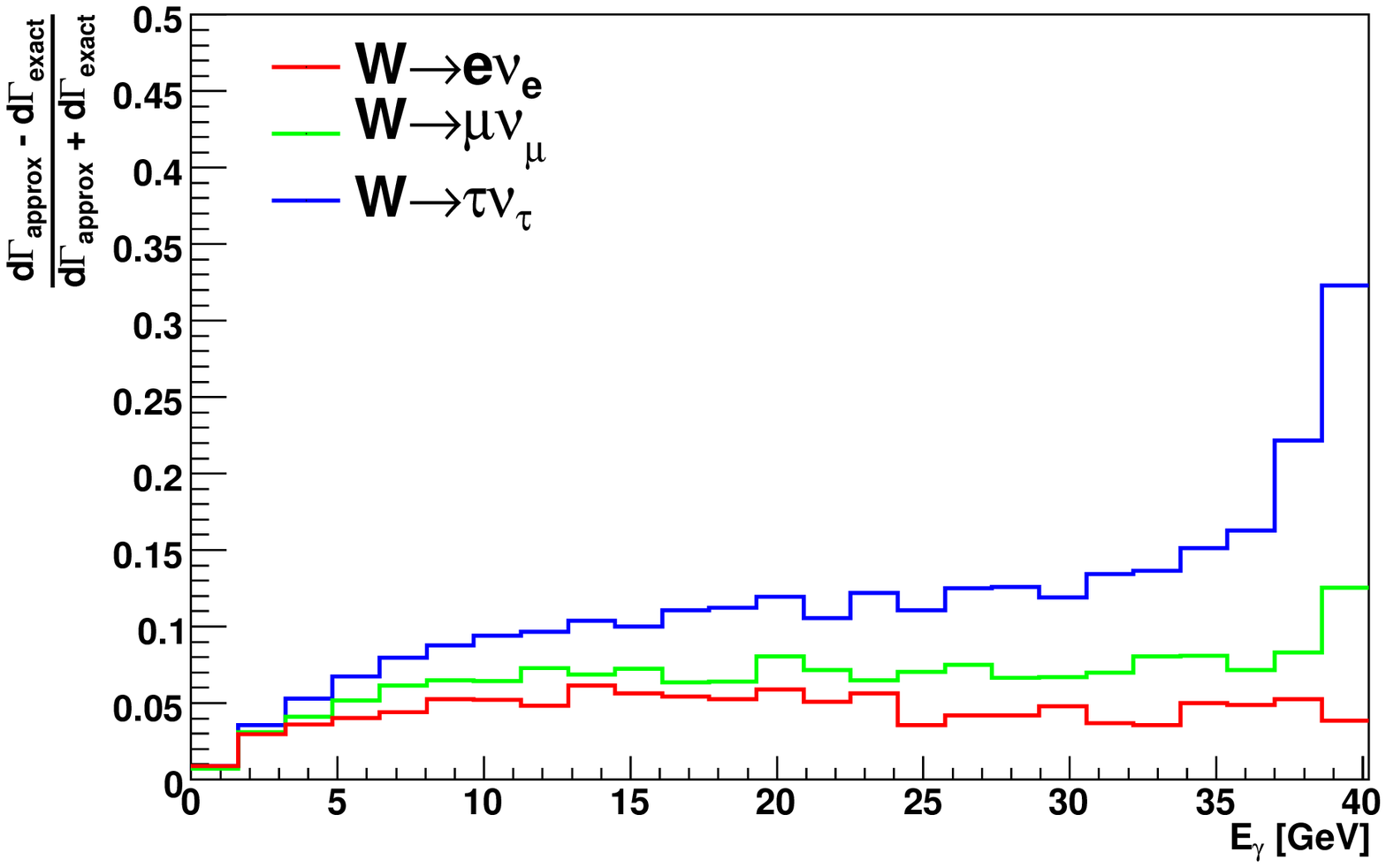}
\caption{The total energy of all photons radiated in $W\to\ell\nu$.  
	Left panel (a): The same plot as  in Fig.~\ref{Z_W_dist}(d), but this 
	time the correction is done by using the {\bf exact} matrix element 
	(solid)	instead of the {\bf approximated} one (dashed). The 
	distribution generated using the eikonals only is shown as a dotted 
	line. Right panel (b): The relative difference of the distributions 
	obtained using the exact and the approximated matrix elements. In both 
	cases again different fermion masses have been used.
	}
\label{Fig:Wln_ME}
 \ec
\end{figure}

\begin{figure}
 \bc
 \includegraphics[width = 230pt]{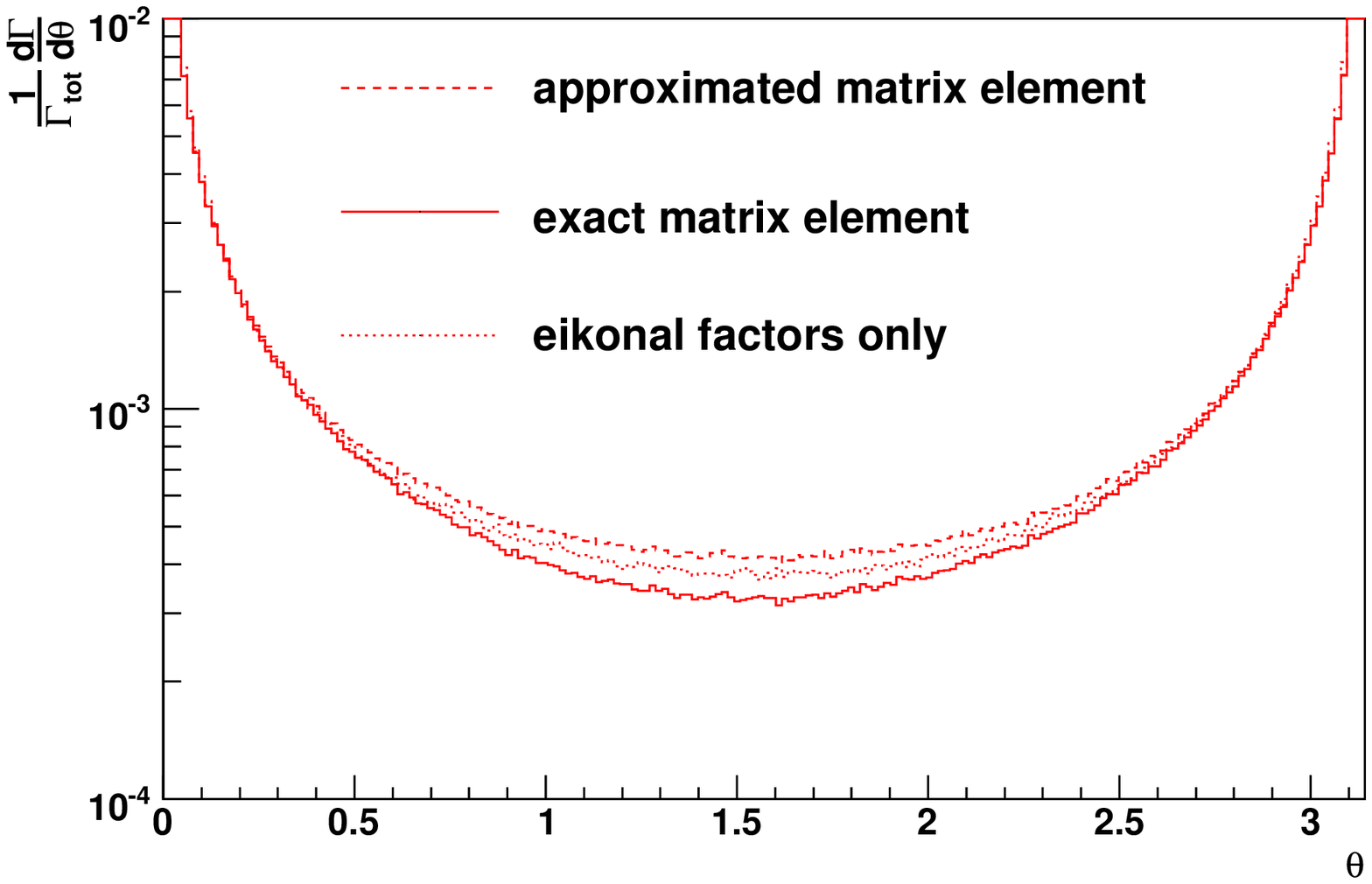}
 \includegraphics[width = 230pt]{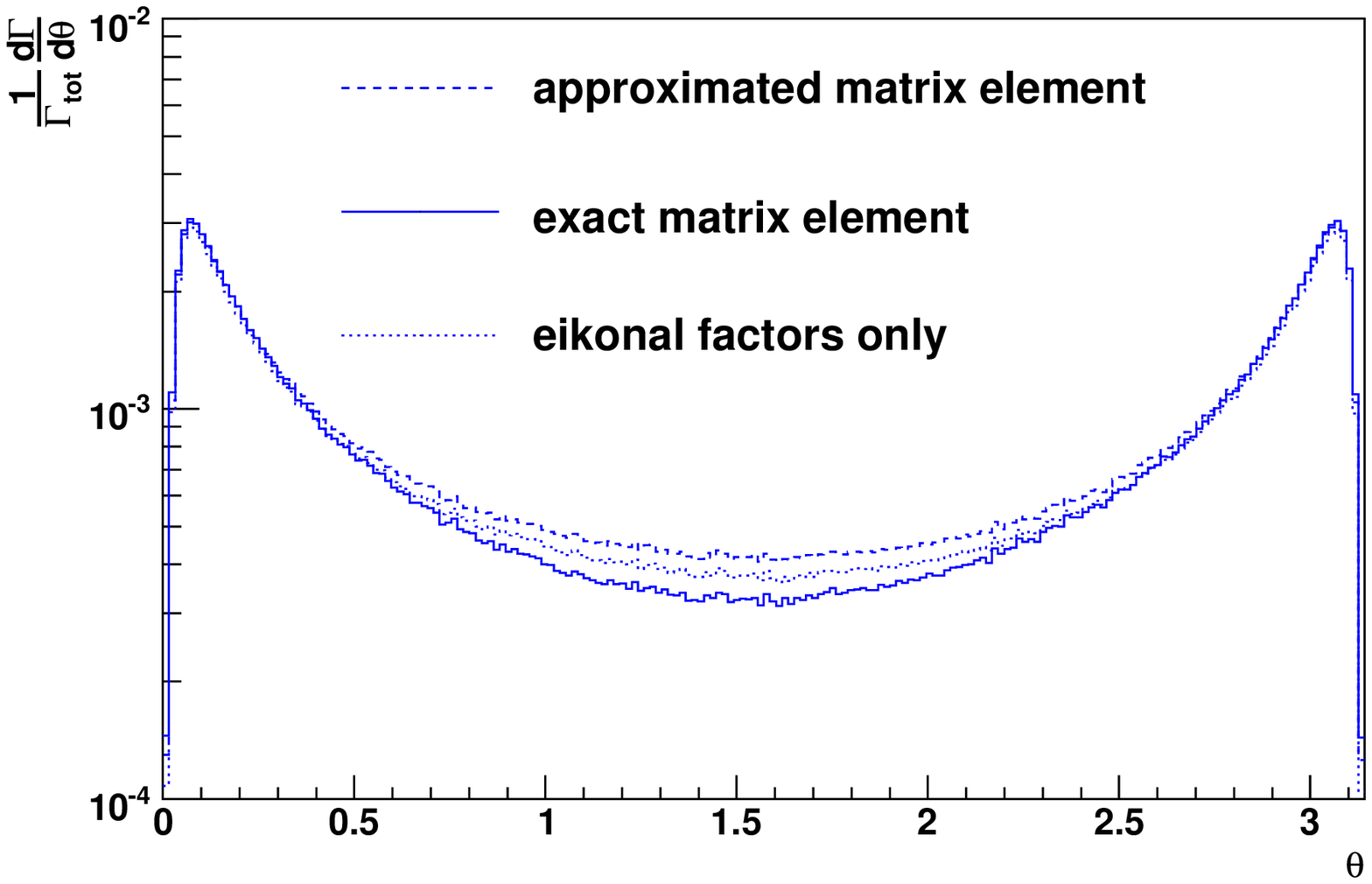}
  \caption{Angular distributions of the emitted photons in $Z\to\ell\bar\ell$,
	using exact and approximated matrix elements.  In the left panel (a) 
	and the right panel (b), the cases $Z\to e e$  and $Z\to \tau \tau$ are 
	exhibited using the eikonals only (dotted lines) and corrections through 
	exact (solid) and approximated matrix elements (dashed). In both plots 
	the leptons sit at $\theta = 0$ and $\theta = \pi$.
	}
\label{Fig:Z_l_ME_angles}
 \ec
\end{figure}

Including exact matrix elements, as discussed in Section \ref{Sec:RealCorrections}, 
further improves the accuracy of the distributions.  This is especially true
away from the singular limits, where considerable differences emerge.  This is 
exemplified in Fig.~\ref{Fig:Z_l_ME_angles}, where the angular distributions 
of photons in $Z\to\ell\bar\ell$ decays is depicted.  Of course, there is also 
an effect on the differential decay rate.  Fig.~\ref{Fig:Z_l_ME_angles} shows
that corrections obtained from the quasi-collinear approximation, i.e.\ from
the approximate matrix element, overestimate the exact matrix element 
resulting in an increased differential decay rate.  Even more so in 
the region of very hard photon emission which, due to the angular constraints 
imposed by the emitter's mass, no longer fulfils the condition 
$(p\cdot k)\to 0$. Here the full matrix element exhibits some 
destructive interference between the two relevant diagrams which is of course
absent in the treatment through the dipole splitting kernels. This leads to a 
slightly earlier drop-off of the differential decay rate w.r.t.\ radiated 
energy than with the approximated matrix elements.

Further, the absence of interference terms in both the eikonals and the 
quasi-collinear approximation leads to an overestimation of radiation at large 
angles. Because of only small correlations between the energy of the photon 
radiated and its angular distribution this overestimation leads to an almost 
constant decline in the differential decay rate w.r.t.~the photon energy when 
corrected by the exact matrix element. 
Of course, while this effect is small in the decay channel $Z\to e^+e^-$, it
increases with the mass of the emitter and when a larger fraction of the radiation 
is radiated at large angles.  Nevertheless, for very high emitter masses (cf.\ 
the fictive $\tau$ with $m_\tau = 40\GeV$ in Fig.~\ref{Fig:Zll_ME}) the 
approximation proves useful again. This is due to the dominance of the soft 
logarithms over the quasi-collinear ones in this limit.

\subsection{Other channels}

Finally, a short overview over other interesting cases is given.  In principle,
\Photonspp can handle any possible final state configuration in single particle 
decays independent of its charge.  Thus, it is well suited to address all $\tau$- 
and hadron decays, which will be the topic of this section.

\subsubsection{$J/\Psi$ decays to leptons}

\begin{figure}
 \bc
  \includegraphics[width = 230pt]{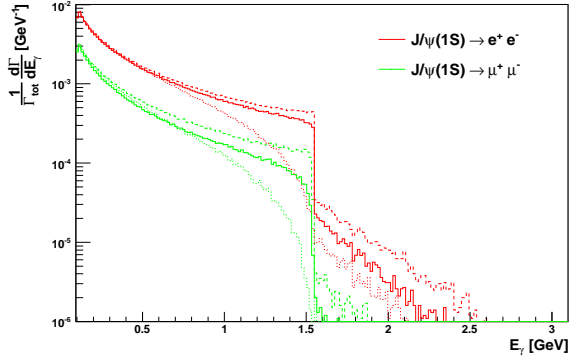}
  \caption{The total energy of the radaited photons in the rest frame of the 
	decaying $J/\psi$ vector meson for different lepton pairs (electrons
	in red, muons in green) in the final state.  $\mathcal{C} = 1$ (dotted) 
	is contrasted with $\mathcal{C} = 1+\tilde{\beta}_1^1/\tilde{\beta}_0^0$, 
	where $\tilde{\beta}_1^1$ is calculated in the quasi-collinear approximation 
	(dashed) and with the complete real emission matrix element (solid).  In
	all cases, the distributions are normalised on the width of the 
	inclusive decay into the respective lepton pair, and the infrared 
	cut-off has been fixed to $\omega=1\MeV$.  
	}
 \label{J_psi_dist}
 \ec
\end{figure}

First of all, consider the case of $J/\Psi\to\ell\bar\ell$, which is 
topologically identical to leptonic $Z$-decay, but nonetheless very important
for the calibration of detectors and as a background source of leptons.  
In Fig.~\ref{J_psi_dist} the decay channels $J/\psi\to e^+e^-$ and 
$J/\psi\to\mu^+\mu^-$ are investigated and the effect of $\order(\alpha)$ 
corrections is scrutinised.  Again, the kinematic limit at half the mass of 
the decaying particle produces a visible and prominent kink.  Due to the much 
smaller mass of the $J/\psi$ compared to the $Z$ mass, the effects of the 
higher muon mass are much more pronounced, both in the sharpness of the kink 
and the quality of the quasi-collinear approximation.

\subsubsection{$B\to D^*+$ pions and semileptonic $B$ decays}

\begin{figure}[t]
 \bc
  \includegraphics[width = 230pt]{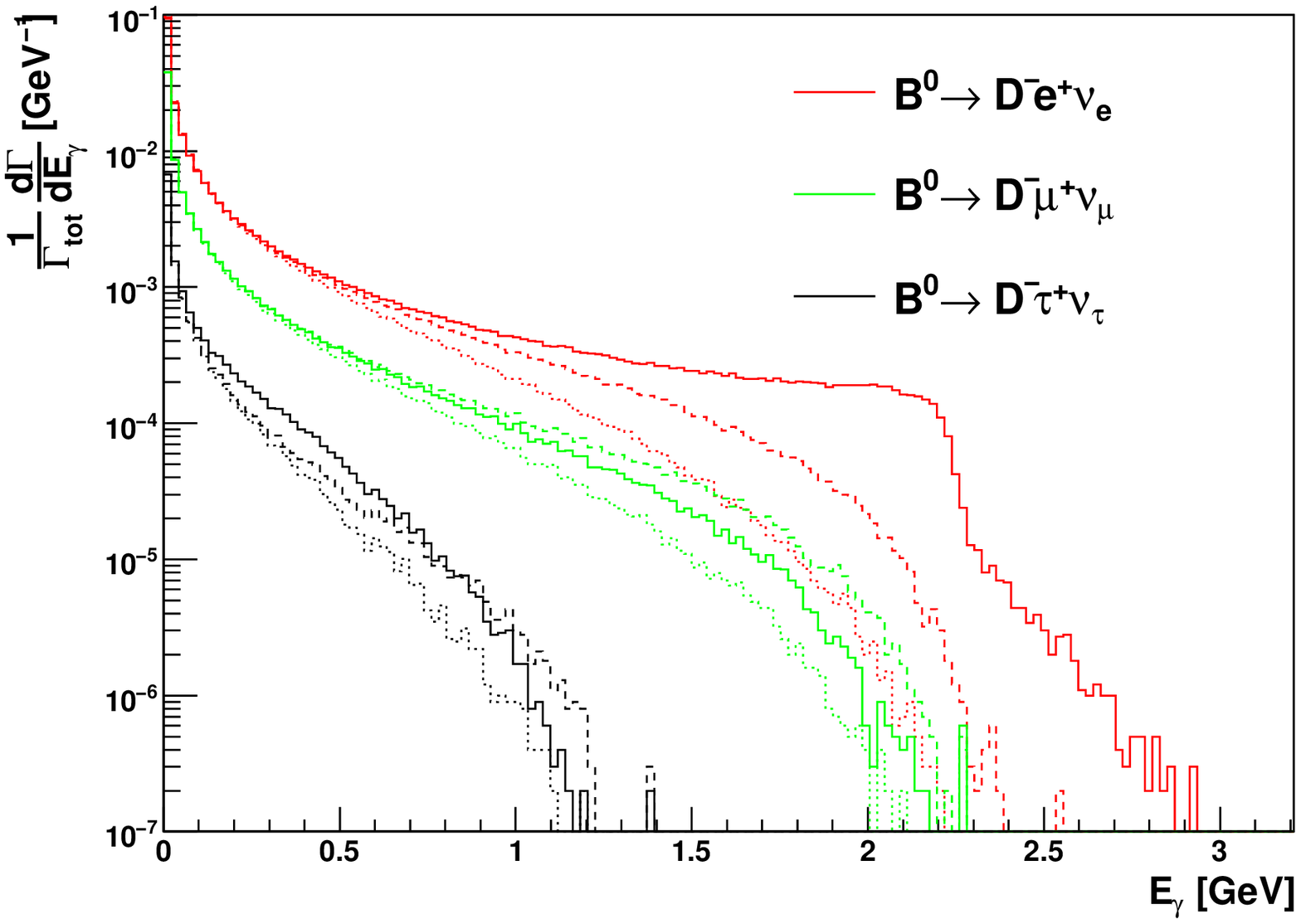}
  \includegraphics[width = 230pt]{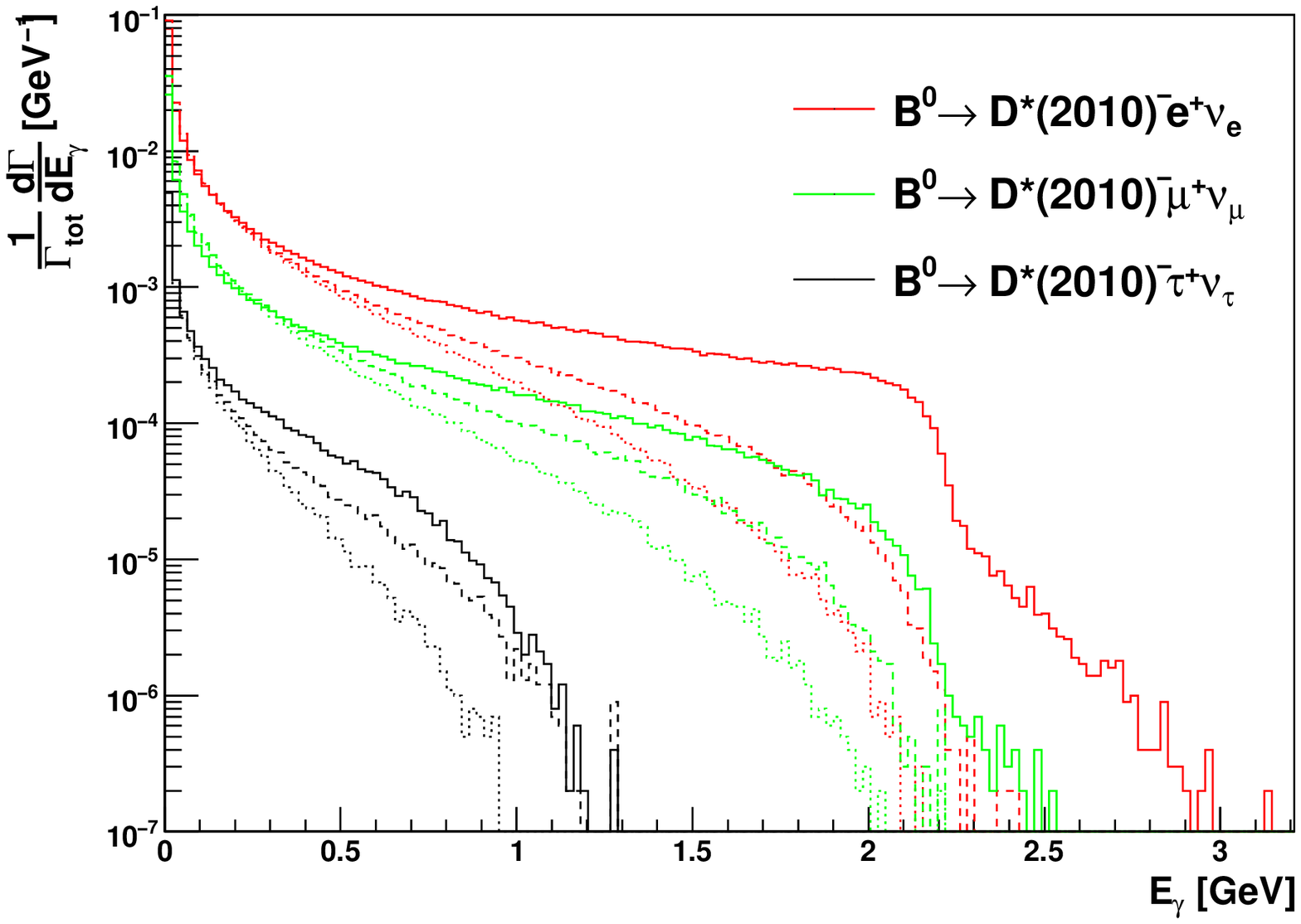} \\
  \includegraphics[width = 230pt]{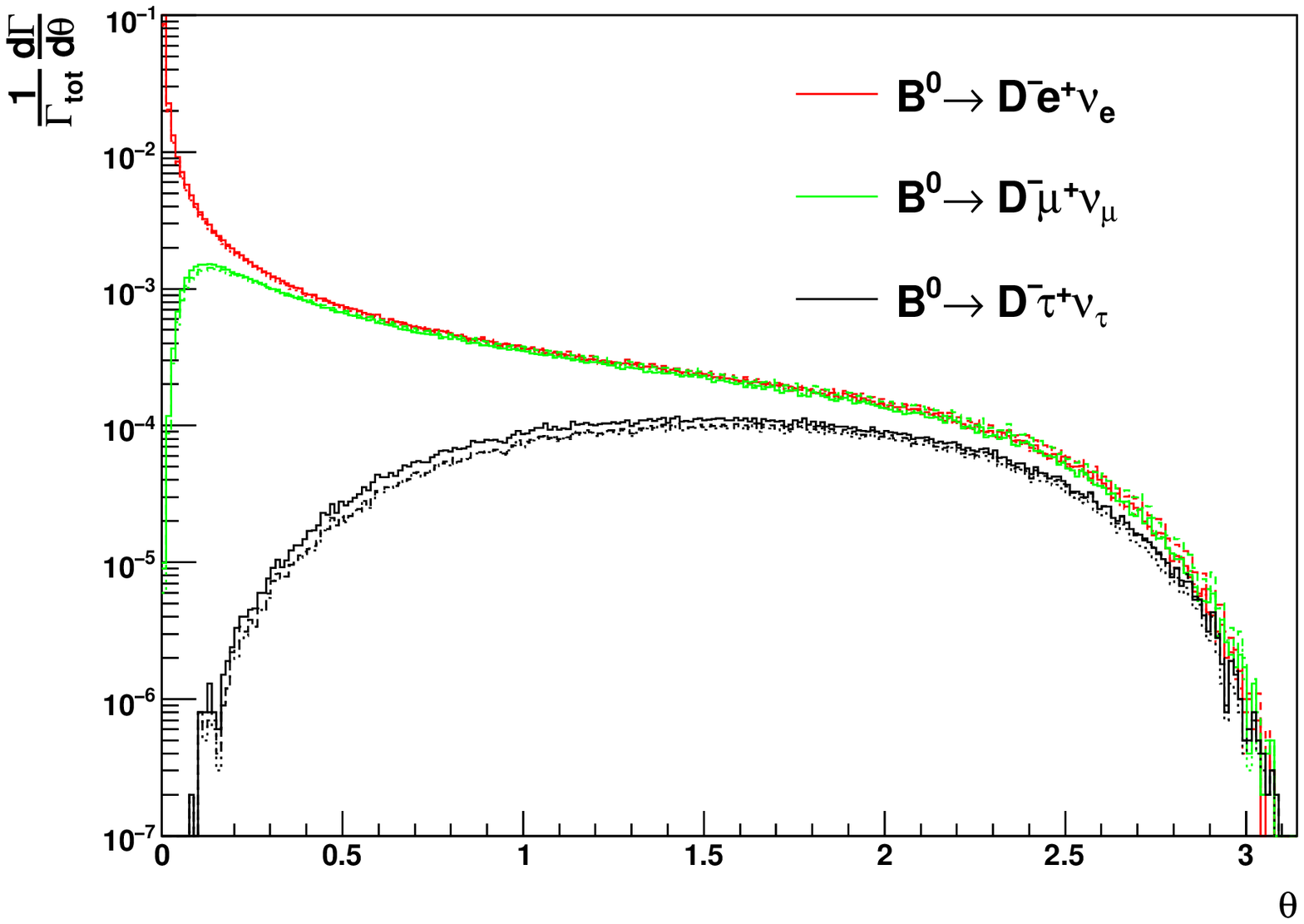}
  \includegraphics[width = 230pt]{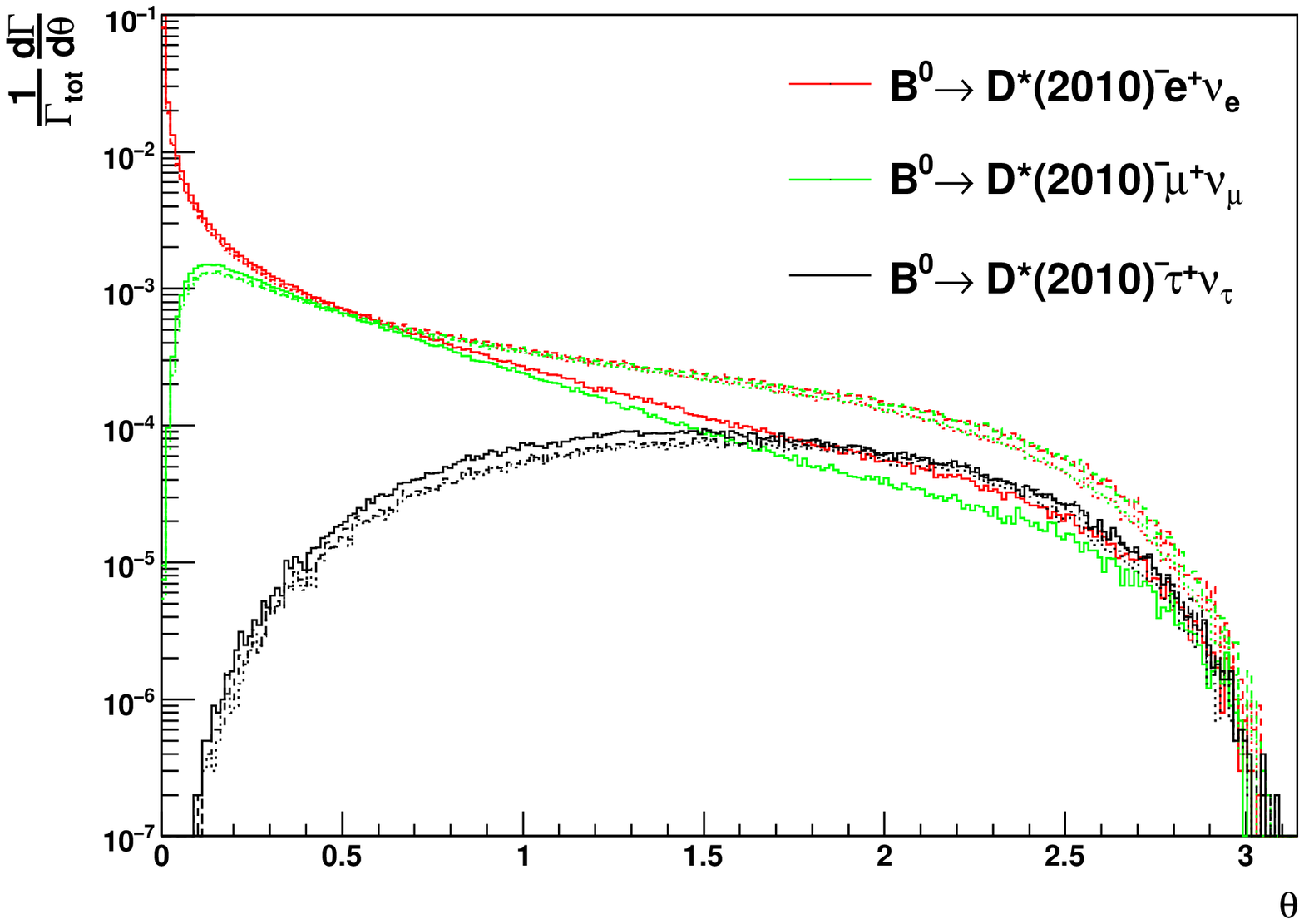}
  \caption{Semi-leptonic decays $B^0\to D^-\ell\nu$ and 
	$B^0\to D^\ast(2010)^-\ell\nu$ for different leptons and with 
	different matrix element corrections. The solid line corresponds 
	to the correction using the full matrix elements in the point-like 
	hadron approximation, the dashed line corresponds to the dipole 
	splitting kernels neglecting interference terms and the dotted 
	line corresponds to using the eikonals only. The angular distributions
	are shown in the $\ell-D^{(\ast)-}$ rest frame with the lepton at
	$\theta=0$. Again, the infrared cut-off was set to $1\MeV$.
	}
\label{Fig:Semi_Lep_B}
 \ec
\end{figure}

Another system to demonstrate the versatility of \Photonspp are $B$-decays 
because of its manyfold topologies in the final state.  

In Figure \ref{Fig:Semi_Lep_B} semi-leptonic decays of $B^0$ mesons into
$D^-$ scalars and $D^{\ast-}$ vectors are displayed.  The resulting 
distributions are similar for $e$ or $\mu$ being the lepton.  This is because 
in both cases the bulk of the radiation is emitted off the lepton and the amount 
of phase space open for bremsstrahlung is of similar magnitude. Only the avarage 
photon multiplicity is noticably affected by the difference in mass between the 
electron and the muon. The $\tau$-channel on the other hand presents itself 
differently due to the mass of the tau being comparable both to the mass of the 
$B^0$-~and the $D^{(\ast)-}$-mesons. This not only leads to the near absence of soft 
bremsstrahlung above the infrared cut-off, as compared to the other semi-leptonic 
channels, it also results in a completely different radiation pattern: The bulk 
of the photons is radiated in between both dipole particles and not primarily 
collinearly.  Furthermore, the absence of interference terms in the radiation off 
the lepton-scalar pair in contrast to the lepton-vector pair is plainly visible
for both $e$ and $\mu$. Again, the relevance of the interference terms is small for 
the $\tau$-mode due to its radiation being dominated by the spin-independent soft 
terms.  The exact matrix element correction also shows the shortcomings of the dipole 
splitting functions in this case as they fail to predict the excess of hard radiation 
for the eletron. This attribute is shielded in the muon case by its already comparable 
large mass.  However, the total radiative decay rate 
is nearly uneffected by this. The spin-dependence of the dipole approximation is 
also suppressed by the large mass of the $D^-$ and $D^{\ast-}$, respectively, 
hence the small difference of both cases in that approximation.

\begin{figure}[thbp]
 \bc
  \begin{picture}(480,300)
   \put(90,110){\includegraphics[width = 300pt]{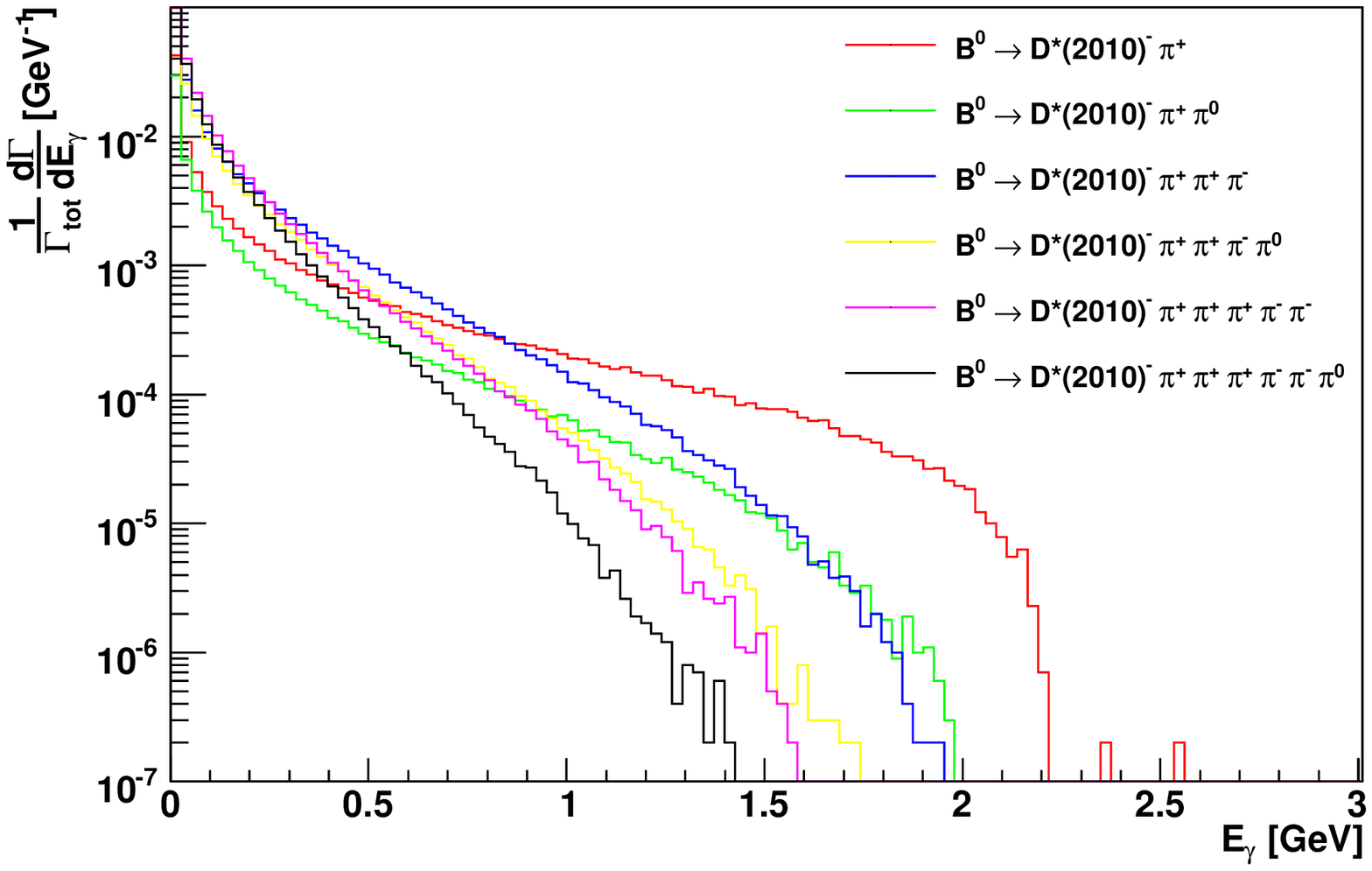}}
   \put(0,0){\includegraphics[width = 160pt]{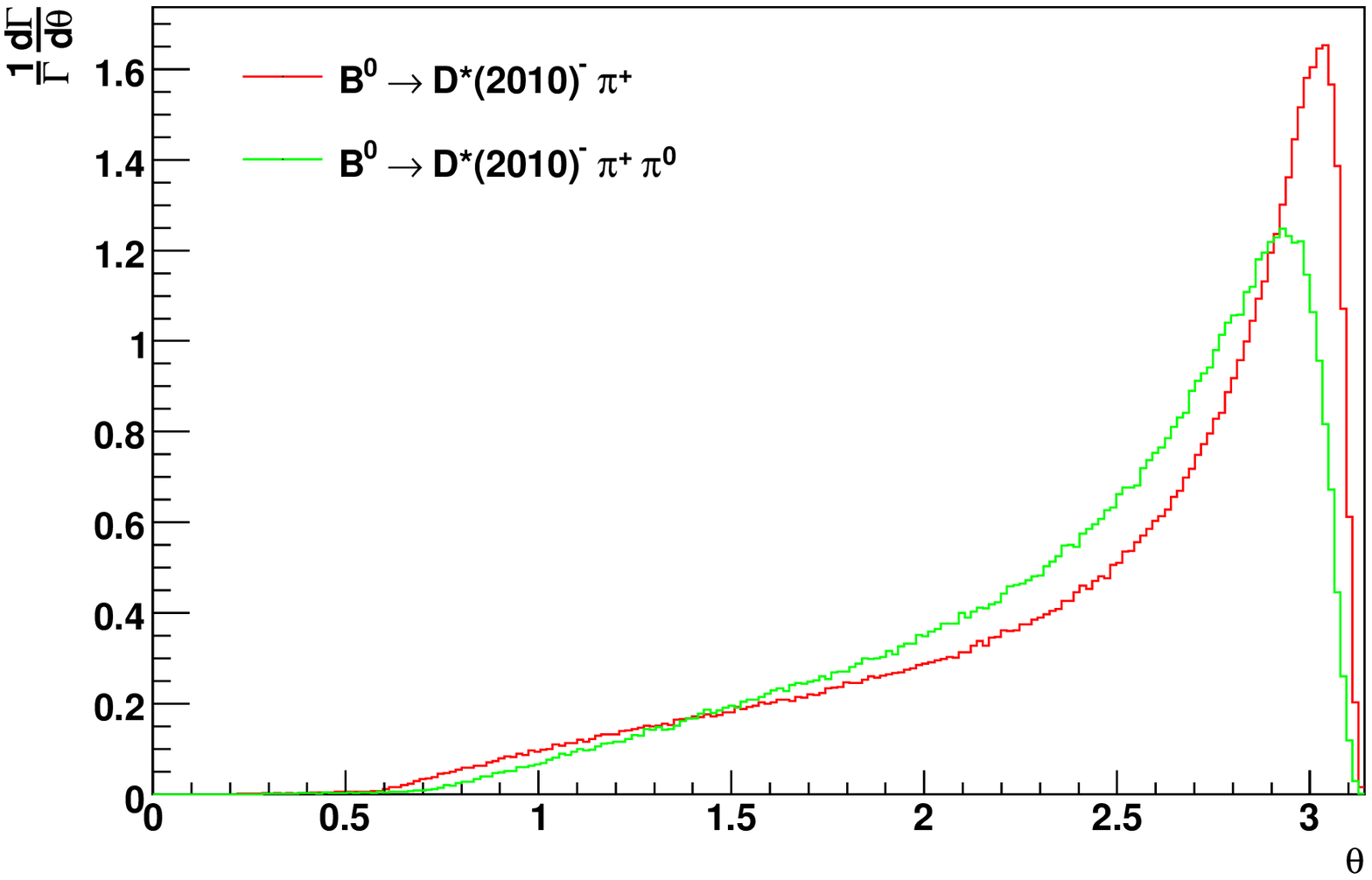}}
   \put(160,0){\includegraphics[width = 160pt]{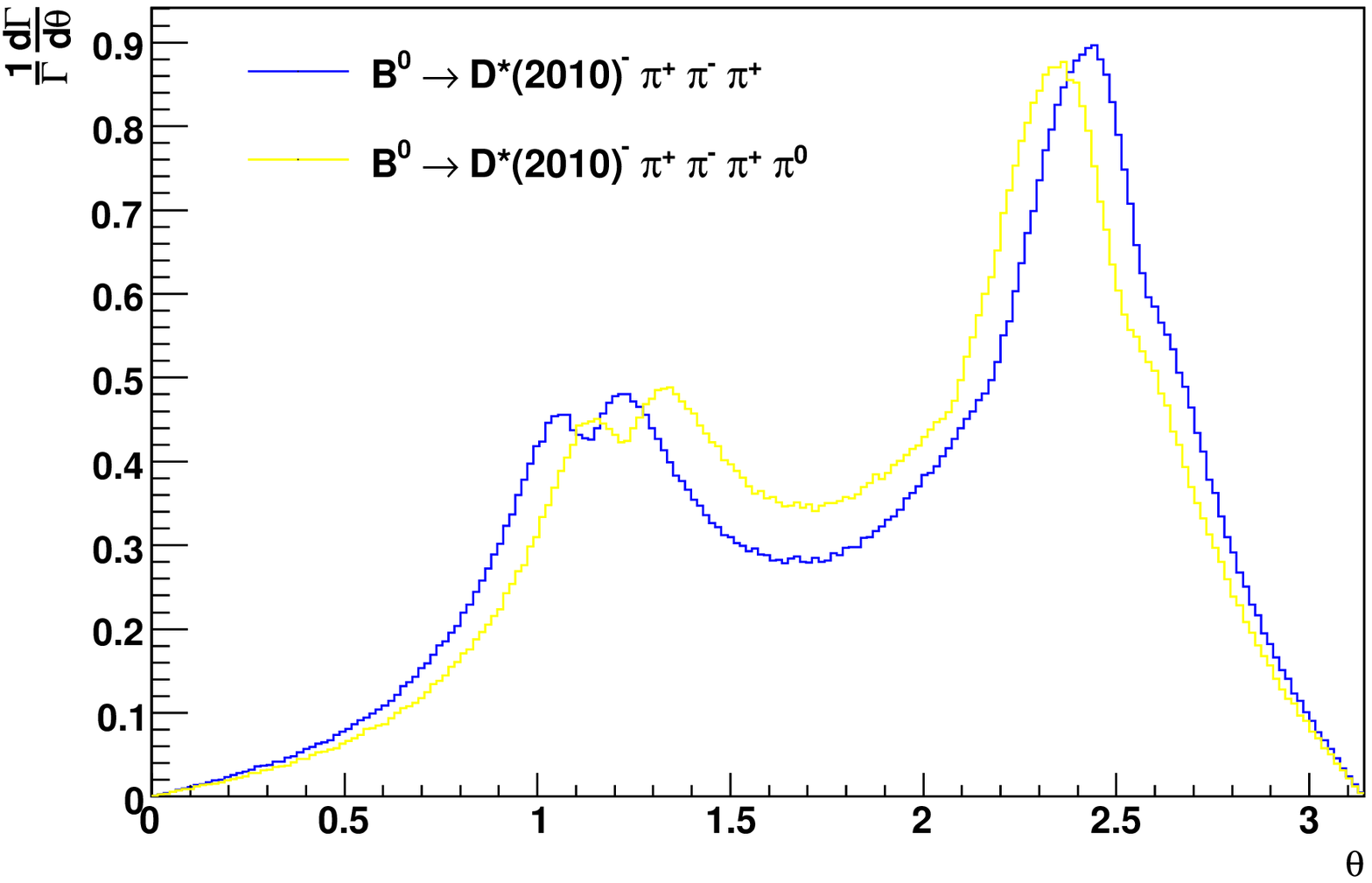}}
   \put(320,0){\includegraphics[width = 160pt]{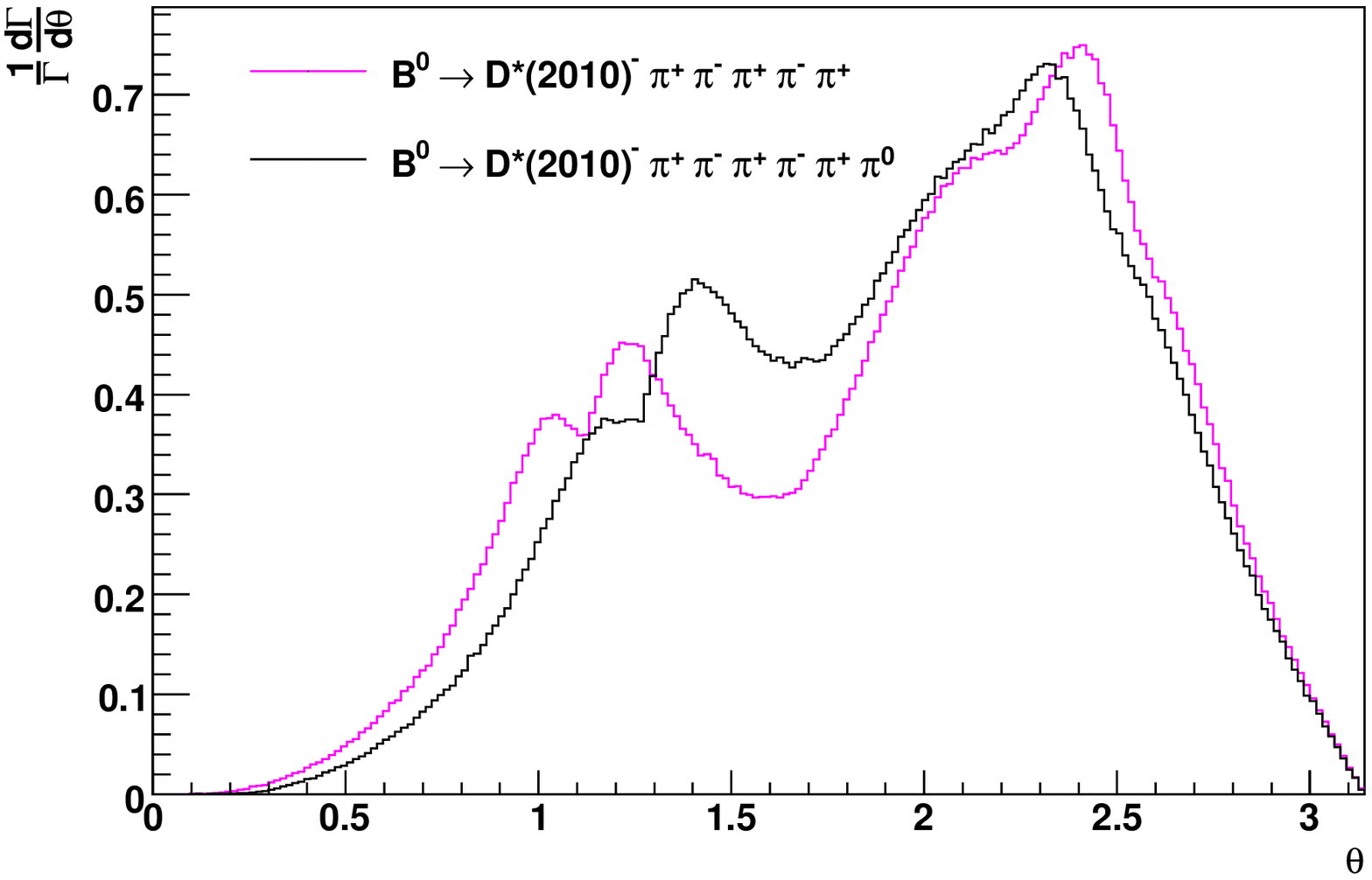}}
  \end{picture}
  \caption{The total photon energy in the rest frame of the decaying $B^0$ 
	meson for different numbers of pions in the final state (upper plot) 
	and the angular distribution of this radiation in the multipole rest 
	frame with the $D^{\ast-}$ at $\theta=0$ (lower panel).  For the 
	multi-body final states the same kinematic configurations have been 
	used, as detailed in the text, to yield easily interpretable results.  
	For identical multipoles similar final state momentum configurations 
	with non-vanishing $\pi^0$ momentum have been chosen to increase 
	comparability.  The infrared cut-off in all cases has been set to 
	$\omega = 1\MeV$.}
 \label{B_pi_dist}
 \ec
\end{figure}

As an example for dealing with multiple charged particles in the final state,
$B^0$ decays into a $D^*$ accompanied with various numbers of charged and neutral 
$\pi$'s have been chosen.  The results are on display in Fig.~\ref{B_pi_dist}, where 
the total radiated photon energy and the angular distribution of the photons are
depicted.  The orientation of the final state momenta has been chosen in such 
a way that configurations of the same multipole structure differing only by a 
neutral pion have a similar momentum distribution within the multipole, but 
still letting the $\pi^0$ have a non-vanishing effect.  The most prominent 
feature in the distriubtion of the total energy of all photons in the $B$ 
meson's rest frame is the receding kinematic limit for the total energy, it is 
independent of the momentum layout within the multipole.  It is due to the 
decreasing amount of phase space open for bremsstrahlung with an increasing 
number of pions.  On the other hand, while the total energy available for
the photon decreases, the amount of Bremsstrahlung increases with the number
of charged particles involved.  Switching from a dipole (two charged  particles)
to a quadrupole (four charged particles), the probability of double hard 
photon emission is increased due to favourable momentum configurations among 
the strongly radiating pions.  Additionally, since the pions are spin-$0$, their 
photon distribution is generated exclusively by a product of eikonal factors. 
Furthermore, the angular distributions of the emitted photons are shown.  There, 
the differential cross-section is integrated over energy and the azimutal 
angle.  For better interpretability these distributions are plotted in the rest 
frame of the multipole.  The $D^\ast(2010)^-$ allways rests at $\theta = 0$. 
Due to its large mass, compared to the pions, very little radiation is emitted 
in its direction.  In contrast, all charged pions are plainly visible as peaks 
in the spectrum.  However, their respective mass cones are hidden due to the 
azimutal integration unless the pion sits at $\theta = \pi$, as is the case 
in the dipole configurations.

\subsubsection{$\Delta^{++}\to p^+\,\pi^+$ decays}

\begin{figure}[t]
 \bc
  \includegraphics[width = 230pt]{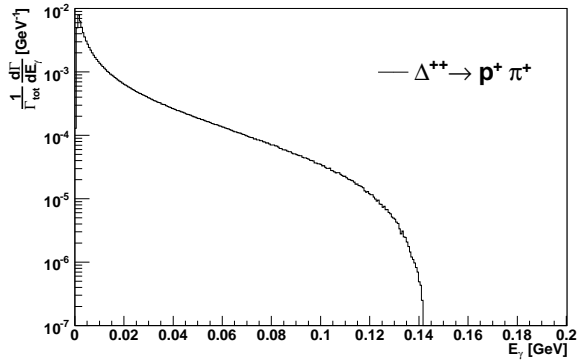}
  \caption{
	The total photon energy in $\Delta^{++}\to p^+\pi^+$ in the rest frame 
	of the decaying $\Delta^{++}$ baryon is exhibited. The infrared cut-off 
	was set to $1\keV$.
	}
 \label{Fig:Delta_dist}
 \ec
\end{figure}

A rather exotic decay for the purpose of this publication is the decay
$\Delta^{++}\to p^+\,\pi^+$, due to its lack of neutral particles.  This case 
is presented in Fig.~\ref{Fig:Delta_dist}, where the total energy and the angular
distribution of the emitted photons are exhibted.  However, this channel leaves 
only very little phase space open for photon radiation. Thus, collinear 
enhancement for the $p^+$ and the $\Delta^{++}$ should be negligible.

\subsubsection{$\tau$ decays}

\begin{figure}[t]
 \bc
  \includegraphics[width = 230pt]{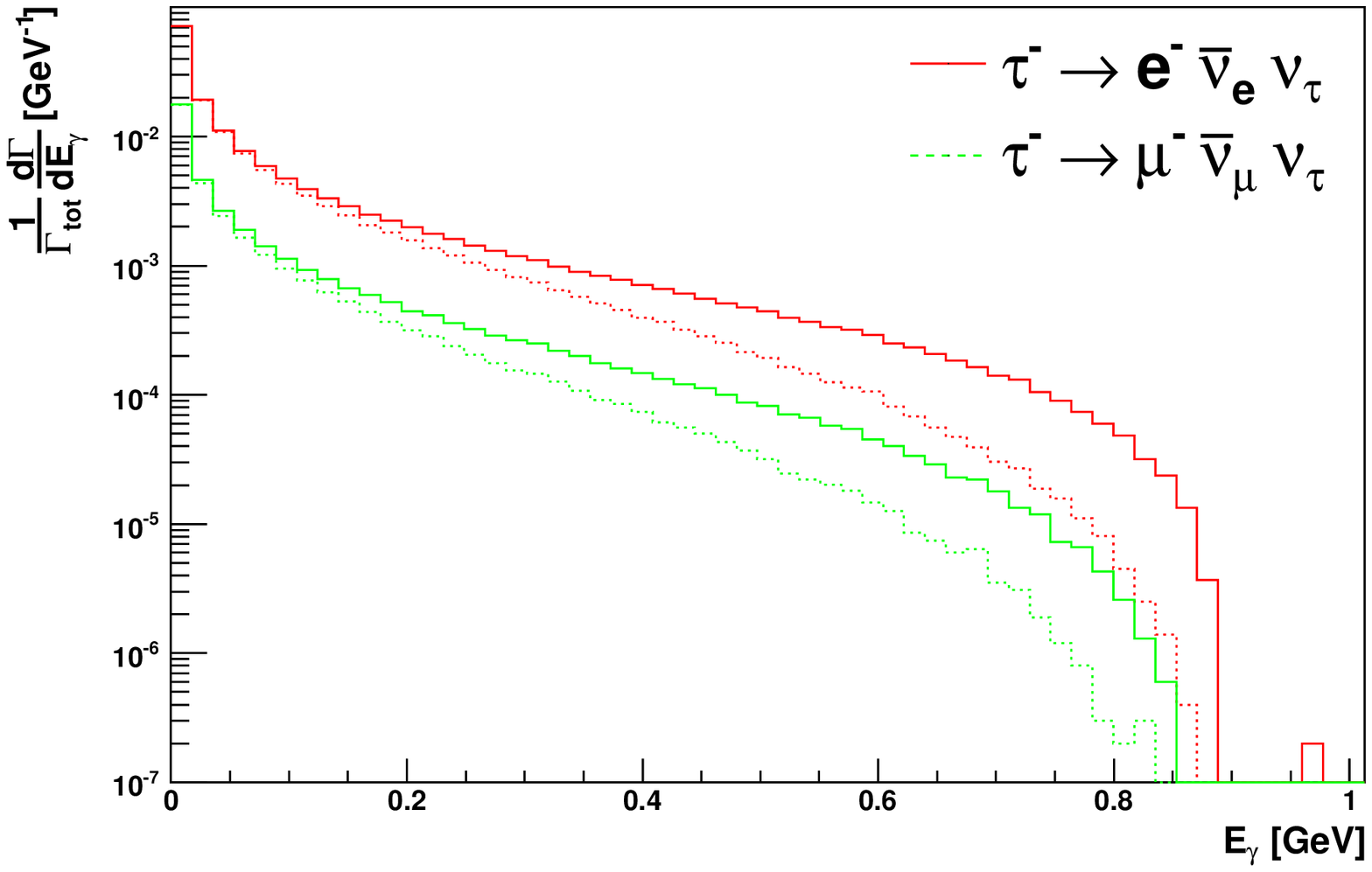}
  \includegraphics[width = 230pt]{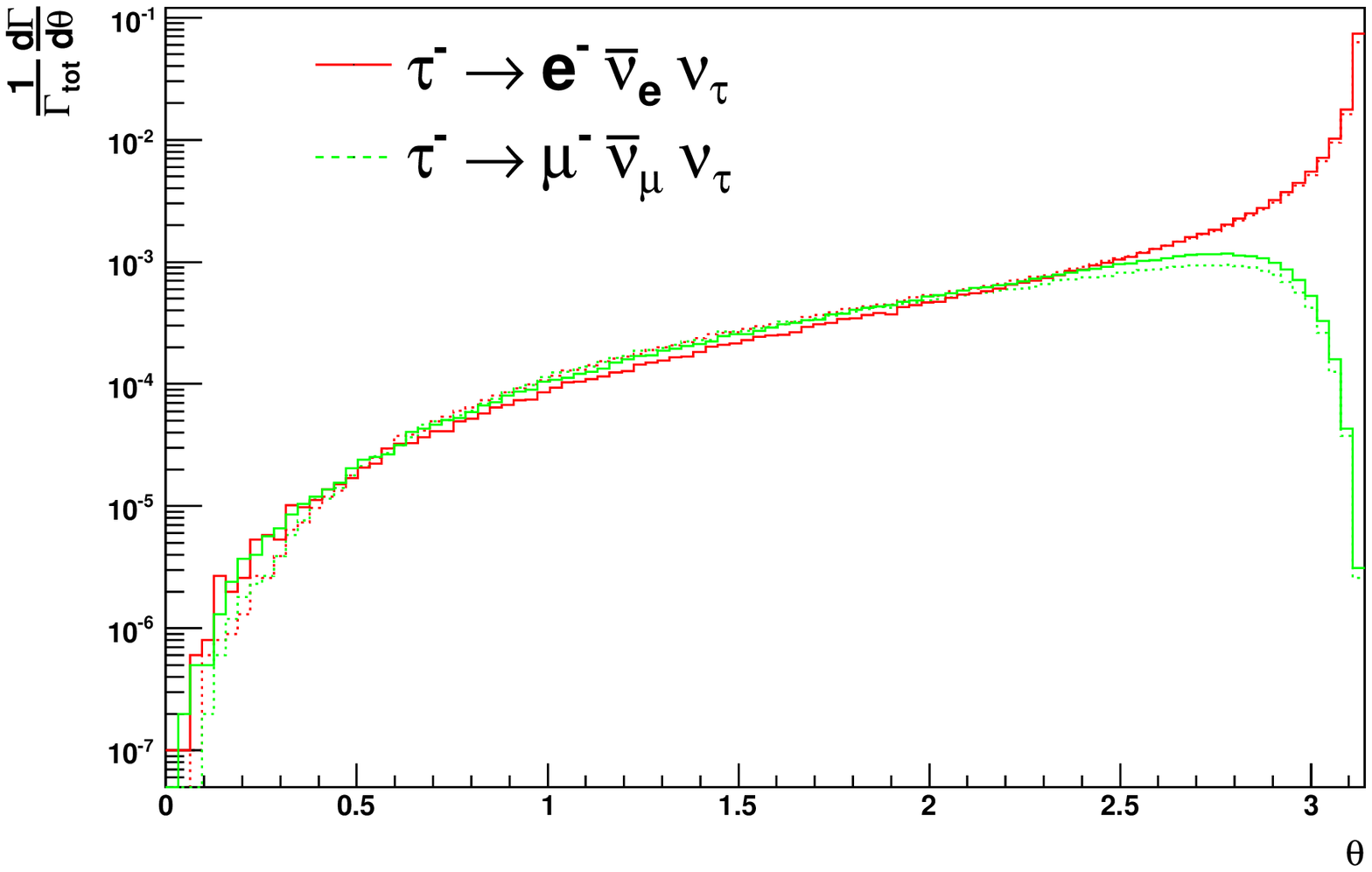}
  \caption{
	The total photon energy in $\tau^-\to \ell^-\,\bar\nu_\ell\,\nu_\tau$ in the 
	rest frame of the decaying $\tau$ lepton is shown in the left panel. 
	In the right panel the distribution of the photons' polar angle is
	shown in the $\tau-\ell$ rest frame with the $\tau$ at	$\theta=0$.
	In both plots, the solid line shows the distribution corrected with 
	the exact matrix element and the dotted line the one using the eikonals 
	only. The infrared cut-off was set to 1MeV.
	}
 \label{Fig:tau_dist}
 \ec
\end{figure}

The leptonic $\tau$ decays are an example of a final state containing multiple 
neutral and massless particles. This has the effect that the leading order 
decays do not have a fixed momentum distribution among the primary decay 
products leading to a smearing out of the sharp kink at $\tfrac{1}{2}m_\tau$,
as depicted in Figure \ref{Fig:tau_dist}.  Because of the relatively small 
$\tau$-mass and the considerable fraction of momentum carried by the neutrinos 
the effects of the different masses of the electron and the muon are plainly 
visible, in the photon energy spectrum as well as the angular distribution.

Furthermore, the branching fraction of radiative leptonic decays in $\mu$ and
$\tau$ decays (with at least one photon with $E_\gamma > 10$MeV) has been checked 
against PDG values \cite{Amsler:2008zz}, cf.\ Tab.~\ref{Tab:tau_BF}.

\begin{table}[thbp]
 \bc
  \begin{tabular}{|c|c|c|c|}
   \hline
   & \multirow{2}{*}{$\frac{\Gamma(\mu\to e\nu_e\nu_\mu\gamma)}{\Gamma(\mu\to e\nu_e\nu_\mu, incl.)}$ }
   & \multirow{2}{*}{$\frac{\Gamma(\tau\to e\nu_e\nu_\tau\gamma)}{\Gamma(\tau\to e\nu_e\nu_\tau, incl.)}$ }
   & \multirow{2}{*}{$\frac{\Gamma(\tau\to \mu\nu_\mu\nu_\tau\gamma)}{\Gamma(\tau\to \mu\nu_\mu\nu_\tau, incl.)}$} \\
   & & & \\
   \hline
    PDG & 0.014(4) & 0.09(1) & 0.021(3) \\
   \hline
    \Photonspp & 0.0147(1) & 0.0999(3) & 0.0233(2) \\
   \hline 
  \end{tabular}
  \caption{
	A comparison of the branching ratios of the radiative leptonic $\mu$ and $\tau$ 
	decay mode ($E_\gamma >$ 10MeV) in relation to their inclusive leptonic mode 
	calculated by \protect\Photonspp and the PDG world avarage. The number in brackets 
	reflects the absolute error on the last digit.
        }
 \label{Tab:tau_BF}
 \ec
\end{table}

\section{Conclusions and outlook}

In this publication a new implementation of the YFS approach to the description of 
higher-order QED corrections in particle processes has been presented in the form 
of a Monte Carlo code.  It is a part of the multi-purpose event generation framework 
\Sherpa since version 1.1 and allows for a simulation of photon radiation in particle 
decays.  This is an important effect with important experimental consequences.  The 
huge advantage of the YFS approach is that it can be systematically improved 
order-by-order in the electromagnetic coupling constant, such that its accuracy can 
be increased to match exact results at in principle any given perturbative order.  
Thus, in terms of possible accuracy, the YFS approach clearly reaches beyond typical 
parton-shower based algorithms.  Some of the effects due the inclusion of exact 
perturbative results have been studied in this publication.  

In contrast to another recent implementation of the YFS approach in \Sophty, here, 
in \Photonspp, there is no constraint in the number of particles produced in the decay, 
i.e.\ \Photonspp stretches beyond the level of $1\to 2$ decays. This is possible due 
to a new method of reconstructing the kinematics after QED radiation has been added 
to a core process, thus shifting its characteristics {\it a posteriori}.  The 
corresponding algorithms have been tested and validated in detail 
through comparison with results from other codes and experimental data. Some of the 
results have also been presented in this paper.  

It is anticipated that in the progress of the further development of
\Sherpa also its modules will be improved; in the case of \Photonspp
this will mainly involve the addition of an increasing number of exact 
higher order results. Some of the most relevant $1\to 2$, such as 
generic $V\to FF$ matrix elements with adjustable couplings, 
cf.\ Sec.~\ref{Sec:ApproximateRealCorrections}, $S\to FF$ and $S\to SS$, 
as well as more dedicated $W\to\ell\nu$, $\tau\to\ell\nu_\ell\nu_\tau$, 
$S\to S\ell\nu$ and $S\to V\ell\nu$ are already present. Others
will need to be added. The structure of the code also permits the inclusion of 
form-factors to take into account the composite nature of hadrons.

\section*{Acknowledgements}

The authors wish to thank K.~Hamilton, S.~Jadach, P.~Richardson and 
Z.~Was for fruitful discussions and C.~M.~Carloni Calame for help with
the correct usage of \Horace.   M.S. wishes to thank the INP for a
pleasant atmosphere and great support during the initial stages of
this project.  M.S.~would also like to thank D.~St\"ockinger for many 
fruitful discussions.

\noindent
Financial support by MCnet (contract number MRTN-CT-2006-035606), DAAD 
and BMBF is gratefully acknowledged.  M.S.\ wishes to thank the DAAD for 
financial support for the scientific visit to INP Cracow.

\newpage

\begin{appendix}
\appendix
\section{The YFS-Form-Factor}\label{Appendix_YFS}

In this appendix, the cancellation of virtual and real soft singularities 
will explicitly be performed and the YFS-Form-Factor will be calculated. 
As already defined in Sections \ref{Section_YFS_Exponentiation} and 
\ref{Sec:Algorithm} the YFS-Form-Factor $Y(\Omega)$ reads
\bea
Y(\Omega) 
= 
2\alpha\sum_{i<j}\left(\mathcal{R}e\;B(p_i,p_j) +
		\tilde{B}(p_i,p_j,\Omega)\right)\,, \nnb
\eea
where the virtual infrared factor is given by
\bea
 B(p_i,p_j) 
&=& 
-\frac{i}{8\pi^3}Z_iZ_j\theta_i\theta_j\int\frac{\done^4k}{k^2}
\left(\frac{2p_i\theta_i-k}{k^2-2(k\cdot p_i)\theta_i}
	+\frac{2p_j\theta_j+k}{k^2+2(k\cdot p_j)\theta_j}\right)^2 \nnb
\eea
and the real infrared factor reads
\bea
\tilde{B}(p_i,p_j,\Omega) 
&=& 
\frac{1}{4\pi^2}Z_iZ_j\theta_i\theta_j\int \done^4k\,
	\delta(k^2)\left(1-\Theta(k,\Omega)\right)
	\left(\frac{p_i}{(p_i\cdot k)}-\frac{p_j}{(p_j\cdot k)}\right)^2 \nnb\,.
\hspace{5mm}
\eea
As before, $Z_i$ and $Z_j$ are the charges of particles $i$ and $j$ in units 
of the positron charge, respectively, and the sign factors 
$\theta_{i,j} = \pm 1$ for final (initial) state particles.  Again, $\Omega$ 
is the ``unresolved'' region of the phase space for the soft photons.  In this
form the divergences need to be regularised, which can be achieved by either
introducng a ficititous small photon mass $\lambda$, as in the original
YFS paper \cite{Yennie:1961ad}, or through dimensional regularisation.
In both cases, however, the limited real emission phase space $\Omega$ will 
lead to potentially large logarithms.  

After performing the momentum integration, the virtual infrared factor can 
be written as
\bea\label{Eq:B}
B(p_i,p_j) 
&=& 
-\frac{Z_iZ_j\theta_i\theta_j}{2\pi}
\left(\ln\frac{m_im_j}{\lambda^2}
     +\tfrac{1}{2}(p_i\cdot p_j)\theta_i\theta_j
	\int\limits_{-1}^{1}\done x
	\frac{\ln\frac{p_x^{\prime 2}}{\lambda^2}}{p_x^{\prime 2}}
     +\tfrac{1}{4}\int\limits_{-1}^1\done x
	\ln\frac{p_x^{\prime 2}}{m_im_j}\right)\,, \nnb
\eea
where 
\bea
p_x^{\prime} = 
\frac{(p_i\theta_i-p_j\theta_j)+x(p_i\theta_i+p_j\theta_j)}{2} \nnb
\eea
and 
\bea
-\ln\lambda^2 = \frac{1}{\epsilon}-\ln\tilde{\mu}^2 \nnb
\eea 
contains the infrared divergence.  Similarly, the real infrared factor reads 
\bea\label{Eq:B_tilde}
\tilde{B}(p_i,p_j,\omega) 
&=& 
\frac{Z_iZ_j\theta_i\theta_j}{2\pi}\left[
\ln\frac{\omega^2}{\lambda^2}+\ln\frac{m_im_j}{E_iE_j}
-\tfrac{1}{2}(p_i\cdot p_j)\int\limits_{-1}^1\done x
	\frac{\ln\frac{p_x^2}{\lambda^2}}{p_x^2}
+\tfrac{1}{2}(p_i\cdot p_j)\int\limits_{-1}^1\done x
	\frac{\ln\frac{E_x^2}{\omega^2}}{p_x^2} \right.\nnb\\
&&\hspace{18mm}\left.{}
-\tilde{G}(1)-\tilde{G}(-1)
+(p_1\cdot p_2)\int\limits_{-1}^1 \done x\frac{\tilde{G}(x)}{p_x^2}\right]\,, \nnb
\hspace{5mm}
\eea
with 
\bea
p_x = \frac{(p_i+p_j)+x(p_i-p_j)}{2} \nnb
\eea
and $\omega$ is the momentum cut-off specifying $\Omega$ in the frame 
$\tilde{B}$ is to be evaluated in.  Furthermore, 
\bea
\tilde{G}(x) = 
\frac{1-\beta_x}{2\beta_x}
\ln\frac{1+\beta_x}{1-\beta_x}+\ln\frac{1+\beta_x}{2}\,. \nnb
\eea
with
\bea
\beta_x 
= \frac{\left|\vec{p}_x\right|}{p_x^0} 
= \frac{\sqrt{(\vec{p}_i+\vec{p}_j)^2 + 2x(\vec{p}_i^2-\vec{p}_j^2) + x^2(\vec{p}_i-\vec{p}_j)^2}}{(E_i+E_j)+x(E_i-E_j)} \nnb\,.
\eea

Combining both terms to the YFS-Form-Factor the divergences cancel and a 
finite result is obtained.  The remaining parameter integrals do not give 
rise to further divergences as long as $p_i^2,p_j^2 > 0$, i.e.\ as long
as the emitting particles are massive.  Thus, taken together, the YFS
form factor reads
\bea
Y(p_i,p_j,\omega) 
&=& 
-\frac{\alpha}{\pi}Z_iZ_j\theta_i\theta_j\left[
\ln\frac{E_iE_j}{\omega^2}
-\tfrac{1}{2}(p_i\cdot p_j)\int\limits_{-1}^1 \done x
	\frac{\ln\frac{E_x^2}{\omega^2}}{p_x^2}
+\tfrac{1}{4}\int\limits_{-1}^1 \done x
	\ln\frac{p_x^{\prime 2}}{m_im_j}
\right. \nnb\\
&& \hspace{23mm}\left.{}
+\tfrac{1}{2}(p_i\cdot p_j)\Theta(\theta_i\theta_j)
	\left(\frac{8\pi^2\Theta(x_1^\prime x_2^\prime)}
			{(x_2^\prime -x_1^\prime)(p_i+p_j)^2}
	     +\int\limits_{-1}^1\done x\frac{\ln x^2}{p_x^2}\right)
\right. \nnb\\
&&\hspace{23mm}\left.{}
+\tilde{G}(1)+\tilde{G}(-1)
-(p_i\cdot p_j)\int\limits_{-1}^1 \done x\frac{\tilde{G}(x)}{p_x^2}\right]\,, \nnb
\eea
where $x_{1,2}^\prime$ are the roots of $p_x^{\prime 2}$ with 
$x_1^\prime < x_2^\prime$.  The general case cannot be evaluated in closed 
form.  This is due to the fact that the term
\bea\label{Eq:complicated}
 \int\limits_{-1}^1 \done x \frac{\tilde{G}(x)}{p_x^2}\, , \nnb
\eea
although completely finite, can only be evaluated analytically for the dipole 
in its rest frame or in the rest frame of one of either of its constituent 
particles.  This can only be achieved if there is one dipole only.  All other 
cases need to be evaluated numerically.

\subsection{Special cases}

\subsubsection[Two-body decay with $(p_i\theta_i+p_j\theta_j)^2<0$]
		{Decay into two particles with $(p_i\theta_i+p_j\theta_j)^2<0$}
If the multipole consists of only two particles in the final state, e.g.\ for
decays of the type $Z\to\ell\bar\ell$, then there is an analytical solution 
in the rest frame of the dipole formed by the two charged particles.  In the 
high-energy limit, given by $E_i\gg m_i$ for both QED corrected charged 
particles, the critical term above can be written as
\bea
(p_i\cdot p_j)\int\limits_{-1}^1 \done x \frac{\tilde{G}(x)}{p_x^2} 
& \cong & \tfrac{1}{6}\pi^2\, . \nnb
\eea
Therefore, in this case, the full YFS form factor reads
\bea
Y(p_i,p_j,\omega) 
& \cong & 
-\frac{\alpha}{\pi}Z_iZ_j\theta_i\theta_j\left[
	\left(1-\ln\frac{2(p_i\cdot p_j)}{m_im_j}\right)
		\ln\frac{E_iE_j}{\omega^2}
	+\ln\frac{E_i}{E_j}\ln\frac{m_i}{m_j}
	-\tfrac{1}{2}\ln^2\frac{E_i}{E_j}
\right. \nnb\\
&&\hspace{20mm}\left.{}
	+\tfrac{1}{2}\ln\frac{(p_i\theta_i+p_j\theta_j)^2}{m_im_j}
	-1-\tfrac{\pi^2}{6}\right]\,. \nnb
\eea
This result in the high-energy limit agrees with the result stated in 
\cite{Yennie:1961ad}. 

\subsubsection[Two-body decay with $(p_i\theta_i+p_j\theta_j)^2=0$]
		{Decay of a charged particle with one charged final state
		 with $(p_i\theta_i+p_j\theta_j)^2=0$}
A similar, but nonetheless different case occurs for the decay of a 
charged particle into a final state involving only one charged particle, 
e.g.\ the case of $W$-decays, $W\to\ell\nu_\ell$.  Then, in the 
corresponding dipole's rest frame neither $m_W \ll E_W$ nor 
$(p_i\theta_i+p_j\theta_j)^2<0$ and therefore this case is different 
from the one above. In this case, for $(p_W-p_l)^2=0$, 
\bea
Y_{\mbox{\tiny\sl W}}(\omega) 
&=& 
\frac{\alpha}{\pi}\left[
2\left(1-\ln\frac{m_W}{m_l}\right)\ln\frac{m_W}{\omega\sqrt{8}}
	+\ln\frac{m_W}{m_l}-\tfrac{1}{2}+\tfrac{3}{2}\ln 2
	-\tfrac{3}{12}\pi^2\right]\, . \nnb
\eea
This result of course differs from the result in \cite{Placzek:2003zg} since 
both results are given in different Lorentz-frames.  Also, if in this 
process a photon is radiated, then $(p_W - p_l)^2=2(p_\nu\cdot p_\gamma)>0$ 
and the YFS-Form-Factor takes a different a form.

\subsection{The full YFS form factor}

Here the complete solutions to analytically integrable parameter integrals 
in the YFS form factor are given. In the following, 
using the invariance of $Y(\Omega)$ under the interchange of 
$p_i\leftrightarrow p_j$, the labels $p_i$ and $p_j$ are chosen such that 
$E_j\geq E_i$. It is useful to define 
\bea
 x_{1,2} 
& = & 
{} - \frac{p_i^2-p_j^2\pm2\sqrt{(p_i\cdot p_j)^2-p_i^2p_j^2}}{(p_i-p_j)^2} \nnb
\eea
as the roots of $p_x^2$ and
\bea
 x_{1,2}^\prime 
& = & 
{} - \frac{p_i^2-p_j^2\pm2\sqrt{(p_i\cdot p_j)^2-p_i^2p_j^2}}{(p_i+p_j)^2} \nnb
\eea
as those of $p_x^{\prime 2}$ in case of $\theta_i\theta_j=+1$, satisfying 
$x_{1,2}\notin[-1,1]$ and $x_{1,2}^\prime\in(-1,1)$, respectively. It holds 
that $x_1,x_2^\prime > 0$ and $x_2,x_1^\prime < 0$ if $(p_i-p_j)^2<0$ and 
$0<x_1<x_2$ and $0<x_1^\prime<x_2^\prime$ if $(p_i-p_j)^2>0$. These difference 
in the relations between $x_1$ and $x_2$ necessitate the differentiation of 
distinct cases in the calculations.

If $(p_i-p_j)^2=0$ then $x_{1,2}$ are not defined. If $\theta_i\theta_j=-1$ 
then $p_x^{\prime 2} = p_x^2$ and $x_{1,2}^\prime$ are meaningless, leading 
to another set of distinct cases.

\vspace{3mm}

When evaluating the first set of the parameter integrals that fact simplifies 
matters a lot resulting in
\bea
\mathcal{R}e\left(
	\theta_i\theta_j\int\limits_{-1}^1\done x
		\frac{\ln\frac{p_x^{\prime 2}}{\lambda^2}}{p_x^{\prime 2}}
	+\int\limits_{-1}^1\done x
		\frac{\ln\frac{p_x^2}{\lambda^2}}{p_x^2}\right)
 \hspace{3mm}\stackrel{\theta_i\theta_j=-1}{=}\hspace{3mm} 0\,. \nnb
\eea
Otherwise, the evaluation is more complicated and involves shifting the poles 
at $x_{1,2}^\prime$ off the real axis. The solution then is
\bea
\lefteqn{\mathcal{R}e\left(
	\theta_i\theta_j\int\limits_{-1}^1\done x
		\frac{\ln\frac{p_x^{\prime 2}}{\lambda^2}}{p_x^{\prime 2}}
	+\int\limits_{-1}^1\done x
		\frac{\ln\frac{p_x^2}{\lambda^2}}{p_x^2}\right)} \nnb\\
& = & \frac{8\pi^2\Theta\left(x_1^\prime x_2^\prime\right)}
		{(x_2^\prime-x_1^\prime)(p_i+p_j)^2}
	+ \frac{8}{(x_1-x_2)(p_i-p_j)^2} \left[
		\ln |x_1|\left(\mbox{Li}_2\left(\tfrac{x_1-1}{x_1}\right)
			-\mbox{Li}_2\left(\tfrac{x_1+1}{x_1}\right)\right)
				\right. \nnb\\
&&\hspace{75mm}\left.{}
		-\ln |x_2|\left(\mbox{Li}_2\left(\tfrac{x_2-1}{x_2}\right)
			-\mbox{Li}_2\left(\tfrac{x_2+1}{x_2}\right)\right)
				\right]\,. \nnb
\eea
In any case, the last piece of the divergence has cancelled, leaving finite 
terms negligible in the high energy limit.

\vspace{3mm}

The other integral containing $p_x^{\prime 2}$ is to be evaluated next. In 
total there are three cases to consider. 
\bit
\item $\theta_i\theta_j = +1$
\bea
\lefteqn{\mathcal{R}e\left(
	\int\limits_{-1}^1\done x\ln\frac{p_x^{\prime 2}}{m_im_j}\right)} \nnb\\
& = & 2\ln\frac{(p_i+p_j)^2}{4m_im_j}
	+\ln\left[(1-x_1^{\prime 2})(1-x_2^{\prime 2})\right]
	-x_1^\prime\ln\left|\frac{1-x_1^\prime}{1+x_1^\prime}\right|
	-x_2^\prime\ln\left|\frac{1-x_2^\prime}{1+x_2^\prime}\right| 
	-4\,. \nnb
\eea
Allthough, there again are poles within the range of integration the integral 
over them is finite. 
\item $\theta_i\theta_j = -1$. The range 
of integration does not comprise any poles and, thus, is real, giving
\bea
\lefteqn{\int\limits_{-1}^1\done x\ln\frac{p_x^2}{m_im_j}} \nnb\\
& = & 2\ln\frac{|(p_i-p_j)^2|}{4m_im_j}
	+\ln\left[(1-x_1^2)(1-x_2^2)\right]
	+x_1\ln\left|\frac{1+x_1}{1-x_1}\right| 
	+x_2\ln\left|\frac{1+x_2}{1-x_2}\right| 
	-4\,. \nnb
\eea
Evidently, the case $(p_i-p_j)^2=0$ has to be treated separately. It yields
\bea
\int\limits_{-1}^1\done x\ln\frac{p_x^2}{m_im_j}
& = & 2\ln\frac{|p_i^2-p_j^2|}{2m_im_j} 
	+\ln|1-x_p^2|
	+x_p\ln\left|\frac{1+x_p}{1-x_p}\right|
	-2\,. \nnb
\eea
where $x_p = -\frac{p_i^2+p_j^2}{p_i^2-p_j^2}$. In decay matrix elements it 
is not kinematically possible to also have $m_i=m_j$.
\eit

\vspace{3mm}

The last integral that is generally solveable analytically differentiates 
even more cases. The easiest to solve is the case of $E_i=E_j$, as it is 
occuring in leptonic $Z$-decays. Here, $E_x$ is independent of $x$, thus 
giving
\bea
\int\limits_{-1}^1\done x\frac{\ln\frac{E_x^2}{\omega^2}}{p_x^2} 
& = & \frac{8}{(x_1-x_2)(p_i-p_j)^2}\,
	\ln\frac{E_i+E_j}{2\omega}\,
	\ln\left|\frac{(1-x_1)(1+x_2)}{(1+x_1)(1-x_2)}\right|\,. \nnb
\eea
For all other dipoles three distinct cases appear:
\bit
\item $(p_i-p_j)^2<0$
\bea
\lefteqn{\int\limits_{-1}^1\done x\frac{\ln\frac{E_x^2}{\omega^2}}{p_x^2}} \nnb\\
& = & \frac{8}{(x_1-x_2)(p_i-p_j)^2}\,\left[
	\ln\frac{E_i}{\omega}\ln\left|\frac{1-x_1}{1+x_1}\right| 
	+\ln|y_1|\ln\left|\frac{1-x_1}{1+x_1}\right|\right. \nnb\\
&&\hspace{20mm}\left.{}
	-\ln\frac{(1+x_2)E_i+(1-x_2)E_j}{2\omega}
		\ln\left|\frac{1-x_2}{1+x_2}\right|\right.  \nnb\\
&&\hspace{20mm}\left.{}
	+\mbox{Li}_2\left(-\tfrac{\zeta(1+x_1)}{y_1}\right)
	-\mbox{Li}_2\left(\tfrac{\zeta(1-x_1)}{y_1}\right)
	-\mbox{Li}_2\left(-\tfrac{1+x_2}{x_E-x_2}\right)
	+\mbox{Li}_2\left(\tfrac{1-x_2}{x_E-x_2}\right)\right.\bigg] \nnb
\eea
with $y_1 = 1+\zeta(1-x_1)$, $\zeta = -\frac{E_i-E_j}{2E_i}$ and 
$x_E = -\frac{E_i+E_j}{E_i-E_j}$. \\
\item $(p_i-p_j)^2>0$
\bea
\lefteqn{\int\limits_{-1}^1\done x\frac{\ln\frac{E_x^2}{\omega^2}}{p_x^2}} \nnb\\
& = & \frac{8}{(x_1-x_2)(p_i-p_j)^2}\,\left[
	\ln\frac{E_i}{\omega}\ln\left|\frac{1-x_1}{1+x_1}\right| 
	+\ln|y_1|\ln\left|\frac{1-x_1}{1+x_1}\right|\right. \nnb\\
&&\hspace{20mm}\left.{}
	+\tfrac{1}{2}\ln^2\left|\frac{y_2}{\xi(1+x_2)}\right| 
	-\tfrac{1}{2}\ln^2\left|\frac{y_2}{\xi(1-x_2)}\right|\right. \nnb\\
&&\hspace{20mm}\left.{}
	-\ln\frac{E_j}{\omega}\ln\left|\frac{1-x_2}{1+x_2}\right| 
	+\ln|y_2|\ln\left|\frac{1-x_2}{1+x_2}\right|\right.  \nnb\\
&&\hspace{20mm}\left.{}
	+\mbox{Li}_2\left(-\tfrac{\zeta(1+x_1)}{y_1}\right)
	-\mbox{Li}_2\left(\tfrac{\zeta(1-x_1)}{y_1}\right)
	-\mbox{Li}_2\left(-\tfrac{y_2}{\xi(1-x_2)}\right)
	+\mbox{Li}_2\left(\tfrac{y_2}{\xi(1+x_2)}\right)\right.\bigg]\nnb
\eea
with $y_2 = 1+\xi(1+x_2)$ and $\xi = \frac{E_i-E_j}{2E_j}$. \\
\item $(p_i-p_j)^2=0$ \\
With the definitions for $x_E$ and $x_p$ from above it allways holds that 
$x_E>x_p>1$, thus
\bea
\lefteqn{\int\limits_{-1}^1\done x\frac{\ln\frac{E_x^2}{\omega^2}}{p_x^2}}  \nnb\\
& = & \frac{4}{p_j^2-p_i^2}\left[
	\ln\frac{E_j-E_i}{2\omega}\ln\left|\frac{1+x_p}{1-x_p}\right|
	+\ln(x_E-x_p)\ln\left|\frac{1+x_p}{1-x_p}\right|\right. \nnb\\
&&\hspace{15mm}\left.{}
	+\mbox{Li}_2\left(\tfrac{x_p-1}{x_p-x_E}\right)
	-\mbox{Li}_2\left(\tfrac{x_p+1}{x_p-x_E}\right)\right.\bigg] \nnb
\eea
\eit
\vspace{3mm}

The last integral can generally only be solved numerically. This is due 
to the complexity of $\beta_x$. If, however, the dipole is in its rest 
frame or in the rest frame of one of its constituents, there are analytical 
solutions. Because \Photonspp allways treats multipoles in their rest 
frames solutions for the integral will only be given in that frame. Two 
important cases are:
\bit
\item $m_i = m_j$
\bea
 \lefteqn{\int\limits_{-1}^1\done x\frac{\tilde{G}(x)}{p_x^2}} \nnb\\
& = & \frac{1}{\beta E^2}\left[
	\tfrac{1}{2}\ln^2\frac{1+\beta}{2}
	+\ln 2\ln(1+\beta) 
	-\tfrac{1}{2}\ln^2 2 
	-\tfrac{1}{2}\ln^2(1+\beta)\right. \nnb\\
&&\hspace{10mm}\left.{} 
	+\mbox{Li}_2\left(\tfrac{1-\beta}{2}\right)
	-\mbox{Li}_2\left(\tfrac{1+\beta}{2}\right)
	+\mbox{Li}_2\left(\beta\right)
	-\mbox{Li}_2\left(-\beta\right)\right.\bigg] \nnb
\eea
with $\beta = \frac{|\vec{p}_i|}{E_i} = \frac{|\vec{p}_j|}{E_j}$ 
and $E = E_i = E_j$.\\
\item Leptonic $W$-decay ($m_i \ll m_j = m_W$)
\bea
\int\limits_{-1}^1\done x\frac{\tilde{G}(x)}{p_x^2} 
& \cong & \frac{2}{m_j^2}\left[
		\tfrac{3}{12}\pi^2
		+\mbox{Li}_2\left(-2\right)\right]\,. \nnb
\eea
\eit

\section{Transforming the phase space elements}

This section details the phase space manipulations neccessary for the
implementation of the YFS algorithm in form of a computer code.  

\subsection{Rewriting the phase space element in other frames}
\label{Sec:RestframeTrafo}

As discussed in Sec.~\ref{Sec:PSTrafo}, the phase space integral with
the phase space element
\bea
\done\Phi 
&=& 
\done\Phi_p\,\done\Phi_k \,
(2\pi)^4\delta\left(p_C+p_N-P_C-P_N-K\right)\nnb\\
&=&
\prod_{i=1}^n\left[\frac{\done^3p_i}{(2\pi)^32p_i^0}\right]\,
\prod_{i=1}^{n_\gamma}\left[\frac{\done^3k}{k^0}\right]\,
(2\pi)^4\delta\left(p_C+p_N-P_C-P_N-K\right)\,, \nnb
\eea
has to be transformed to explicitely be in the chosen frame, the multipole 
rest frame.  This can be achieved by using the identities
\bea
1 & = & 
\frac{2}{M^2}\int\done^4(p_C+p_N)\,\done^4P_C\,\done m_{M,p}^2\,
	\delta\left(\frac{1}{M}(\vec{p}_C+\vec{p}_N)\right)\,
	\delta((p_C+p_N)^2-M^2) \nnb\\
&&\hspace{15mm}\times\,\delta^4\left(P_C+P_N-\sum p_i\right)
	\delta((p_C+P_C)^2-m_{M,p}^2)\,\Theta\left((p_C+p_N)^0\right)\,. \nnb
\eea
and
\bea\label{id_Jadach}
1 & = & 
\frac{2}{m_{M,p}^4}\int\done^4x\,\delta\left(\frac{x^2}{m_{M,p}^2}-1\right)
\delta^3\left(\frac{1}{m_{M,p}}L^{-1}(p_C+P_C)\right). 
\eea

Here, $m_{M,p}$ is the invariant mass of $P_M = p_C+P_C$ and $M$ is the 
invariant mass of the initial state $p_C+p_N$. As before, $p_C$ and $p_N$ 
and $P_C$ and $P_N$ are the sums of the initial and final state charged and 
neutral particles' momenta.  The first identity basically amounts to extending 
the integration to an integration over the full phase space including the 
initial particles.  The second identity, taken from \cite{Jadach_Torino},
involves a Lorentz-transformation, denoted by $L^{-1}$, being the boost into 
the rest frame of $x$.  Applying this boost on the phase space integral of 
course is a valid operation, since the full expression at this point is 
formulated in a Lorentz-invariant way.  The result of this 
Lorentz-transformation, after inserting both identities, reads
\bea
\done\Phi &=& 
(2\pi)^4\done\Phi_p\done\Phi_k 
	\int\done^4(p_C+p_N)\,\done^4P_C\,\done m_{M,p}^2\,\done^4x\, 
	\frac{2}{M^2}\frac{2}{m_{M,p}^4}\,\delta^4(p_C+p_N-P_C-P_N-K) \nnb\\
&&\hspace{15mm}\times\,
	\delta^3\left(L(p_C+p_N)\right)\,\delta((p_C+p_N)^2-M^2)\,
	\delta^4\left(P_C+P_N-\sum p_i\right)\nnb\\
&&\hspace{15mm}\times\,
	\delta((p_C+P_C)^2-m_{M,p}^2)\,\Theta\left((p_C+p_N)^0\right) \nnb\\
&&\hspace{15mm\times}\,
	\delta\left(\frac{x^2}{m_{M,p}^2}-1\right)
	\delta^3\left(\frac{1}{m_{M,p}}(\vec{p}_C+\vec{P}_C)\right). \nnb
\eea
Reordering and using the identity
\bea
\delta\left(\frac{x^2}{m_{M,p}^2}-1\right) 
= \int\done M^2\delta\left(\frac{x^2}{M^2}-1\right)\delta(M^2-m_{M,p}^2) \nnb
\eea
yields
\bea
\done\Phi &=& 
(2\pi)^4 \done\Phi_p\done\Phi_k 
	\int\done^4(p_C+p_N)\,\done^4P_C\,\done m_{M,p}^2 
	\frac{2}{M^2}\frac{2}{m_{M,p}} \,\delta((p_C+p_N)^2-M^2) \nnb\\ 
&&\hspace{15mm}\times\,
	\delta^3(\vec{p}_C+\vec{P}_C)\,\delta^4(p_C+p_N-P_C-P_N-K)\,
	\delta^4\left(P_C+P_N-\sum p_i\right) \nnb\\
&&\hspace{15mm}\times\,
	\delta((p_C+P_C)^2-m_{M,p}^2)\,\Theta\left((p_C+p_N)^0\right) \nnb\\
&&\hspace{15mm}\times\,
	\int\done^4x\,\done M^2\,\delta\left(\frac{x^2}{m_{M,p}^2}-1\right)
	\delta^3\left(\frac{1}{M}L(p_C+p_N)\right)\delta(M^2-m_{M,p}^2)\,. \nnb
\eea
The last line can be further simplified by using the identity of 
Eq.~(\ref{id_Jadach}) again and by integrating over $M^2$.  Now, 
the other integrations can be performed, first over $(p+p_N)$, 
then over $P$ and finally over $m_{M,p}^2$.  This results in
\bea
\done\Phi = 
(2\pi)^4 \done\Phi_p\done\Phi_k \frac{2m_{M,p}^3}{M^2}\, 
\delta^3(2\sum\vec{p}_i-\vec{P}_N+\vec{K}-\vec{p}_N)
\,\delta\left(\left(\sum p_i+K\right)^2-M^2\right)\,, \nnb
\eea
where $m_{M,p}^2 = (p_C+P_C)^2 = \left(2\sum p_i-P_N+K-p_N\right)^2 = P_M^2$ is
the invariant mass of the QED-corrected multipole. 

Finally, the identity
\bea\label{Eq:PS_trafo}
\delta\left(\left(\sum p_i+K\right)^2-M^2\right) = 
\frac{1}{2(P_C^0+P_N^0+K^0)}\delta(P_M^0 - P_{M,0}^0) \nnb
\eea
will be used, where $P_{M,0}^0 = P_C^0+p_C^0 = m_{M,p}$ and where all 
zero-components are taken in the rest frame of $P_M = p_C + P_C$. Therefore,
\bea
\done\Phi = 
(2\pi)^4 \frac{m_{M,p}^3}{M^2(P_C^0+P_N^0+K^0)}\,\done\Phi_p\,\done\Phi_k \,
\delta^3(\vec{P}_M)\,\delta\left(P_M^0 - P_C^0 - p_C^0\right) \,.\nnb
\eea
The phase space element $\done\Phi$ has thus been explicitely rewritten in 
the rest frame of the multipole, at the cost of a Jacobian.

Similarily, the zeroth order uncorrected cross section can be transformed to
\bea
\done \Phi_0 & = & (2\pi)^4\done\Phi_q\,\delta^4\left(p_C+p_N-Q_C-Q_N\right) \nnb\\
& = & (2\pi)^4 \frac{m_{M,q}^3}{M^2(Q_C^0+Q_N^0)}\,\done\Phi_q\, 
\delta^3(\vec{Q}_M)\,\delta\left(Q_M^0-Q_C^0-p_C^0\right)\,. \nnb
\eea
where $m_{M,q}$ is the invariant mass of the uncorrected multipole and the 
$Q_C^0$ and $Q_N^0$ are taken in the $Q_M$ rest frame.

\subsection{Rewriting the phase space element in terms of the undressed momenta}

In both cases the manipulations can be done in close analogy to the unitary 
algorithm of \cite{Kleiss:1991rn}.  The neccessary manipulations are easiest 
done backwards, starting with the phase space integral in terms of the $q_i$ 
and defining $n=n_C+n_N$ to be the number of final state particles.  

\subsubsection{Mixed multipoles}

In this case the starting point reads
\bea
\lefteqn{\int\prod_{i=1}^n\frac{\done^3q_i}{2q_i^0}\,
\delta^3(\vec{Q_M})\delta(Q_M^0-Q_C^0-p_C^0)} \nnb\\
&=& 
\int\prod_{i=1}^{n}\left[\done^4q_i\,
\delta\left(q_i^2-m_i^2\right)\Theta(q_i^0)\right]\,
\delta^3\left(\sum_C\vec{q}_i+\vec{p}_C\right)\,
\delta\left(Q_M^0-\sum_C q_i^0-p_C^0\right)\,.  \nnb
\eea
This can be recast into a better form by inserting the identity
\bea\label{Eq:id_p}
1 &=& 
\int \prod_{i=1}^{n}
\left[\done^4p_i\,
\delta^3\left(\vec{p}_i-u\vec{q}_i+\frac{1}{2n_C+n_N}\vec{K}\right)\,
\delta\left(p_i^0-\sqrt{\vec{p}_i^2+m_i^2}\right)\right]\nnb\\
&=&
\int \prod_{i=1}^{n}
\left[\done^4p_i\,
\delta^3\left(\vec{p}_i-u\vec{q}_i+\vec\kappa\right)\,
\delta\left(p_i^0-\sqrt{\vec{p}_i^2+m_i^2}\right)\right] 
\eea
with the abbreviation
\bea
\vec\kappa = \frac{\vec K}{2n_C+n_N}\,, \nnb
\eea
by using the definition of $u$ written as
\bea\label{Eq:id_u}
1 &=& 
\int \done u\,
\delta\left[\sqrt{M^2+\left(u\sum_C\vec{q}_i-
n_C\vec\kappa\right)^2}\right.  \nnb\\
&&\left.\hspace*{16mm} 
- \sum_{C,N}\sqrt{m_i^2+
\left(u\vec{q}_i-\vec\kappa\right)^2}-K^0\right]
\left(\frac{\vec{p}_C^\prime\vec{p}_C}{p_C^{\prime 0}}-
\sum_{C,N}\frac{\vec{p}_i\vec{q}_i}{p_i^0}\right)
\eea
and by expressing the $\delta$-function fixing $Q_M^0$ in terms of 
the kinematically relevant variables $q_i^0$ and $p_C^0$.  This then
yields
\bea
\lefteqn{\int\prod_{i=1}^n\frac{\done^3q_i}{2q_i^0}\,
\delta^3(\vec{Q}_M)\delta(Q_M^0-Q_C^0-p_C^0)} \nnb\\
&=& 
\int\done u\prod_{i=1}^{n}
\left[\done^4q_i\,\done^4p_i\,\delta\left(q_i^2-m_i^2\right)
\Theta(q_i^0)\,\delta^3\left(\vec{p}_i-u\vec{q}_i+\vec\kappa\right)
\delta\left(p_i^0-\sqrt{\vec{p}_i^2+m_i^2}\right)\right]
\nnb\\
&&\times\,
\delta\left(\sqrt{M^2+\left(u\sum_C\vec{q}_i-n_C\vec\kappa\right)^2} -
\sum_{C,N}\sqrt{m_i^2+\left(u\vec{q}_i-\vec\kappa\right)^2} -
K^0\right) \nnb\\
&&\times 
\delta^3\left(\sum_C\vec{q}_i+\vec{p}_C\right)\,
\delta\left(\sqrt{M^2+\left(\sum_C\vec{q_i}\right)^2}-
\sum_{C,N}q_i^0\right)\times
\left[\frac{\vec{p}_C^\prime\vec{p}_C}{p_C^{\prime 0}}-
\sum_{C,N}\frac{\vec{p}_i\vec{q}_i}{p_i^0}\right]\,. \nnb
\eea
Integrating over $\done^3q_i$ and $\done q_i^0$, using 
$\delta\left(x^2-x_0^2\right)\Theta(x) = \frac{1}{2x_0}\delta(x-x_0)$, 
and integrating over $u$ yields
\bea 
\lefteqn{\int\prod_{i=1}^n\frac{\done^3q_i}{2q_i^0}\,
\delta^3(\vec{Q_M})\delta(Q_M^0-Q_C^0-p_C^0)} \nnb\\
&=& 
\int\done u\,\prod_{i=1}^{n}
\left[\done^4p_i\delta\left(p_i^0-\sqrt{\vec{p}_i^2+m_i^2}\right)
\frac{1}{u^3}\frac{1}{2\sqrt{\frac{1}{u^2}
\left(\vec{p}_i+\vec\kappa\right)^2+m_i^2}}\right]
\nnb \\
&&\times\,
\delta^3\left(\frac{1}{u}
\left[\sum_C\vec{p}_i+n_C\vec\kappa+u\vec{p}_C\right]\right) \nnb\\
&&\times\,
\delta\left(\sqrt{M^2+\frac{1}{u^2}
\left[\sum_C\vec{p}_i+n_C\vec\kappa\right]^2}-
\sum_{C,N}\sqrt{m_i^2+
\frac{1}{u^2}\left[\vec{p}_i+\vec\kappa\right]^2}\right) 
\nnb\\
&&\times\,
\delta\left(\sqrt{M^2+\left[\sum_C\vec{p}_i\right]^2} - 
\sum_{C,N}\sqrt{m_i^2+\vec{p}_i^2}-K^0\right)
\times\left[\frac{\vec{p}_C^\prime\vec{p}_C}{p_C^{\prime 0}}-
\sum_{C,N}\frac{\vec{p}_i\left(\vec{p}_i+\vec\kappa\right)}{up_i^0}\right]\,. 
\nnb
\eea
\bea
&=& 
\int\prod_{i=1}^{n}\left[\done^4p_i
\delta\left(p_i^0-\sqrt{\vec{p}_i^2+m_i^2}\right)
\frac{1}{u^3}
\frac{1}{2\sqrt{\frac{1}{u^2}
\left(\vec{p}_i+\vec\kappa\right)^2+
m_i^2}}\right] \nnb\\
&&\times \,
u^3\delta^3\left(\sum_C\vec{p}_i+\vec{p}_C^\prime\right)
\delta\left(\sqrt{M^2+\left[\sum_C\vec{p}_i\right]^2} - 
\sum_{C,N}\sqrt{m_i^2+\vec{p}_i^2}-K^0\right) \nnb\\
&&\times\,
\left[\frac{\vec{p}_C^\prime\vec{p}_C}{p_C^{\prime 0}}-
\sum_{C,N}\frac{\vec{p}_i
\left(\vec{p}_i+\vec\kappa\right)}{up_i^0}
\right]
\left[\frac{u}{\frac{\vec{p}_C^2}{p_C^0}-
\sum_{C,N}\frac{\vec{q}_i^2}{q_i^0}}\right]\,, \nnb
\eea
where in the integration over $u$ the second last $\delta$-function
of the line above has been used.  Furthermore, in this transformation,
an identity similar to (\ref{Eq:id_u}), arising when defining $u$ in 
terms of $p_i$, has been employed. A rearrangement of terms and a 
suitable transformation of the last 
$\delta$-function in terms of $P_M$ yields
\bea
\lefteqn{\int\prod_{i=1}^n\frac{\done^3q_i}{2q_i^0}\,
\delta^3(\vec{Q_M})\delta(Q_M^0-Q_C^0-p_C^0)} \nnb\\
&=& 
\int\prod_{i=1}^{n}\left[\done^4p_i
\delta\left(p_i^2-m_i^2\right)\Theta(p_i^0)
\frac{1}{u^3}
\frac{\sqrt{\vec{p}_i^2+m_i^2}}
{\sqrt{\frac{1}{u^2}
\left(\vec{p}_i+\vec\kappa\right)^2+m_i^2}}
\right]
\nnb\\
&&\times\,
u^4\delta^3\left(\sum_C\vec{p}_i+\vec{p}_C^\prime\right)
\delta\left(p_C^{\prime 0}-P_C^0-P_N^0-K^0\right)
\frac{\frac{\vec{p}_C^\prime\vec{p}_C}{p_C^{\prime 0}}
-\sum_{C,N}\frac{\vec{p}_i\vec{q}_i}{p_i^0}}
{\frac{\vec{p}_C^2}{p_C^0}-\sum_{C,N}\frac{\vec{q}_i^2}{q_i^0}} 
\nnb
\eea
\bea
&=& 
\int\prod_{i=1}^n\left[\done^4p_i
\delta\left(p_i^2-m_i^2\right)\Theta(p_i^0)\right]
\delta^3\left(\vec{P}_M\right)
\delta\left(P_M^0-P_C^0-p_C^{\prime 0}\right) \nnb\\
&&\times\,\frac{1}{u^{3n-4}}
\frac{\frac{\vec{p}_C^\prime\vec{p}_C}{p_C^{\prime 0}}-
\sum_{C,N}
\frac{\vec{p}_i\vec{q}_i}{p_i^0}}
{\frac{\vec{p}_C^2}{p_C^0}-\sum_{C,N}\frac{\vec{q}_i^2}{q_i^0}}
\prod_{i=1}^n
\left[\frac{p_i^0}{q_i^0}\right]\,. \nnb
\eea
Here, the identity
\bea
q_i^0 = 
\sqrt{\frac{1}{u^2}
\left(\vec p_i+\vec\kappa\right)^2+m_i^2}  \nnb
\eea
has been used.  Reversing the procedure allows to express the phase 
space element through the undressed final state momenta as
\bea
\done\Phi 
&=& 
(2\pi)^4\done\Phi_q\done\Phi_k
\delta^3\left(\vec{Q}_M\right)
\delta\left(Q_M^0-Q_C^0-p_C^0\right)
\frac{m_M^3}{M^2\left(P_C^0+P_N^0+K^0\right)} \nnb\\ 
&&
\times\,u^{3n-4}\,
\frac{\frac{\vec{p}_C^2}{p_C^0}-\sum_{C,N}\frac{\vec{q}_i^2}{q_i^0}}
{\frac{\vec{p}_C^\prime\vec{p}_C}{p_C^{\prime 0}}-\sum_{C,N}
\frac{\vec{p}_i\vec{q}_i}{p_i^0}}\,
\prod_{i=1}^n \left[\frac{q_i^0}{p_i^0}\right]\,. \nnb
\eea

\subsubsection{Final state multipoles}

The transformation will be done using the same techniques as above. Starting 
from
\bea
\lefteqn{\int\prod_{i=1}^n \frac{\done^3q_i}{2q_i^0}\,
			\delta^3(\vec{Q}_M)\delta(Q_M^0-Q_C^0)} \nnb\\
& = & \int\prod_{i=1}^n\left[\done ^4q_i\delta(q_i^2-m_i^2)\Theta(q_i^0)\right]
\delta^3\left(\sum_C\vec{q_i}\right)\delta\left(Q_M^0-\sum_C q_i^0\right)\,. \nnb
\eea
Again, similar identities to (\ref{Eq:id_p}) and (\ref{Eq:id_u}) will be used, 
but due to the different mapping scheme they now read
\bea
 1 = \int\prod_{i=1}^n
\left[\done^4p_i\,\delta^3(\vec{p_i}-u\vec{q_i})\,
		\delta\left(p_i^0-\sqrt{\vec{p_i}^2+m_i^2}\right)\right] 
\eea
and
\bea\label{Eq:id_u_2}
 1 = \int \done u\,
\delta\left[\sqrt{M^2+\left(u\sum_N\vec{q}_i+\vec{K}\right)^2}
			 	- \sum_{C.N}\sqrt{m_i^2+u^2\vec{q}_i^2}
				 - K^0\right]
\left(\frac{\vec{p}_N^\prime\vec{p}_N}{p_N^{\prime 0}}
			-\sum_{C,N}\frac{\vec{p}_i\vec{q}_i}{p_i^0}\right)\, .
\eea
And, as before, the $\delta$-function over $Q_M^0$ is expressed in the 
kinematically relevant variables $q_i^0$. This then yields
\bea
\lefteqn{\int\prod_{i=1}^n \frac{\done^3q_i}{2q_i^0}\,
				\delta^3(\vec{Q}_M)\delta(Q_M^0-Q_C^0)}\nnb\\
 & = & \int \done u\prod_{i=1}^n
\left[\done^4q_i\done^4p_i\,\delta(q_i^2-m_i^2)\Theta(q_i^0)\,
		\delta^3(\vec{p_i}-u\vec{q_i})\,
		\delta\left(p_i^0-\sqrt{\vec{p_i}^2+m_i^2}\right)\right] \nnb\\
&&\times\,\delta\left(\sqrt{M^2+\left(u\sum_N\vec{q}_i+\vec{K}\right)^2}
		 - \sum_{C,N}\sqrt{m_i^2+u^2\vec{q}_i^2} - K^0\right) \nnb\\
&&\times\,\delta^3\left(\sum_C\vec{q_i}\right)
		\delta\left(\sqrt{M^2+\left(\sum_N\vec{q}_i\right)^2}
		 - \sum_{C,N}q_i^0\right)
	\times\left[\frac{\vec{p}_N^\prime\vec{p}_N}{p_N^{\prime 0}}
			-\sum_{C,N}\frac{\vec{p}_i\vec{q}_i}{p_i^0}\right]\,. \nnb
\eea
Integrating over $\done^4q_i$ and $u$ yields
\bea
\lefteqn{\int\prod_{i=1}^n \frac{\done^3q_i}{2q_i^0}\,
				\delta^3(\vec{Q}_M)\delta(Q_M^0-Q_C^0)}\nnb\\
 & = & \int\done u \prod_{i=1}^n
	\left[\done^4p_i\delta\left(p_i^0-\sqrt{\vec{p_i}^2+m_i^2}\right)
	\frac{1}{u^3}\frac{1}{2\sqrt{\frac{1}{u^2}\vec{p_i}^2+m_i^2}}\right]\,
	\delta^3\left(\frac{1}{u}\sum_C\vec{p}_i\right)  \nnb\\
&&\times\,\delta\left(\sqrt{M^2+\frac{1}{u^2}\left[\sum_N\vec{p}_i\right]^2}
		 - \sum_{C,N}\sqrt{\frac{1}{u^2}\vec{p_i}^2+m_i^2}\right)  \nnb\\
&&\times\,\delta\left(\sqrt{M^2+\left[\sum_N\vec{p}_i+\vec{K}\right]^2}
		 - \sum_{C,N}\sqrt{m_i^2+\vec{p}_i^2}
		 - K^0\right)
	\times\left[\frac{\vec{p}^\prime\vec{p}}{p^{\prime 0}}
			-\sum_{C,N}\frac{\vec{p}_i^2}{up_i^0}\right] \nnb\\
 & = & \int\prod_{i=1}^n
\left[\done^4p_i\delta(p_i^2-m_i^2)\Theta(p_i^0)\frac{1}{u^3}
	\frac{\sqrt{\vec{p_i}^2+m_i^2}}
			{\sqrt{\frac{1}{u^2}\vec{p_i}^2+m_i^2}}\right]\,
	u^3\,\delta^3\left(\sum_C\vec{p_i}\right)  \nnb\\
&&\times\,\delta\left(\sqrt{M^2+\left[\sum_N\vec{p}_i+\vec{K}\right]^2}
		 - \sum_{C,N}\sqrt{m_i^2+\vec{p}_i^2} - K^0\right) \nnb\\
&&\times\,\left[\frac{\vec{p}^\prime\vec{p}}{p^{\prime 0}}
		-\sum_{C,N}\frac{\vec{p}_i^2}{up_i^0}\right]
	\left[\frac{u}{\frac{\vec{p}^2}{p^0}
		-\sum_{C,N}\frac{\vec{q}_i^2}{q_i^0}}\right] \nnb
\eea
where, again, the second last $\delta$-function has been used in the 
integration over $u$. Additionally, an identity similar to (\ref{Eq:id_u_2}), 
arising when defining $u$ in terms of $p_i$, has been used. Rearranging terms 
leads to
\bea
\lefteqn{\int\prod_{i=1}^n \frac{\done^3q_i}{2q_i^0}\,
				\delta^3(\vec{Q}_M)\delta(Q_M^0-Q_C^0)}\nnb\\
 & = & \int\prod_{i=1}^n
\left[\done^4p_i\,\delta(p_i^2-m_i^2)\Theta(p_i^0)\,\frac{1}{u^3}\,
	\frac{\sqrt{\vec{p_i}^2+m_i^2}}
		{\sqrt{\frac{1}{u^2}\vec{p_i}^2+m_i^2}}\right] \nnb\\
&&\times\,u^4\,\delta^3(\vec{P}_C)\,
		\delta\left(p_N^{\prime 0} - P_C^0 - P_N^0 - K^0 \right)\,
	\frac{\frac{\vec{p}_N^\prime\vec{p}_N}{p_N^{\prime 0}}
			-\sum_{C,N}\frac{\vec{p}_i\vec{q}_i}{p_i^0}}
	{\frac{\vec{p}_N^2}{p_N^0}-\sum_{C,N}\frac{\vec{q}_i^2}{q_i^0}}  \nnb\\
& = & \int\prod_{i=1}^n
\left[\done^4p_i\,\delta(p_i^2-m_i^2)\Theta(p_i^0)\right]\,
	\delta^3(\vec{P_M})\,\delta(P_M^0 - P_C^0)\,
	\frac{1}{u^{3n-4}}\,
	\frac{\frac{\vec{p}_N^\prime\vec{p}_N}{p_N^{\prime 0}}
		-\sum_{C,N}\frac{\vec{p}_i\vec{q}_i}{p_i^0}}
	{\frac{\vec{p}_N^2}{p_N^0}-\sum_{C,N}\frac{\vec{q}_i^2}{q_i^0}}\,
	\prod_{i=1}^n\left[\frac{p_i^0}{q_i^0}\right]\hspace{7mm} \nnb
\eea
where the identity
\bea
q_i^0 = \sqrt{\tfrac{1}{u^2}\vec{p}_i^2+m_i^2} \nnb
\eea
has been used. Reversing the procedure allows to express the phase space 
element through the undressed final state momenta as
\bea
 \done\Phi & = & \done\Phi_q \,\done\Phi_k\,(2\pi)^4\,\delta^3(\vec{Q}_M)\,
	\delta(Q_M^0-Q_C^0)\, \frac{m_M^3}{M^2(P^0+P_N^0+K^0)} \nnb\\
&&\times\, u^{3n-4}\,
	\frac{\frac{\vec{p}_N^2}{p_N^0}-\sum_{C,N}\frac{\vec{q}_i^2}{q_i^0}}
		{\frac{\vec{p}_N^\prime\vec{p}_N}{p_N^{\prime 0}}
			-\sum_{C,N}\frac{\vec{p}_i\vec{q}_i}{p_i^0}}\,
	\prod_{i=1}^n\left[\frac{q_i^0}{p_i^0}\right]\, . \nnb
\eea

\section{Details on the photon generation}\label{Appendix_Generation}

In this section the generation of the photon distribution is detailed.

\subsection{Avarage photon multiplicity}

The average photon multiplicity $\bar{n}$ is the avarage of the Poisson 
distribution before it is corrected by the various weights. It is therefore 
not immediately connected to the true avarage photon multiplicity of the 
final event.  Nonetheless, it is an integral part of the generation procedure. 
An analytical result in closed form is available for both dipoles and multipoles. 
However, the calculations for multipoles are more involved as the integrations 
do not nicely seperate as they do in the dipole case in the chosen frame. Thus, 
as a starting point the analytical result for the dipole in its rest frame will 
be given. It reads
\bea
\label{n_bar_dipole}
\bar{n} 
= 
\int_{\omega_{\mbox{\tiny min}}}^{\omega_{\mbox{\tiny max}}}
	\frac{\done^3k}{k^0}\tilde{S}_q(k)  \nnb
= 
- \frac{\alpha}{\pi}Z_1Z_2\theta_1\theta_2 \; 
	\ln\frac{\omega_{\mbox{\scriptsize max}}}
		{\omega_{\mbox{\scriptsize min}}} \;
	 \left(\frac{1+\beta_1\beta_2}{\beta_1+\beta_2}
		\ln\frac{(1+\beta_1)(1+\beta_2)}
		        {(1-\beta_1)(1-\beta_2)}\;-2\right)\,,\nnb
\eea
where $\omega_{\mbox{\scriptsize min}}$ is the infrared cut-off and 
$\omega_{\mbox{\scriptsize max}}$ is the maximal kinematically allowed 
photon energy.  The latter can be determined by setting the rescaling 
parameter $u$ to zero in Eqs.~(\ref{def_u_ff}) and (\ref{def_u_if}), 
respectively, and by assuming single photon emission. Additionally, 
$\beta_i = \frac{|\vec{p}_i|}{E_i}$.

In the case of a multipole, the integral over the photon energy can still
be separated, as long as the soft photon region is sufficiently well-behaved. 
This is the case, if $\Theta(k,\Omega)$ forms an isotropic hypersurface in 
the frame of the integration.  However, the angular integration still remains 
to be done:
\bea
\bar{n} 
&=& 
\int\frac{\done^3 k}{k^0}\Theta(k,\Omega)\tilde{S}_q(k) \nnb\\
& = & \frac{\alpha}{4\pi^2}\sum_{i<j}Z_iZ_j\theta_i\theta_j
	\int\frac{\done^3k}{k^0}\Theta(k,\Omega)
	\left(\frac{q_i}{(q_i\cdot k)}-\frac{q_j}{(q_j\cdot k)}\right)^2 \nnb\\
&=& 
\frac{\alpha}{4\pi^2}\,
	\ln\frac{\omega_{\mbox{\scriptsize max}}}
		{\omega_{\mbox{\scriptsize min}}}\;
	\sum_{i<j}Z_iZ_j\theta_i\theta_j 
	\left(8\pi-\int\done\Omega
	\frac{2(q_i\cdot q_j)}{(q_i\cdot e_k)(q_j\cdot e_k)}\right)\,. \nnb
\label{Eq:decomp_integral}
\eea

\begin{figure}
\bc
 \includegraphics[width = 200pt]{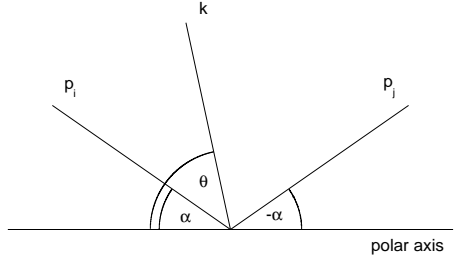}
\caption{Sketch of how the axes are chosen
	in the angular integration in multipoles.}
\label{multipole_integration} 
\ec
\end{figure}
Choosing different orientations of the polar axes for each interference 
term of every constituent dipole, all angular integrations can be done 
analytically.  Although this may sound like quite an {\it ad-hoc} procedure,
it is completely valid and simplyfies the integration immensely. The 
orientation for each of the interference terms is thus chosen to be such that 
both momenta lie symmetrically in the unit sphere, both forming an angle 
$\alpha_{ij}$ with the polar axis, see Fig.~\ref{multipole_integration}. 
Therefore, by this choice,
\bea
 (q_i\cdot q_j) & = & E_iE_j\left(1-a_ia_j+b_ib_j\right) \nnb\\
 (q_i\cdot e_k) & = & E_i\left(1-a_i\sin\varphi\sin\theta-b_i\cos\theta\right)  \nnb\\
 (q_j\cdot e_k) & = & E_j\left(1-a_j\sin\varphi\sin\theta+b_j\cos\theta\right) \,,\nnb
\eea
where $e_k^\mu$ again is $\frac{1}{k^0}k^\mu$ with $e_k^2 = 0$, cf.\ Eq.\
(\ref{Eq:e_k}), and the further parameters are given by
\bea
a_{i,j} = \beta_{i,j}\sin\alpha_{ij} \;\;\;\mbox{\rm and}\;\;\;  \nnb
b_{i,j} = \beta_{i,j}\cos\alpha_{ij} \,. \nnb
\eea
With these choices the last integral reads
\bea\label{num_int_decomp} 
\lefteqn{\int\done\Omega\frac{E_iE_j}{(q_i\cdot e_k)(q_j\cdot e_k)}} \nnb\\
& = & \int\limits_0^{2\pi}\done\varphi\int\limits_0^\pi \done\theta\,\sin\theta\,
	\frac{1}
	{\left(1-a_i\sin\varphi\sin\theta-b_i\cos\theta\right)
	 \left(1-a_j\sin\varphi\sin\theta+b_j\cos\theta\right)} \,. \nnb
\eea
Using the decomposition
\bea\label{Eq:decomp_int_term}
\lefteqn{\frac{1}{b_j\left(1-a_i\sin\varphi\sin\theta-b_i\cos\theta\right)}
	 -\frac{1}{b_i\left(1-a_j\sin\varphi\sin\theta+b_j\cos\theta\right)}}\nnb\\
& = & \frac{(b_i-b_j)+2b_ib_j\cos\theta}
	{b_ib_j\left(1-a_i\sin\varphi\sin\theta-b_i\cos\theta\right)
		\left(1-a_j\sin\varphi\sin\theta+b_j\cos\theta\right)} \nnb
\eea
and $a_ib_j = a_jb_i$, this can be easily integrated giving
\bea
\lefteqn{\int\done\Omega\frac{E_iE_j}{(q_i\cdot e_k)(q_j\cdot e_k)}} \nnb\\
& = & 2\pi\left[\frac{b_i}{\sqrt{B^2C_i-ABD_i+A^2E_i}}\,
		\ln\frac{A+B}{A-B}
		   \frac{\sqrt{C_i-D_i+E_i}+\frac{B(2C_i-D_i)-A(D_i-2E_i)}
						 {2\sqrt{B^2C_i-ABD_i+A^2E_i}}}
			{\sqrt{C_i+D_i+E_i}+\frac{B(2C_i+D_i)-A(D_i+2E_i)}
						 {2\sqrt{B^2C_i-ABD_i+A^2E_i}}}\right. \nnb\\
&&\hspace{7mm}\left.{}-
		\frac{b_j}{\sqrt{B^2C_j-ABD_j+A^2E_j}}\,
		\ln\frac{A+B}{A-B}
		   \frac{\sqrt{C_j-D_j+E_j}+\frac{B(2C_j-D_j)-A(D_j-2E_j)}
						 {2\sqrt{B^2C_j-ABD_j+A^2E_j}}}
			{\sqrt{C_j+D_j+E_j}+\frac{B(2C_j+D_j)-A(D_j+2E_j)}
						 {2\sqrt{B^2C_j-ABD_j+A^2E_j}}}\right]\,,\nnb
\eea
with
\bea
A & = & b_i-b_j \nnb\\
B & = & 2b_ib_j \nnb\\
C_{i,j} & = & 1-a_{i,j}  \nnb\\
D_{i,j} & = & \mp\,b_{i,j} \nnb\\
E_{i,j} & = & a_{i,j}^2 + b_{i,j}^2 \nnb\,.
\eea
Upon closer examination it can be seen that for $\alpha_{ij}\to 0$ 
the result of the dipole case is recovered.

\subsection{Photon energy}

Due to the decompostion of the integration over the photon energy and the 
integration over the unit sphere, the photon energy distribution and the 
photon angular distribution can be generated seperately.  Of course, this 
independence of distributions is no longer true after the reweighting 
procedure, but it alleviates the generation of the crude distribution.  

In the imlementation presented here, the photon energy is distributed 
according to $\frac{1}{k^0}$, generated through
\bea
k^0 &=& \omega_{\mbox{\scriptsize min}}
	\left(\frac{\omega_{\mbox{\scriptsize max}}}
			{\omega_{\mbox{\scriptsize min}}}\right)^{\mathcal{R}} \nnb
\eea
where $\mathcal{R}$ is a uniformly distributed random number on the interval 
$[0,1]$.

\subsection{Photon angles}

Similar to all other parts of the photon distribution, the photon angles 
are also generated according to $\tilde{S}_q(k)$. For this, the relevant
function is recast into the form 
\bea
 \lefteqn{-\left(\frac{q_i}{(q_i\cdot e_k)}-\frac{q_j}{(q_j\cdot e_k)}\right)^2}\nnb\\ 
& = & {}-\frac{1-\beta_i^2}{(1-\beta_i \cos\theta)^2}
	+\frac{2(1+\beta_i\beta_j)}{(1-\beta_i \cos\theta)(1+\beta_j \cos\theta)}
	-\frac{1-\beta_j^2}{(1+\beta_j \cos\theta)^2}\,, \nnb
\eea
where $\theta$ is some polar angle w.r.t.\ the dipole axis in the dipole rest 
frame.  In this frame, the generation of the azimuthal is trivial - it just 
follows a flat distribution in $[0,\,2\pi]$.  The polar distribution above can
be bound from above through the interference term.  This allows to generate
the true distribution by generating the angle according to the interference
term and applying a hit-or-miss rejection.  The interference term can be 
decomposed analogously to the general case above into two independent 
terms according to
\bea
 \frac{1}{(1-\beta_i \cos\theta)(1+\beta_j \cos\theta)} 
& = & \frac{\beta_i\beta_j}{\beta_i+\beta_j}
	\left(\frac{1}{\beta_j(1-\beta_i \cos\theta)}
		-\frac{1}{\beta_i(1+\beta_j \cos\theta)}\right). \nnb
\eea
The cosine of the polar angle, $\cos\theta$, is then generated to either
of the two terms, i.e.\ it is generated according to 
$(1-\beta_i\cos\theta)^{-1}$ with probability
\bea
P_i 
& = & \frac{\ln\frac{1+\beta_i}{1-\beta_i}}
	{\ln\frac{1+\beta_i}{1-\beta_i}	+\ln\frac{1+\beta_j}{1-\beta_j}} \nnb
\eea
and according to $(1+\beta_j\cos\theta)^{-1}$ with probability $P_j = 1-P_i$,
selected through a random number. These angles can be generated by
\bea
 \cos \theta 
& = & \frac{1}{\beta_i}\left[1-(1+\beta_i)\left(\frac{1-\beta_i}{1+\beta_i}\right)^\mathcal{R}\right]
 \nnb
\eea
in the former case and
\bea
 \cos \theta 
& = & -\frac{1}{\beta_j}\left[1-(1-\beta_j)\left(\frac{1+\beta_j}{1-\beta_j}\right)^\mathcal{R}\right]
 \nnb
\eea
in the latter. $\mathcal{R}$ again is a uniformly distributed random 
number on $[0,1]$.
The correction weight for obtaining the full distribution reads
\bea
W 
&=& 
\frac{-\frac{1-\beta_i^2}{(1-\beta_i \cos\theta)^2}
	+\frac{2(1+\beta_i\beta_j)}{(1-\beta_i \cos\theta)(1+\beta_j \cos\theta)}
	-\frac{1-\beta_j^2}{(1+\beta_j \cos\theta)^2}}
     {\frac{2(1+\beta_i\beta_j)}{(1-\beta_i \cos\theta)(1+\beta_j \cos\theta)}} \leq 1\,. \nnb
\eea
The azimutal angle $\varphi$ is distibuted uniformly.

\subsection{Photons from multipoles}
In a multipole configuration again the photons are generated according to 
$\tilde{S}_q(k)$. The integral over photon energies can still be seperated
from the angular integrations, decoupling the generation of the energy of 
a single photon as above. However, its angular distribution is very complex. 
But due to 
\bea
\tilde{S}_q(k) & = & \sum_{i<j} \tilde{S}(q_i,q_j,k) \nnb
\eea
the photon angles are distributed according 
to
\bea
 -\sum_{i<j}\left|Z_iZ_j\theta_i\theta_j\right|
	\left(\frac{q_i}{(q_i\cdot e_k)}-\frac{q_j}{(q_j\cdot e_k)}\right)^2\,. \nnb
\eea
This is nothing but a sum of angular distributions of different dipoles which 
are not in their respective rest frame.

Subsequently, one of those constituent dipoles is chosen with the probability
\bea
P_{ij} 
= \frac{\left|\bar{n}_{ij}\right|}{\sum\limits_{i<j}\left|\bar{n}_{ij}\right|}  \nnb
= \frac{\left|\int\frac{\done^3k}{k^0}\tilde{S}(q_i,q_j,k)\right|}{\sum\limits_{i<j}\left|\int\frac{\done^3k}{k^0}\tilde{S}(q_i,q_j,k)\right|}\,. \nnb
\eea
Then, photon angle generation can proceed as above in the rest frame of the 
dipole. To obtain the right distribution in the rest frame of the overall 
multipole, a null-vector of unit length is created in the rest frame of the 
dipole using the newly generated angles $\varphi$ ant $\theta$. Then this null 
vector is boosted into the rest frame of the multipole. It now 
has the angular distribution according to its constituent dipole in this 
frame. Since it is a null vector it has the properties of a photon and only 
needs to be rescaled to the energy generated earlier.

\section{Massive dipole splitting functions}\label{Appendix_dsf}

The massive dipole splitting functions are needed for the calculation of the 
approximation to the infrared subtracted single hard photon emission matrix 
element $\tilde{\beta}_1^1$. They are taken directly from 
\cite{Dittmaier:1999mb} for spin-$\frac{1}{2}$ emitters and are generalised 
from \cite{Catani:2002hc} for all other cases. Problems arising during this 
generalisation are related to the fact that these splitting functions for 
spin-1 particles are only given for massless gluons and that all initial 
states are considered massless as well. The extension to radiation off massive 
spin-1 particles is rather straight forward by augmentation with a simple mass 
term. The extension to massive initial states is less clear since decay matrix 
element are far off the massless initial state limit. However, the decaying 
particle is allways much more massive than its decay products when those are 
supposed to emit hard bremsstrahlung. Thus, photons are predominantly emitted 
at large angles to the initial state resulting in negligible contributions 
from these splitting functions. Hence, they can safely be omitted. 

Also, velocity factors from \cite{Catani:2002hc} have been omitted. 
They were introduced to facilitate the analytic integration and 
change neither the infrared nor the quasi-collinear limit. They only result in 
a different interpolation inbetween. The same is true for the factor $R_{ij}$ 
in the massive fermion splitting function of \cite{Dittmaier:1999mb}. 
Nonetheless, here this factor is kept because of the direct applicability of 
these splitting functions to the completely massive splitting. 

Three cases need to be differentiated regarding the state, initial or final, 
the emitter and spectator are in. The fourth case where both emitter and 
spectator are in the initial state lies outside the present applicability of 
this program, it will therefore be omitted.

To repeat the notation, $p_i$ is the 4-momtentum of the emitter, $p_j$ that of 
the spectator and $k$ is the emitted photon. All massive dipole splitting 
functions will be given, in that order, for spin-$0$, spin-$\tfrac{1}{2}$ and 
spin-$1$ emitters. Since there are no massive dipole splitting functions 
available for emitters of spin-$\frac{3}{2}$ or spin-$2$, their emissions 
have to be described by the soft limit only. Of course, it is allways possible 
to implement exact process specific matrix elements.

\subsection*{Final State Emitter, Final State Spectator}

\bea
 \mbox{\sl g}_{ij}(p_i,p_j,k) 
& = & \mbox{\sl g}_{ij}^{\mbox{\tiny (soft)}}(p_i,p_j,k) \nnb\\
& = & \frac{1}{(p_i\cdot k)R_{ij}(y_{ij})}
	\left[\frac{2}{1-z_{ij}(1-y_{ij})}-1-z_{ij}
					-\frac{m_i^2}{(p_i\cdot k)}\right] \nnb\\
& = & \frac{1}{(p_i\cdot k)}
	\left[\frac{2}{1-z_{ij}(1-y_{ij})}+\frac{2}{1-z_{kj}(1-y_{ij})}
					+2z_{ij}z_{kj}-4
					-\frac{m_i^2}{(p_i\cdot k)}\right] \nnb
\eea
with
\bea
y_{ij} & = & \frac{p_i k}{p_ip_j + p_i k + p_j k}  \nnb\\
z_{ij} & = & \frac{p_ip_j}{p_ip_j+p_jk}  \nnb\\
z_{kj} & = & 1-z_{ij}  \nnb\\
v_{ik,j} & = & \tfrac{1}{2}R_{ij}(y_{ij})\,
		\frac{\sqrt{\lambda(P_{ij}^2,m_i^2,m_j^2)}}{(p_i+k)\cdot p_j} \nnb
\eea
and
\bea
 R_{ij}(x) 
& = & \frac{\sqrt{\left(2m_j^2+\bar{P}_{ij}^2(1-x)\right)^2-4P_{ij}^2m_j^2}}
	{\sqrt{\lambda(P_{ij}^2,m_i^2,m_j^2)}}  \nnb
\eea
with
\bea
 P_{ij} 	& = & p_i+ p_j + k  \nnb\\
\bar{P}_{ij}^2 	& = & P_{ij}^2-m_i^2-m_j^2 
			= 2\left(p_ip_j+p_i k+p_j k\right) \nnb
\eea
wherein the photon is massless, $\lambda(x,y,z)$ is the Kallen-function.

\subsection*{Final State Emitter, Initial State Spectator}
\bea
 \mbox{\sl g}_{ij}(p_i,p_j,k) 
& = & \mbox{\sl g}_{ij}^{\mbox{\tiny (soft)}}(p_i,p_j,k) \nnb\\
& = & \frac{1}{(p_i\cdot k)x_{ij}}\left[\frac{2}{2-x_{ij}-z_{ij}}-1-z_{ij}
				-\frac{m_i^2}{(p_i\cdot k)}\right] \nnb\\
& = & \frac{1}{(p_i\cdot k)x_{ij}}\left[\frac{2}{2-x_{ij}-z_{ij}}
					+\frac{2}{2-x_{ij}-z_{kj}}
					+2z_{ij}z_{kj}-4
				-\frac{m_i^2}{(p_i\cdot k)}\right] \nnb
\eea
with
\bea
x_{ij} & = & \frac{p_ip_j + p_j k - p_i k}{p_ip_j + p_j k}  \nnb\\
z_{ij} & = & \frac{p_ip_j}{p_ip_j + p_j k}  \nnb\\
z_{kj} & = & 1-z_{ij} \nnb
\eea

\subsection*{Initial State Emitter, Final State Spectator}

The emitting particle is allways assumed to be much heavier than its decay 
products resulting in its contributions to the real emission corrections to 
be negligible. Thus, 
\bea
 \mbox{\sl g}_{ij}(p_i,p_j,k) & = & \mbox{\sl g}_{ij}^{\mbox{\tiny (soft)}}(p_i,p_j,k) \nnb
\eea
is set irrespective of the emitter's spin.

\section{Basic building blocks For matrix element calculations}
\label{App:XYZ-functions}

In this Appendix a short summary on the defintions of the basic building blocks 
(cf.\ \cite{Krauss:2001iv,Gleisberg:2003ue}) for the calculations of exact 
matrix elements will be given. Additionally, techniques to incorporate 
propagators into that scheme will be reviewed.

\subsection*{$X$-Function}

The $X$-function is a contraction over a ferimonic current coupled to a vector 
with an arbitrary structure of the vertex.
\bea
X\left(p_1,s_1;p;p_2,s_2;c_L,c_R\right) 
& = & \bar u(p_1,s_1)\;\dsl p\left[c_L P_L + c_R P_R\right] u(p_2,s_2)\,, \nnb
\eea
where $u(p_i,s_i)$ may be a particle or anti-particle spinor, 
$P_L=\frac{1-\gamma^5}{2}$ and $P_R=\frac{1+\gamma^5}{2}$. The vector $p^\mu$ 
dotted into the $\gamma$-matrix may be a momentum vector or a polarisation 
vector. For the explicite calculation of the $X$-Function see Table 
\ref{Table:X_Func}.

\begin{table}
 \begin{center}
  \begin{tabular}{|c|c|}
   \hline 
   $s_1 s_2$ & $X(p_1,s_1;p;p_2,s_2;c_L,c_R)$ \\
   \hline 
   $++$      & $\mu_1\mu_2\eta^2c_L + \mu^2\eta_1\eta_2c_R
		+ c_RS(+;p_1,p)S(-;p,p_2)$ \\
   $+-$      & $c_L\mu_1\eta S(+;p,p_2) + c_R\mu_2\eta S(+;p_1,p)$ \\
   \hline
  \end{tabular}
  \caption{$X$-Functions for different helicity combinations. Missing 
	   combinations can be obtained using the simultaneous replacements 
	   $+\leftrightarrow-$ and $L\leftrightarrow R$.\label{Table:X_Func}}
 \end{center}
\end{table}

\subsection*{$Y$-Function}

The $Y$-function is the pendant of the $X$-function when the fermionic current 
is coupling to a scalar rather than a vector.
\bea
Y\left(p_1,s_1;p_2,s_2;c_L,c_R\right) 
& = & \bar u(p_1,s_1)\left[c_L P_L + c_R P_R\right] u(p_2,s_2)\,. \nnb
\eea
Its explicit calculation is shown in Table \ref{Table:Y_Func}.

\begin{table}
 \begin{center}
  \begin{tabular}{|c|c|}
   \hline 
   $s_1 s_2$ & $Y(p_1,s_1;p_2,s_2;c_L,c_R)$ \\
   \hline 
   $++$      & $c_R\mu_1\eta_2 + c_L\mu_2\eta_1$ \\
   $+-$      & $c_L S(+;p_1,p_2)$ \\
   \hline
  \end{tabular}
  \caption{$Y$-Functions for different helicity combinations. Missing 
	   combinations can be obtained using the simultaneous replacements 
	   $+\leftrightarrow-$ and $L\leftrightarrow R$.\label{Table:Y_Func}}
 \end{center}
\end{table}

\subsection*{$Z$-Function}

The $Z$-function is a contraction over two ferionic currents connected by a 
massless gauge boson (cf.\ Table \ref{Table:Z_Func}).
\bea
\lefteqn{Z\left(p_1,s_1;p_2,s_2;p_3,s_3;p_4,s_4;c_L^{12},c_R^{12};c_L^{34},c_R^{34}\right)} \nnb\\
& = & \bar u(p_1,s_1)\gamma^\mu\left[c_L^{12} P_L + c_R^{12} P_R\right] u(p_2,s_2)
	\bar u(p_3,s_3)\gamma_\mu\left[c_L^{34} P_L + c_R^{34} P_R\right] u(p_4,s_4)\,. \nnb
\eea

\begin{table}
 \begin{center}
  \begin{tabular}{|c|c|}
   \hline 
   $s_1 s_2 s_3 s_4$ & $Z(p_1,s_1;p_2,s_2;p_3,s_3;p_4,s_4;c_L^{12},c_R^{12};c_L^{34},c_R^{34})$ \\
   \hline 
   $++++$ & $2\left[S(+;p_3,p_1)S(-;p_2,p_4)c_R^{12}c_R^{34}
			+\mu_1\mu_2\eta_3\eta_4c_L^{12}c_R^{34}
			+\mu_3\mu_4\eta_1\eta_2c_R^{12}c_L^{34}\right]$ \\
   $+++-$ & $2\eta_2c_R^{12}\left[S(+;p_1,p_4)\mu_3c_L^{34}
				  +S(+;p_1,p_3)\mu_4c_R^{34}\right]$ \\
   $++-+$ & $2\eta_1c_R^{12}\left[S(-;p_3,p_2)\mu_4c_L^{34}
				  +S(-;p_4,p_2)\mu_3c_R^{34}\right]$ \\
   $++--$ & $2\left[S(+;p_4,p_1)S(-;p_2,p_3)c_R^{12}c_L^{34}
			+\mu_1\mu_2\eta_3\eta_4c_L^{12}c_R^{34}
			+\mu_3\mu_4\eta_1\eta_2c_R^{12}c_R^{34}\right]$ \\
   $+-++$ & $2\eta_4c_R^{34}\left[S(+;p_1,p_3)\mu_2c_R^{12}
				  +S(+;p_2,p_3)\mu_1c_L^{12}\right]$ \\
   $+-+-$ & $0$ \\
   $+--+$ & $-2\left[\mu_1\mu_4\eta_2\eta_3c_L^{12}c_L^{34}
			+\mu_2\mu_3\eta_1\eta_4c_R^{12}c_R^{34}
			-\mu_1\mu_3\eta_2\eta_4c_L^{12}c_R^{34}
			-\mu_2\mu_4\eta_1\eta_3c_R^{12}c_L^{34}\right]$ \\
   $+---$ & $2\eta_3c_R^{34}\left[S(+;p_4,p_2)\mu_1c_L^{12}
				  +S(+;p_1,p_4)\mu_2c_R^{12}\right]$ \\
   \hline
  \end{tabular}
  \caption{$Z$-Functions for different helicity combinations. Missing 
	   combinations can be obtained using the simultaneous replacements 
	   $+\leftrightarrow-$ and $L\leftrightarrow R$.\label{Table:Z_Func}}
 \end{center}
\end{table}

\subsection*{$S$-Function}

For the calculation of the above spinoral products it is useful to define the 
$S$-Function
\bea
S(s;p_1,p_2) 
& = & \bar{u}(p_1,s) u(p_2,-s)\,. \nnb
\eea
Its two possible forms for given $p_1$ and $p_2$ are
\bea
S(+;p_1,p_2) 
& = & 2\,\frac{(p_1\cdot k_0)(p_2\cdot k_1)
		-(p_1\cdot k_1)(p_2\cdot k_0)
		-i\epsilon_{\alpha\beta\gamma\delta}
			p_1^\alpha p_2^\beta k_0^\gamma k_1^\delta}
		{\eta_1 \eta_2}\nnb\\
S(-;p_1,p_2)
& = & -2\,\frac{(p_1\cdot k_0)(p_2\cdot k_1)
		-(p_1\cdot k_1)(p_2\cdot k_0)
		+i\epsilon_{\alpha\beta\gamma\delta}
			p_1^\alpha p_2^\beta k_0^\gamma k_1^\delta}
		{\eta_1 \eta_2}\,, \nnb
\eea
where $k_0$ is an arbitrary null vector ($k_0^2 = 0$) and $k_1$ satisfies the 
relations $k_1^2 = -1$ and $(k_0\cdot k_1) = 0$. Furthermore,
\bea
\eta_i & = & \sqrt{2(p_i\cdot k_0)}\,.  \nnb
\eea
It is also useful to define the quantity 
\bea
\mu_i & = & \pm \frac{m_i}{\eta_i}\,, \nnb
\eea
where $\pm$ refers to particles/anti-particles.

\subsection*{Fermionic Propagators}

These propagators can be incorporated using the following identity:
\bea
(\,\dsl p\pm m) 
& = & \tfrac{1}{2}\sum_s\left[\left(1\pm\frac{m}{\sqrt{p^2}}\right)u(p,s)\bar u(p,s) 
	+ \left(1\mp\frac{m}{\sqrt{p^2}}\right)v(p,s)\bar v(p,s)\right]\,. \nnb
\eea 
This allows to cut the line and replace it with a sum of external particles

\subsection*{Bosonic Propagators}

Bosonic propagators can be incorporated by writing out their Lorentz-structure 
explicitely. This is trivial in Feynman gauge, if the vector is massless . 
Massive propagators are best included in unitary gauge, since then no 
additional goldstone boson exchange has to be included.

\end{appendix}
\nocite{Low:1958sn}
\nocite{Feynman:1949zx}
\nocite{Gribov:1966hs}
\nocite{Chaichian:1995kq}
\nocite{Marciano:1974vg}
\nocite{Berends:1987ab}
\nocite{Jadach:1987ik}
\nocite{Jadach:1987ii}
\nocite{Golonka:2005pn}


\bibliographystyle{amsunsrt_mod_m1}
\bibliography{References}

\end{document}